\documentclass[%
 reprint,
superscriptaddress,
nofootinbib,
 amsmath,amssymb,
 aps,
]{revtex4-2}

\usepackage{aps_macros}
\usepackage{graphicx}% Include figure files
\usepackage{subfigure}%
\usepackage{dcolumn}% Align table columns on decimal point
\usepackage{bm}% bold math
\usepackage{hyperref}% add hypertext capabilities
\usepackage{float}
\usepackage{threeparttable}% 
\usepackage{leftindex}

\hypersetup{
     colorlinks   = true,
     citecolor    = blue
}

\hypersetup{colorlinks=true, citecolor=blue, linkcolor = magenta}

\begin{document}

\preprint{APS/123-QED}

\title{Magnetized tori with magnetic polarization around Kerr black holes: variable angular momentum disks}

\author{Sergio Gimeno-Soler}
\email{sergiogimenosoler@ua.pt}
\affiliation{Departamento de Matem\'atica da Universidade de Aveiro and
Center for Research and Development in Mathematics and Applications (CIDMA),
Campus de Santiago, 3810-183 Aveiro, Portugal.}
\affiliation{Departamento de Astronom\'{\i}a y Astrof\'{\i}sica, Universitat de Val\`encia, Dr. Moliner 50, 46100, Burjassot (Val\`encia), Spain.}

\author{Oscar M. Pimentel}
\email{oscar.pimentel@correo.uis.edu.co}
\affiliation{Grupo de Investigación en Relatividad y Gravitación, Escuela de Física, Universidad Industrial de Santander, A. A. 678, Bucaramanga 680002, Colombia.}

\author{Fabio D. Lora-Clavijo}
\email{fadulora@uis.edu.co}
\affiliation{Grupo de Investigación en Relatividad y Gravitación, Escuela de Física, Universidad Industrial de Santander, A. A. 678, Bucaramanga 680002, Colombia.}
 
\author{Alejandro Cruz-Osorio}
\email{aosorio@astro.unam.mx}
\affiliation{Instituto de Astronom\'{\i}a, Universidad Nacional Aut\'onoma de M\'exico, AP 70-264, Ciudad de M\'exico 04510, M\'exico}
%\affiliation{Institut f{\"u}r Theoretische Physik, Max-von-Laue-Stra{\ss}e 1, 60438 Frankfurt, Germany.}

\author{Jos\'e A.~Font}
\email{j.antonio.font@uv.es}
\affiliation{Departamento de Astronom\'{\i}a y Astrof\'{\i}sica, Universitat de Val\`encia, Dr. Moliner 50, 46100, Burjassot (Val\`encia), Spain.}
\affiliation{Observatori Astron\`omic, Universitat de Val\`encia, C/ Catedr\'atico Jos\'e Beltr\'an 2, 46980, Paterna (Val\`encia), Spain.}

\date{\today}% It is always \today, today,

%%%%%%%%%%%%%%%%%%%%%%%%%%%%%%%%%%%%%%%%%%%%%%%%%%%%%%%%%%%%%%%%%%%%%%%%%%%%%%%%%%%%%%%%%%%
\begin{abstract}
Analytical models of magnetized, geometrically thick disks are relevant to understand the physical conditions of plasma around compact objects and to explore its emitting properties. This has become increasingly important in recent years in the light of the Event Horizon Telescope observations of Sgr A$^*$ and M87. Models of thick disks around black holes usually consider constant angular momentum distributions and do not take into account the magnetic response of the fluid to applied magnetic fields. We present a generalization of our previous work on stationary models of magnetized accretion disks with magnetic polarization~\citep{Pimentel:2018}. This extension is achieved by accounting for non-constant specific angular momentum profiles, done through a two-parameter ansatz for those distributions. We build a large number of new equilibrium solutions of thick disks with magnetic polarization around Kerr black holes, selecting suitable parameter values within the intrinsically substantial parameter space of the models. We study the morphology and the physical properties of those solutions, finding  qualitative changes with respect to the  constant angular momentum tori solutions~\cite{Pimentel:2018}. However, the dependences found on the angular momentum distribution or on the black hole spin do not seem to be strong.
Some of the new solutions, however, exhibit a local maximum of the magnetization function, absent in standard magnetized tori. Due to the enhanced development of the magneto-rotational instability as a result of  magnetic susceptibility, those models might be particularly well-suited to investigate jet formation through general-relativistic MHD simulations. The new equilibrium solutions reported here can be used as initial data in numerical codes to assess the impact of magnetic susceptibility in the dynamics and observational properties of black hole--thick disk systems.
\end{abstract}
%%%%%%%%%%%%%%%%%%%%%%%%%%%%%%%%%%%%%%%%%%%%%%%%%%%%%%%%%%%%%%%%%%%%%%%%%%%%%%%%%%%%%%%%%%%


\maketitle

%\tableofcontents

%%%%%%%%%%%%%%%%%%%%%%%%%%%%%%%%%%%%%%%%%%%%%%%%%%%%%%%%%%%%%%%%%%%%%%%%%%%%%%%%%%%%%%%%%%%
\section{\label{sec:Intro} Introduction}
%%%%%%%%%%%%%%%%%%%%%%%%%%%%%%%%%%%%%%%%%%%%%%%%%%%%%%%%%%%%%%%%%%%%%%%%%%%%%%%%%%%%%%%%%%%
The accretion of hot plasma onto black holes is of great interest in relativistic astrophysics. At galactic scales this process is associated with the growth of supermassive black holes (SMBHs) and their emission of relativistic jets and outflows, and, at stellar scales, with the appearance of short-lived, highly energetic phenomena such as X-ray transients and gamma-ray bursts (GRBs)~\citep{Blandford:1977,Rees:1984,Begelman:1984,Meier:2003,Meszaros:2006}. Observations at distances close to SMBHs have recently become possible thanks to the Event Horizon Telescope (EHT), a Very Long Baseline Interferometry (VLBI) array consisting of radio telescopes devoted to observing the immediate environment of a black hole with an angular resolution comparable to the compact object's event horizon. The EHT collaboration has imaged for the first time the SMBHs at the core of the galaxy M\,87 \citep{EHT_M87_PaperI} and at the galactic center, Sagittarius A* \citep{EHT_SgrA_PaperI}. These scenarios are perfect laboratories to test fundamental plasma and black hole physics, accretion processes, and general relativity in the strong field regime, as well as alternative theories of gravity \citep{EHT_M87_PaperV,EHT_M87_PaperVI,EHT_SgrA_PaperVI,Kocherlakota2021}. These recent developments are opening a new window to perform direct comparisons between theory and observations, further constraining existing and new models.

The accretion of matter onto a black hole leads to the efficient conversion of gravitational energy into heated plasma, particle acceleration and electromagnetic radiation~\citep{Frank-book}. If the black hole rotates, the available energy in the near-horizon region increases significantly.  In astrophysical accreting systems,
the gravitational energy is stored in a disk-shaped plasma which rotates around a compact object, typically a Kerr black hole. For a maximally rotating Kerr black hole, the binding energy per unit mass a test particle can reach is $0.42\,c^2$ and the innermost stable circular orbit shrinks to coincide with the horizon (in areal coordinates)~\citep{Bardeen:1972}. Accretion is regulated through the (outward) transport of angular momentum in the disk originated by dissipative processes triggered by turbulent viscous stresses. It has long been recognized that magnetic fields play a prominent role in the accretion process. Indeed, random (seed) perturbations in a magnetized disk can trigger the magneto-rotational instability (MRI), which destabilizes the delicate balance between magnetic and rotational forces and initiates the energy conversion process (see~\cite{Balbus:1998,Abramowicz:2013}). 

The shape and extension of accretion disks are determined by their angular momentum distribution and the pressure gradients. In particular, it is believed that geometrically thick disks (or tori) are present in active galactic nuclei and quasars, micro-quasars, X-ray binaries, and in systems leading to GRBs, either through black-hole-forming, collapsing massive stars or via mergers of compact binaries comprising at least one neutron star~\citep{Rees:1982,Begelman:1984,Andrew:1999,Abramowicz:2013,Just:2015}. Thick disks, which are the accretion model we adopt in this paper, have been used in diverse astrophysical contexts such as e.g.~the central engine of short GRBs and kilonovae~\citep{Troja:2010,Berger:2014,Metzger:2017}, in semi-analytic studies of super-Eddington accretion \citep{jiang2019super}, or in  calculations of low-amplitude quasi-periodic oscillations in X-ray binaries~\citep{Zanotti:2005,Blaes:2006,Montero:2007,Ingram:2019, Rezzolla:2013, deAvellar:2017}. Magnetized thick disks have also been used to compute images of SgrA* and to fit its millimetre and radio spectrum~\citep{Vincent:2015}. Within the geometrically thick disk formalism, models with a Keplerian angular momentum distribution are infinitely thin and have infinite extension, while disks with non-Keplerian profiles have finite thickness and their radial extent depends on the specific angular momentum distribution.

First numerical models of geometrically thick disks around black holes were developed by~\cite{Fishbone:1976,Abramowicz:1978,Kozlowski:1978}. Those models were constructed for both isentropic and barotropic matter distributions supported by pressure gradients and centrifugal forces. A constant specific angular momentum law was adopted for the material in the disk as the main simplifying assumption. 
These thick tori (also commonly referred to as ``Polish doughnuts'') are routinely used as initial states for general relativistic magneto-hydrodynamical (GRMHD) simulations to study their nonlinear stability and their dynamics in connection with investigations of MRI-driven turbulence, angular momentum transport and accretion (see~\cite{Abramowicz:2013,Porth2019} and references therein). Since the original thick disk models are purely hydrodynamical, a weak poloidal magnetic field needs to be superimposed on top of this initial state to seed the MRI. A self-consistent solution for a magnetized thick disk around a Kerr black hole was derived by~\cite{Komissarov:2006}, assuming a purely toroidal (i.e.~azimuthal) magnetic field. This disk model has been shown to be MRI-unstable under non-axisymmetric perturbations by~\cite{2015MNRAS.447.3593W}, being otherwise stable in axisymmetry. Dynamical differences between  Komissarov's self-consistent solution and a standard Polish doughnut dressed with an ``ad hoc'' magnetic field have been studied by \cite{Cruz-Osorio2020} through GRMHD simulations. Highly  magnetized disks were found to be unstable (and hence prone to be accreted or expelled) unless the initial data incorporate the magnetic field in a self-consistent way. 
More recently~\cite{liska2020large} have shown that a purely toroidal initial field can generate a large-scale poloidal flux, which is necessary to power jets.

In~\cite{Pimentel:2018} we extended the Komissarov solution to include the response of the fluid to an applied magnetic field, considering the relativistic magnetic properties of the plasma. This response, parameterized by the magnetic susceptibility, $\chi_m$, leads to equilibrium structures with different sizes and distinct magnetic behaviour.  It was found that paramagnetic disks ($\chi_m>0$) are smaller (and more compact) but more strongly magnetized near the black hole than the Komissarov solution ($\chi_m=0$), and even more than diamagnetic disks ($\chi_m<0$). Accounting for the magnetic susceptibility in Komissarov's solution leads to a number of additional features worth mentioning: (a) Numerical simulations of paramagnetic disks show that small magnetic stresses grow rapidly, especially near the black hole, indicating that MRI seems to be more effective to drive accretion in such disks ~\citep{pimentel2021magneto}. MRI generates turbulence that acts as an effective viscosity in the disk, producing the dissipation of energy necessary for the accretion of matter. (b) Models with a non-constant profile of the magnetic susceptibility, decreasing with radius, present a maximum of the magnetization parameter $\beta_{\mathrm{m}} = p/p_m$ (where $p$ is the thermal pressure and $p_m$ is the magnetic pressure) near the inner edge of the disk. Interestingly, this particular radial dependence of $\beta_{\mathrm{m}}$ seems to be consistent with results from numerical simulations \citep{shiokawa2011global,Porth2019, liska2020large}. (c) Magnetic-field amplification through the Kelvin-Helmholtz instability is more efficient and effective when the susceptibility of the matter is included in the model~\citep{pimentel2019linear}. (d) Finally,  magnetic polarization also induces effects on the physical properties of the plasma that impact the radiation from the disks~\citep{velasquez2022synchrotron}.

The susceptibility values used in this work are motivated by the theory of Langevin, in which, for an electron gas, the paramagnetism is associated with the intrinsic magnetic moment of electrons and the alignment of the spins due to magnetic torques. In contrast, the diamagnetism is related to the electron orbital motion around the magnetic field lines. For a detailed description, see our recent paper \cite{velasquez2022synchrotron}. Based on this work, 
diamagnetism depends only on particle number. Hence, for hydrogen gas, we calculate the susceptibility in a range from densities in accretion disks ($10^{5} [kg/m^3] - 10^{7} [kg/m^3]$, \cite{cold_plasma})  to densities in neutron star densities ($10^{19} [kg/m^3]$, \cite{neutron-star}). For accretion disks, the scales for magnetic susceptibility are of the order of $10^{-3} - 10^{-1}$, while for more compact objects like neutron stars, the theory would give values of the order of $10^{9} - 10^{11}$. These values for Neutron stars may not be realistic. However, it is important to note that there are alternative theories, such as Landau diamagnetism and Pauli paramagnetism and that, in the specific case of neutron stars, other theories may also need to be considered, including the possibility of a ferromagnetic phase transition. On the other hand, density, temperature, and magnetic field affect the magnetic susceptibility of paramagnetic materials.  For accretion disks, temperatures for cool accretion in some astrophysical scenarios are in the range of $10^{2}[K] - 10^{4} [K]$, while hot gas can reach temperatures of $10^7[K] - 10^9 [K]$ \cite{cold_plasma}. For example, it is believed that the nucleus of the X-ray source Cygnus X-1 is a stellar-mass black hole surrounded by a gas with a temperature of $\sim 10^4[K]$. Furthermore, we show the influence of the magnetic field considering scales from $10^{-2}[T] - 10^{-1} [T]$. 
%Magnetic fields of these orders of magnitude may start the accretion process, for example, O-type stars feeding $SgrA*$ or in X-ray binaries systems \citep{mag_field}. 
We found that the magnetic field does not significantly affect the magnitude of the magnetic susceptibility in the range between $10^{-2}[T]$ and $10[T]$, but amplifies the range of temperatures by which it is possible to calculate the magnetic susceptibility. Nevertheless, if the magnetic field is stronger, for example in the range of $10^{2} [T]- 10^{3} [T]$, the magnetic field modifies the magnitude of the susceptibility. In general, for low temperatures, it is possible to obtain susceptibilities in order of $10^{1} - 10^{-1}$, but for high temperatures and densities we found orders of $10^{-1} - 10^{-2}$.

One of the main simplifications of the magnetically polarized disk solutions obtained by~\cite{Pimentel:2018} is that they assume a constant distribution of specific angular momentum. While this choice facilitates the computation of the models it is, however, an academic choice. Moreover, this choice may be even unsuitable to describe systems where the black hole mass grows through accretion due to the appearance of a runaway instability on dynamical time scales, as shown by~\cite{Font:2002} for unmagnetized disks. A possible solution to this problem is to include a non-constant distribution of angular momentum by e.g.~considering that the angular momentum in the equatorial plane increases with the radial distance as a positive power law. This was shown by~\cite{Font:2002b,Daigne:2004} to have a highly stabilizing effect. Another possibility was presented by~\cite{Qian:2009} who combined different distributions of angular momentum in the disks (Keplerian and constant) to build sequences of purely hydrodynamical thick disks. Magnetized thick disk models with non-constant angular momentum distributions have been obtained by~\cite{2015MNRAS.447.3593W} and~\cite{2017A&A...607A..68G}. On the one hand, \cite{2015MNRAS.447.3593W} extended
Komissarov’s original solution   for  the  particular  case  of a power-law  distribution  of  angular  momentum. On the other hand, \cite{2017A&A...607A..68G} combined the two approaches for the angular momentum distribution considered in~\cite{Komissarov:2006} and~\cite{Qian:2009} to build magnetized Polish doughnuts around Kerr black holes.  Additional stationary models of magnetized thick disks have been constructed for hairy black hole spacetimes~\citep{Gimeno-Soler_etal_2019,Gimeno-Soler_etal:2021}, a Yukawa black hole potential~\citep{Cruz-Osorio:2021}, viscous disks~\citep{Sayantani:2021} and self-gravitating disks~\citep{Shibata:2007,Mach:2019}.

The present paper aims to build new sequences of magnetically polarized disk solutions for non-constant angular momentum distributions. To this aim, we merge the approach laid out in~\cite{Pimentel:2018} with that of~\cite{Gimeno-Soler_etal:2021} and discuss the modifications that the new rotation law introduces on the properties of stationary magnetically polarized disks. The paper is organized as follows: Section~\ref{framework} presents the mathematical and computational framework used to construct the models. In particular, we discuss the GRMHD equations with magnetic polarization, the angular momentum distribution used in our models as well as the space of the parameters we span in our study. The models and their main properties are discussed in Section~\ref{results}. Finally, Section~\ref{conclusions} summarizes our conclusions. The paper closes with one appendix where the conditions for the appearance of a local maximum in the magnetization parameter are investigated. Geometrized units ($G = c = 1$) are used throughout the paper. Additionally, this choice is complemented by $\mu_0 = \epsilon_0 = 1$ (i.e. Heaviside-Lorentz units).

%=============================================
\section{Framework}
\label{framework}
%=============================================

\subsection{\label{sec:equations} General relativistic magnetohydrodynamics equations with magnetic polarization}

\begin{figure}
\centering
\includegraphics[scale=0.20]{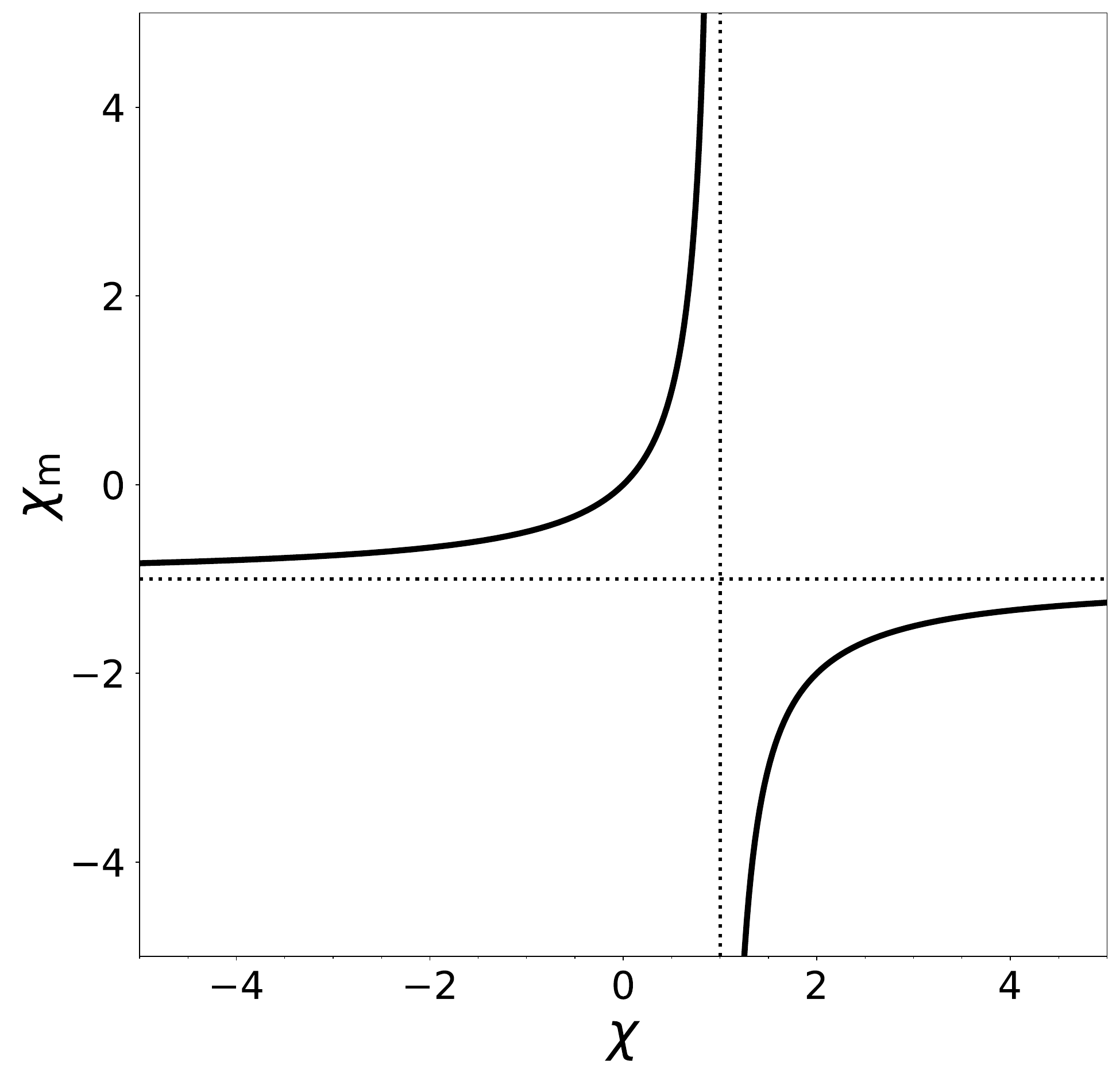}
\caption{Dependence of the magnetic susceptibility $\chi_{\mathrm{m}}$ as a function of $\chi$.}
\label{chim_chi}
\end{figure}

The dynamics of an ideal fluid with magnetic polarization in a magnetic field is described by the  conservation laws
\begin{eqnarray}
\nabla_{\mu}T^{\mu\nu}=0,\label{momentum_conservation}\\
\nabla_{\mu}(\rho u^{\mu})=0,\label{mass_conservation}
\end{eqnarray}  
and by the relevant Maxwell equations in the ideal GRMHD limit,
\begin{equation}
\nabla_{\mu} \leftindex^*F^{\mu\nu}= \nabla_{\mu}(u^{\mu}b^{\nu}-b^{\mu}u^{\nu})=0\,.
\label{relevant_maxwell}
\end{equation}
In these equations $T^{\mu\nu}$ is the energy-momentum tensor, $\rho$ is the rest-mass density, $\leftindex^*F^{\mu\nu}$ is the dual of the Faraday tensor, and $b^{\mu}$ is the magnetic field measured in a reference frame that moves with the same four-velocity of the fluid, $u^{\mu}$. The magnetic polarization can be characterized macroscopically through the magnetization vector $m^{\mu}$, which is defined as the magnetic dipole moment per unit volume. From now on, we will concentrate on the physically important case in which $m^{\mu}$ and $b^{\mu}$ are related by means of a linear constitutive equation, $m^{\mu}=\chi b^{\mu}$, where $\chi=\chi_m/(1+\chi_m)$, and $\chi_m$ is the magnetic susceptibility\footnote{In our units, $\mu_r = 1+\chi_m$ is the relative magnetic permeability. Then, $\chi = \chi_m/\mu_r$. It is relevant to note that the magnetic field $b^{\mu}$ also includes the magnetization.} (see Fig.~\ref{chim_chi}). When $\chi_m <0$ the fluid is diamagnetic and when $\chi_m >0$ the fluid is paramagnetic. In the first case the polarization results from induced orbital dipole moments in a magnetic field \citep{griffiths2005introduction}, and in the second case the polarization is generated by magnetic torques in substances whose atoms have a non-zero spin dipole moment.

The total energy-momentum tensor, $T^{\mu\nu}$, for a magnetically polarized fluid was computed in \cite{Maugin:1978tu} and more recently in \cite{2015MNRAS.447.3785C} following a different approach. The resulting tensor takes the form
\begin{eqnarray}
T^{\mu\nu}&=&\left[w+b^{2}(1-\chi)\right]u^{\mu}u^{\nu}+\left[p+\frac{1}{2}b^{2}(1-2\chi)\right]g^{\mu\nu}\nonumber\\
&&-(1-\chi)b^{\mu}b^{\nu},
\label{tensor_general}
\end{eqnarray}
in the linear media approximation. Here, $w \equiv \rho h$ is the enthalpy density, $h$ is the specific enthalpy, $p$ is the thermodynamic pressure, $b^{2}=b^{\mu}b_{\mu}$, and $g^{\mu\nu}$ is the metric tensor.

Following previous works \citep{Komissarov:2006, 2015MNRAS.447.3593W, 2017A&A...607A..68G, Pimentel:2018}, we assume the test-fluid approximation (i.e.~neglect the fluid's self-gravity) and the gravitational field as given by the Kerr metric in Boyer-Lindquist coordinates $(t,R,\theta,\phi)$. We also consider that the fluid is axisymmetric and stationary, so the physical variables depend neither on the azimuthal angle $\phi$ nor on the time $t$. Finally, we restrict the movement of the fluid in such a way that $u^{R}=u^{\theta}=0$, and the magnetic field topology to a purely toroidal one, so $b^{R}=b^{\theta}=0$. With these assumptions, the equation for baryon number conservation (\ref{mass_conservation}) and the relevant Maxwell equations (\ref{relevant_maxwell}) are identically satisfied, and the equilibrium structure of the tori is given by the Euler equation as follows 
\begin{equation}
(\ln|u_{t}|)_{,i}-\frac{\Omega}{1-l\Omega}l_{,i}+\frac{p_{,i}}{w}-\frac{(\chi p_m)_{,i}}{w}+\frac{[(1-\chi)\mathcal{L} p_m]_{,i}}{\mathcal{L}w}=0,
\label{eulerv2}
\end{equation}
where $p_{m}=b^{2}/2$ and $\mathcal{L}=g_{t\phi}^{2}-g_{tt}g_{\phi\phi}$.
In this last expression 
\begin{equation}
\Omega=\frac{u^{\phi}}{u^{t}}=-\frac{g_{\phi t}+lg_{tt}}{g_{\phi\phi}+lg_{t\phi}},
\label{omega}
\end{equation} 
and 
\begin{equation}
l=-\frac{u_{\phi}}{u_{t}}=-\frac{g_{\phi t}+\Omega g_{\phi\phi}}{g_{tt}+\Omega g_{t\phi}},
\label{angular}
\end{equation}
correspond to the angular velocity and specific angular momentum, respectively. As it can be seen, equation (\ref{eulerv2}) reduces to the one obtained by \cite{Komissarov:2006} when $\chi=0$.

Following the procedure presented in \cite{Komissarov:2006} and assuming that $\chi = \chi(\mathcal{L})$, Euler's equation can be solved in the form 
\begin{equation}
\ln|u_{t}|+\int_{0}^{p}\frac{dp}{w}-\int_{0}^{l}\frac{\Omega dl}{1-l\Omega}+(1-2\chi)\frac{\eta}{\eta-1}\frac{p_{m}}{w}=\text{const.},
\label{euler_exact_final}
\end{equation} 
where, for the particular case where $\chi$ takes the form~\citep{Pimentel:2018}
\begin{equation}\label{eq:chi}
    \chi = \chi_{0}+\chi_{1}\mathcal{L}^{\sigma}\,
\end{equation}
the magnetic pressure can be expressed as follows
\begin{equation}
p_{m}=K_{m}\mathcal{L}^{\tilde{\lambda}}w^{\eta}\tilde{f},
\label{mag_pres}
\end{equation}
with,
\begin{equation}
\tilde{\lambda}=\frac{1-\chi_{0}}{1-2\chi_0}(\eta-1),\hspace{5mm}\tilde{f}=(1-2\chi)^{\textstyle{\frac{1-\eta}{2\sigma(1-2\chi_{0})}-1}},
\label{eq:functions_case2}
\end{equation}
here $\chi_{0}$, $\chi_{1}$, and $\sigma$ are constants and $\eta$ is the exponent of the magnetic EoS in Komissarov's model.

In the following we depart from the procedure followed by~\citet{Pimentel:2018} and introduce the same equation of state as it was done in~\citet{Montero:2007} and~\citet{Gimeno-Soler_etal_2019},
\begin{equation}
p = K \rho^{\Gamma}\,
\end{equation}
where $K$ and $\Gamma$ are constants. Integrating Eq.~\eqref{euler_exact_final} as in~\citet{Pimentel:2018}, we arrive at the final equation we will need to solve
\begin{widetext}
\begin{equation}\label{eq:final_eq}
W - W_{\mathrm{in}} +\ln\left(1 + \frac{\Gamma K}{\Gamma-1}\rho^{\Gamma-1}\right) + (1-2\chi)\frac{\eta}{\eta-1}K_{\mathrm{m}}\mathcal{L}^{\tilde{\lambda}}(\rho h)^{\eta-1}\tilde{f}= 0\,,
\end{equation}
\end{widetext}
where $W - W_{\mathrm{in}}$ is the relativistic (gravitational + centrifugal) potential and is defined as
\begin{equation}
W(R, \theta) - W_{\mathrm{in}} = \ln|u_t| - \ln|u_{t,\mathrm{in}}| - \int^{l}_{l_{\mathrm{in}}}\frac{\Omega \mathop{dl}}{1-\Omega l},
\end{equation}
where the subscript ``in'' means that the quantity is evaluated at the inner edge of the disk $R_{\mathrm{in}}$.

\subsection{\label{sec:ang_mom}Angular momentum distribution}

Next, we turn to describe the specific angular momentum distribution $l(R, \theta)$ we use in this work. We will follow the procedure described in~\citet{Gimeno-Soler_etal:2021} (see also references therein for the full description) in which the specific angular momentum distribution at the equatorial plane, $l(R, \pi/2)$, is defined as
\begin{equation}
l \left(R,\frac{\pi}{2}\right) = \left\{ \label{eq:ansatz} 
  \begin{array}{lr}
    l_0 \left(\frac{l_{\mathrm{K}}(R)}{l_{\mathrm{K}}(R_{\mathrm{ms}})}\right)^{\alpha} &  \text{for } R \geq R_{\mathrm{ms}}\\
    l_0 & \text{for } R < R_{\mathrm{ms}}
  \end{array}
\right.
\end{equation}
where $l_{\mathrm{K}}(R)$ is the Keplerian angular momentum function, $R_{\mathrm{ms}}$ is the radius of the innermost (marginally) stable circular orbit (ISCO), $l_0$ is the parameter that controls the constant part of the distribution, and $\alpha$ changes the slope of the angular momentum profile (from $\alpha = 0$ for a constant profile to $\alpha = 1$ for a Keplerian profile).

The specific angular momentum outside the equatorial plane is computed using the so-called von Zeipel cylinders (see~\cite{Daigne:2004} and~\cite{Gimeno-Soler_etal:2021} for a detailed explanation of the exact procedure\footnote{See also~\cite{Cassing:2023} for a different approach to the computation of the von Zeipel cylinders applied to non-magnetized disks around parametrized spherically symmetric black holes.} ) where, once the specific angular momentum distribution at the equatorial plane $l_{\mathrm{eq}}(R) = l(R, \pi/2)$ is prescribed, a value of $l$ is assigned to each point outside the equatorial plane by solving the following equation
\begin{eqnarray}\label{eq:vonzeipel}
[g_{tt}(R, \theta)\tilde{g}_{t\phi}(R_0)-\tilde{g}_{tt}(R_0)g_{t\phi}(R, \theta)]l^2_{\mathrm{eq}}(R_0)
\nonumber \\ 
+[g_{tt}(R, \theta)\tilde{g}_{\phi\phi}(R_0)-\tilde{g}_{tt}(R_0)g_{\phi\phi}(R, \theta)]l_{\mathrm{eq}}(R_0)
\nonumber \\ 
+[g_{t\phi}(R, \theta)\tilde{g}_{\phi\phi}(R_0)-\tilde{g}_{t\phi}(R_0)g_{\phi\phi}(R, \theta)] = 0\,,
\end{eqnarray}
where the tilde is a short-hand notation for metric components evaluated at the equatorial plane. The von Zeipel cylinders are surfaces of constant angular velocity $\Omega$ and constant angular momentum $l$, so if a cylinder passing through a generic point $(R, \theta)$ also passes through a point $(R_0, \pi/2)$ at the equatorial plane, then $l(R, \theta) = l_{\mathrm{eq}}(R_0)$ and the potential at said point can be computed as
\begin{equation}
W(R, \theta) = W_{\mathrm{eq}}(R_0) + \ln\left[\frac{-u_t(R, \theta)}{-u_t(R_0, \pi/2)}\right]\,.
\end{equation}

\begin{figure*}
\centering
\includegraphics[scale=0.20]{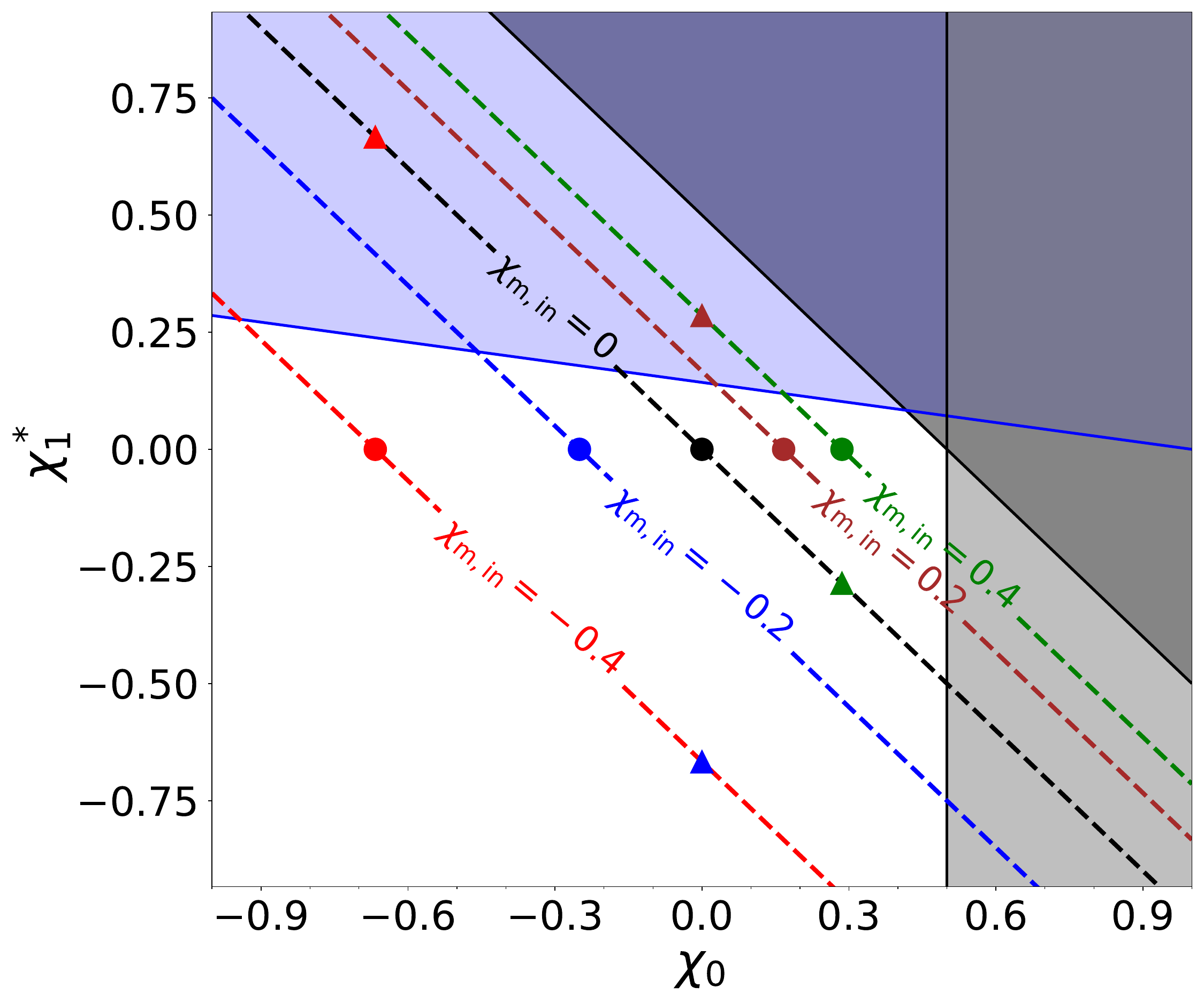}
\hspace{-0.1cm}
\includegraphics[scale=0.20]{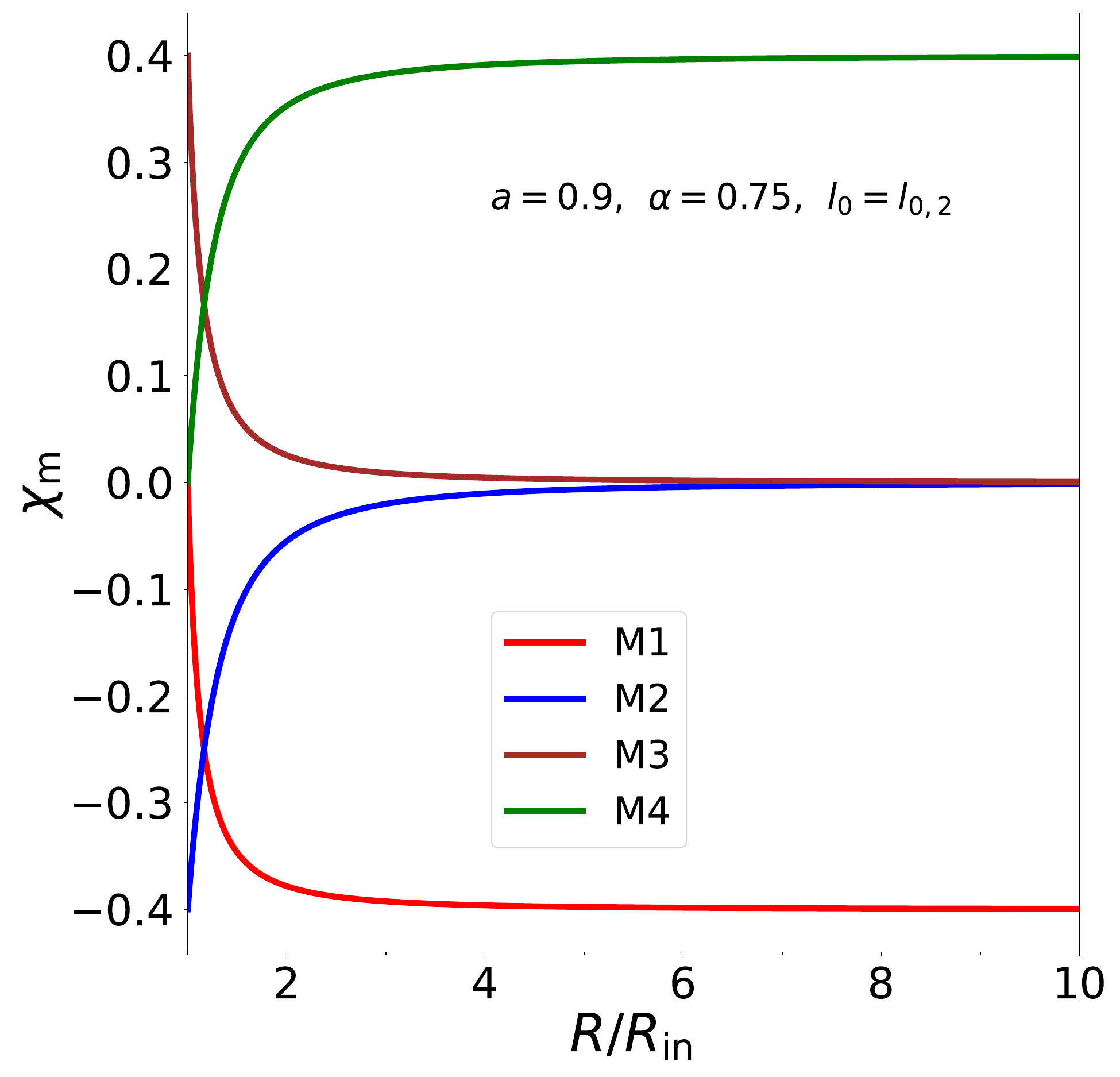}
\caption{Left panel: existence diagram for our different magnetic susceptibility models evaluated at $R = R_{\mathrm{in}}$. In red, blue, brown and green circles we show the constant models, namely C1, C2, C3 and C4. In the same colors, but in triangles, we show the non-constant susceptibility models, namely M1, M2, M3 and M4. The black circle marks the model with $\chi_{\mathrm{m}} = 0$. The grey shaded regions mark the forbidden regions of the parameter space (see discussions in Section~\ref{magsus_subsubsection}). The blue-shaded area represents the region of the parameter space that allows the existence of a maximum of the magnetization function $\beta_{\mathrm{m}}(R)$ for some value of the radial coordinate $R$ (see Appendix~\ref{partial_beta_0}) and the dashed lines represent the values of the parameters $\chi_0$ and $\chi_1^*$ that yield the same value of the magnetic susceptibility $\chi_{\mathrm{m}}$. Right panel: Radial profiles of $\chi_{\mathrm{m}}(R)$ for the non-constant susceptibility models we have considered for $a = 0.9$, $\alpha = 0.75$ and $l_0 = l_{0, 2}$ The $\chi_{\mathrm{m}}$ radial profiles are similar for all the values of $(a, \alpha, l_0)$ we have considered.}
\label{existence}
\end{figure*}

\begin{figure*}
\centering
\includegraphics[scale=0.42]{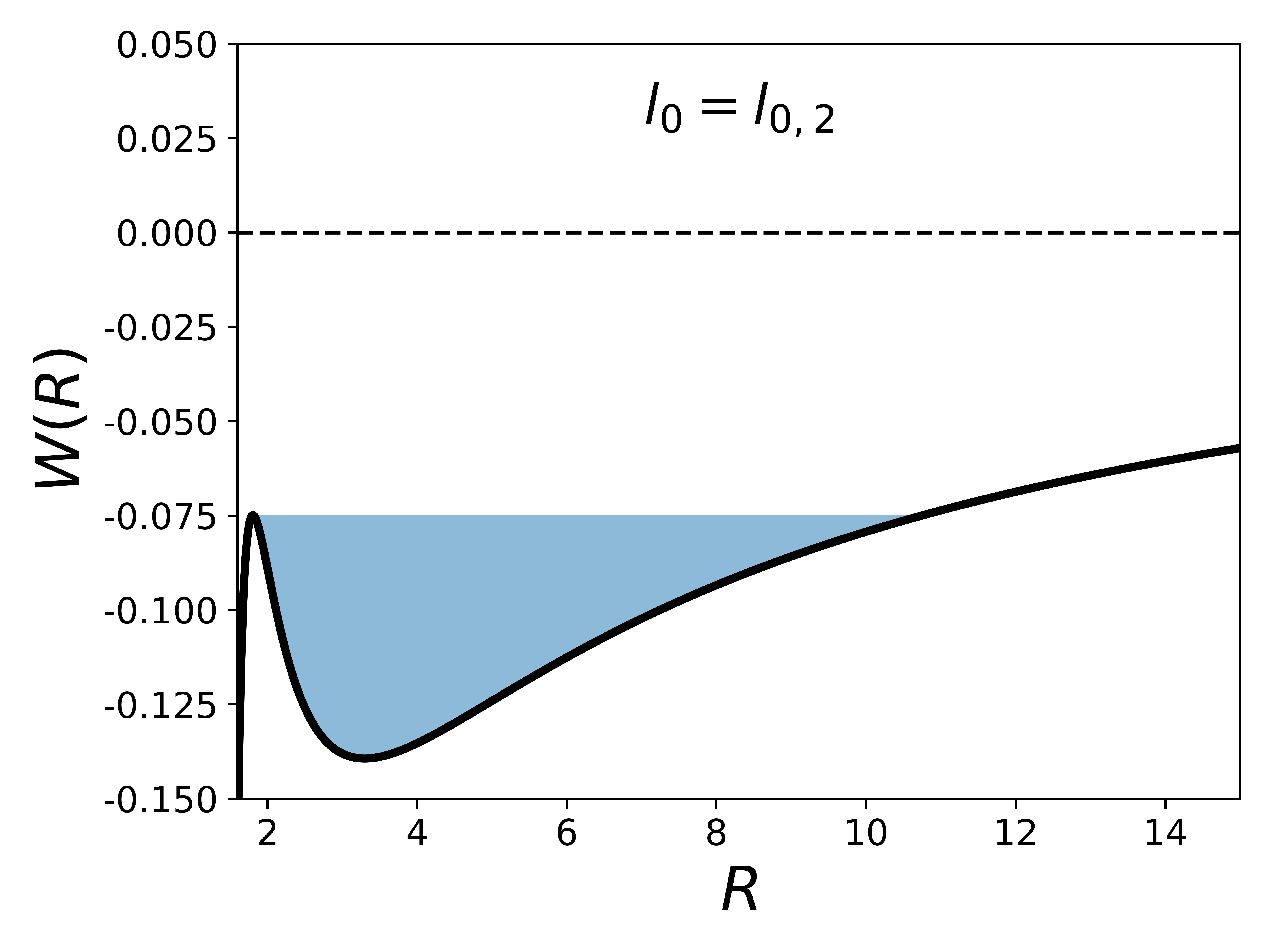}
%\hspace{-0.39cm}
\hspace{-0.2cm}
\includegraphics[scale=0.42]{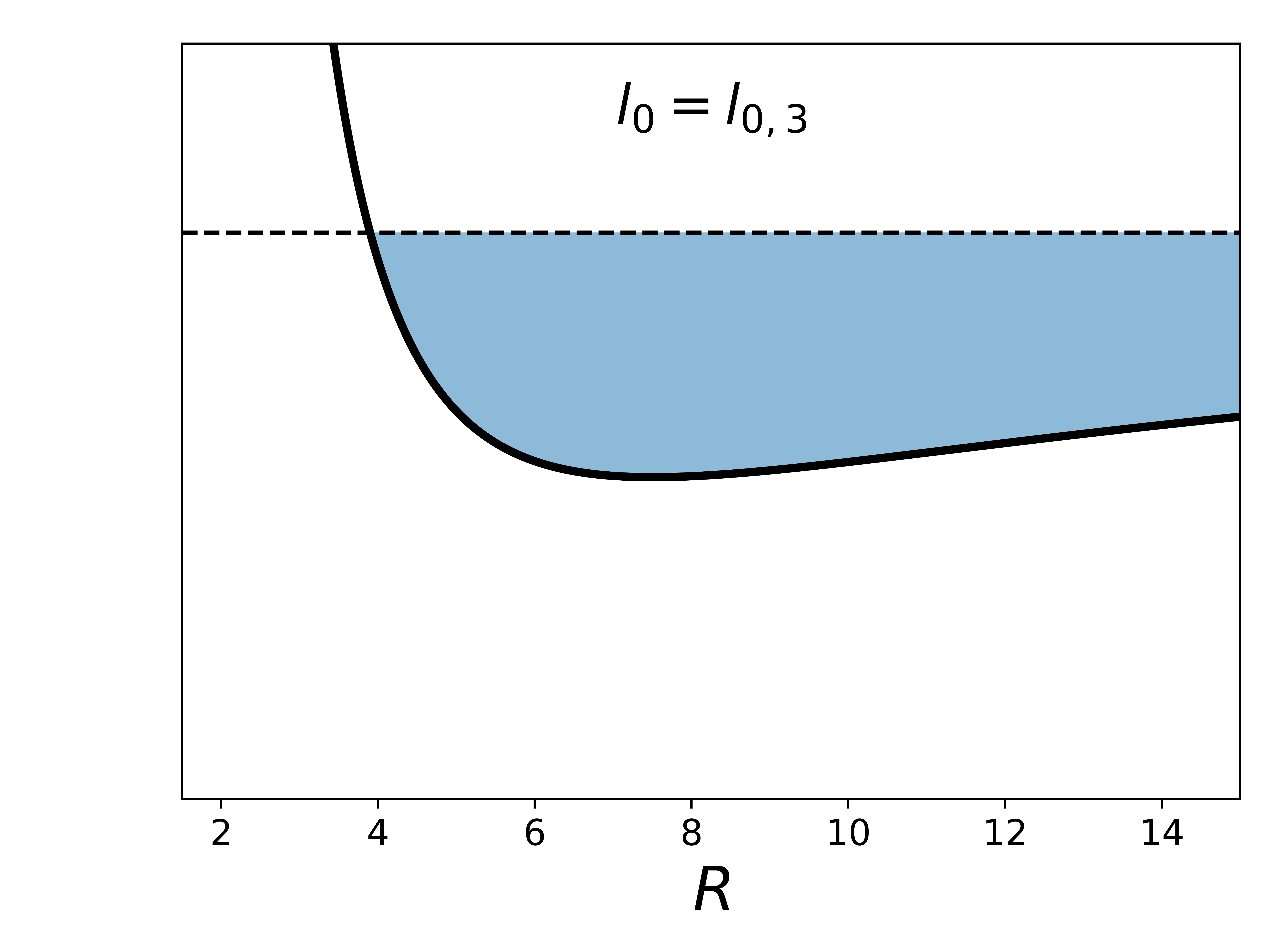}
\caption{Radial profiles of the potential $W(R)$ at the equatorial plane for the two models of the constant part of the angular momentum distribution $l_{0, 2}$ and $l_{0, 3}$, in this case for $a = 0.9$ and $\alpha = 0$. The blue-shaded region shows the region of the potential well that our disks fill. It must be noted that the depth of the potential well $\Delta W_{\mathrm{c}} (\equiv W_{\mathrm{in}} - W_{\mathrm{c}})$ is the same for both models.}
\label{angular_momentum_models}
\end{figure*}

\begin{figure*}
\includegraphics[scale=0.1]{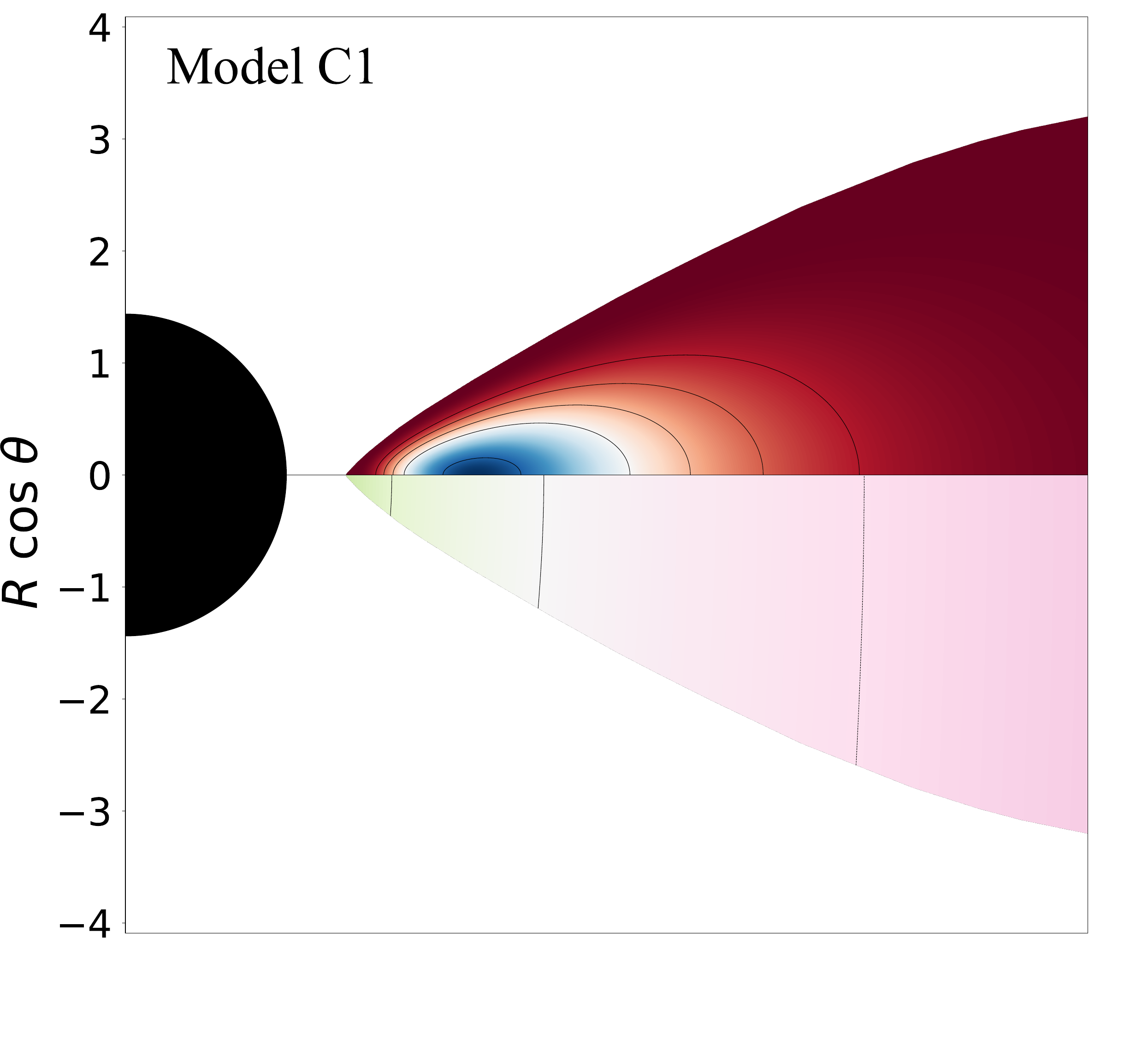}
\hspace{-0.9cm}
\includegraphics[scale=0.1]{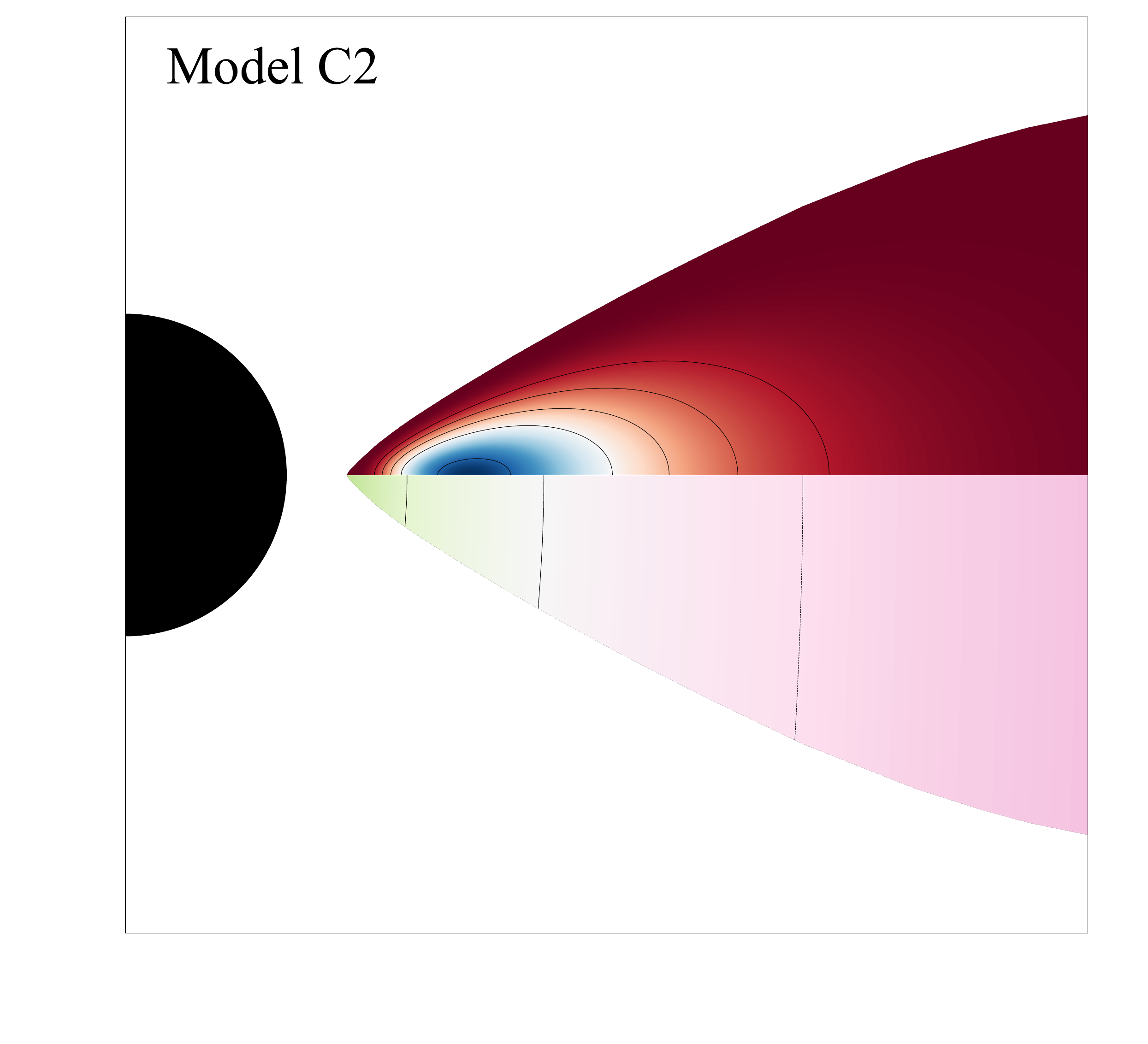}
\hspace{-0.9cm}
\includegraphics[scale=0.1]{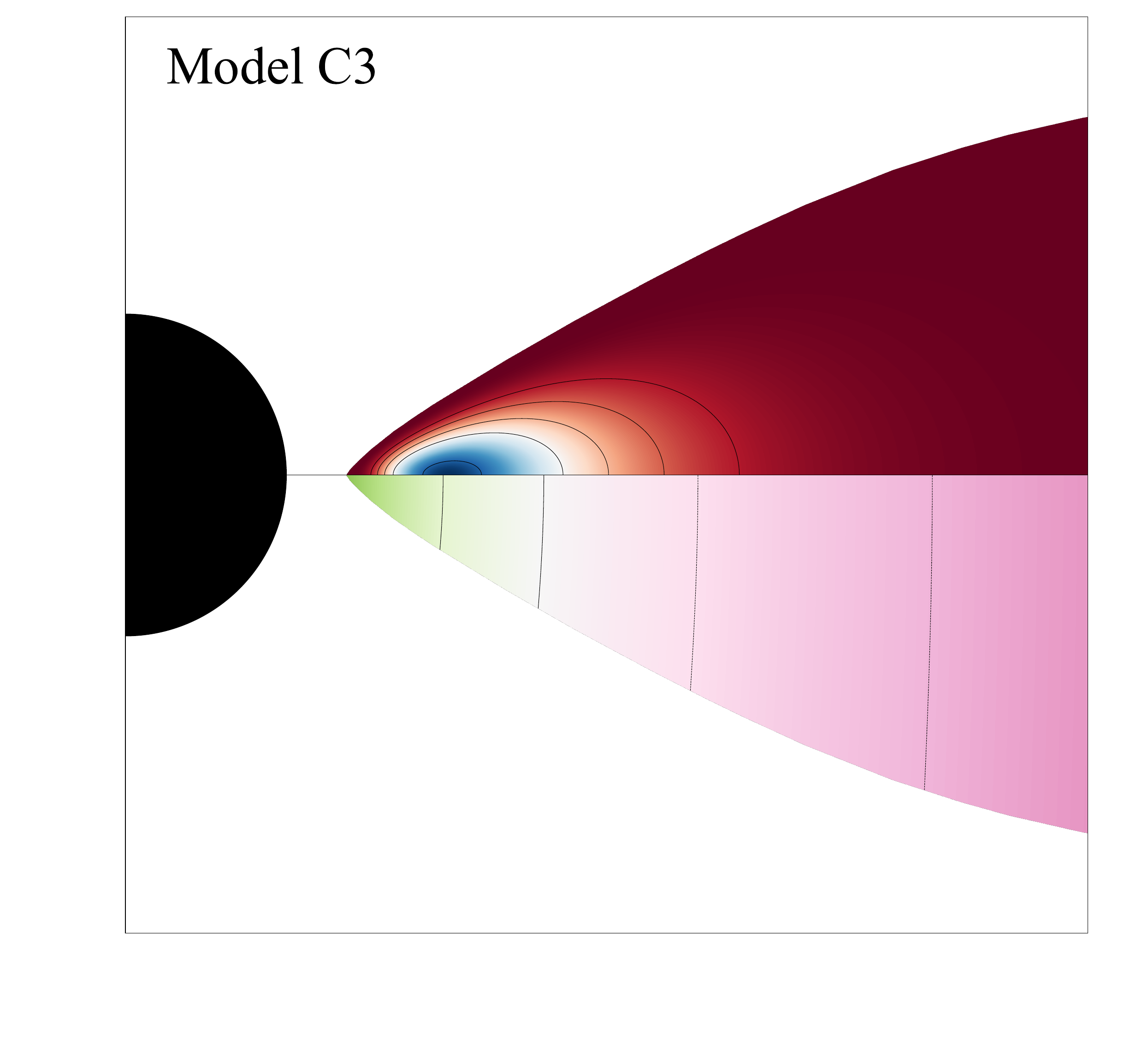}
\hspace{-1.135cm}
\includegraphics[scale=0.1]{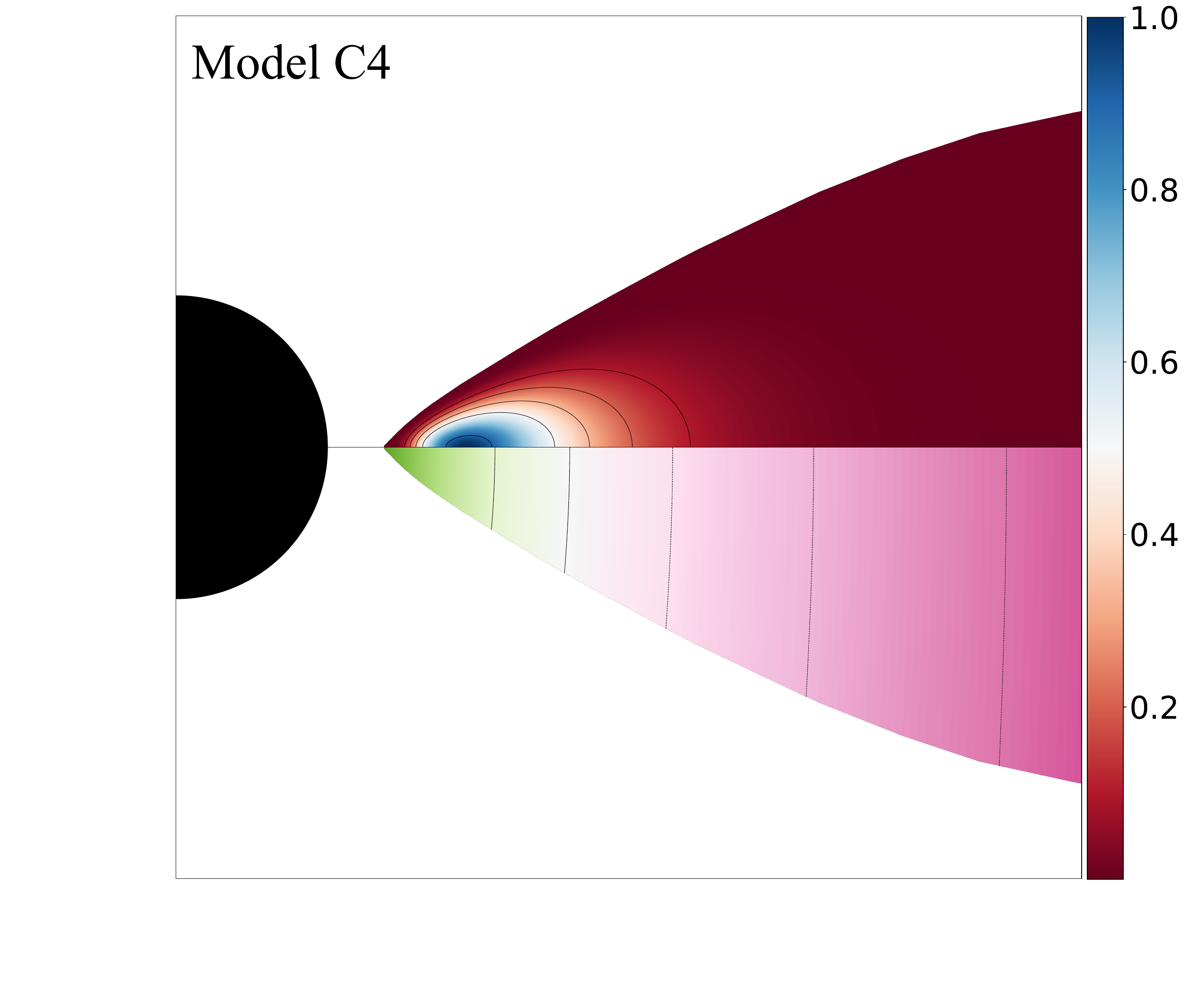}
\vspace{-0.3cm}
\\
\hspace{-0.2cm}
\includegraphics[scale=0.1]{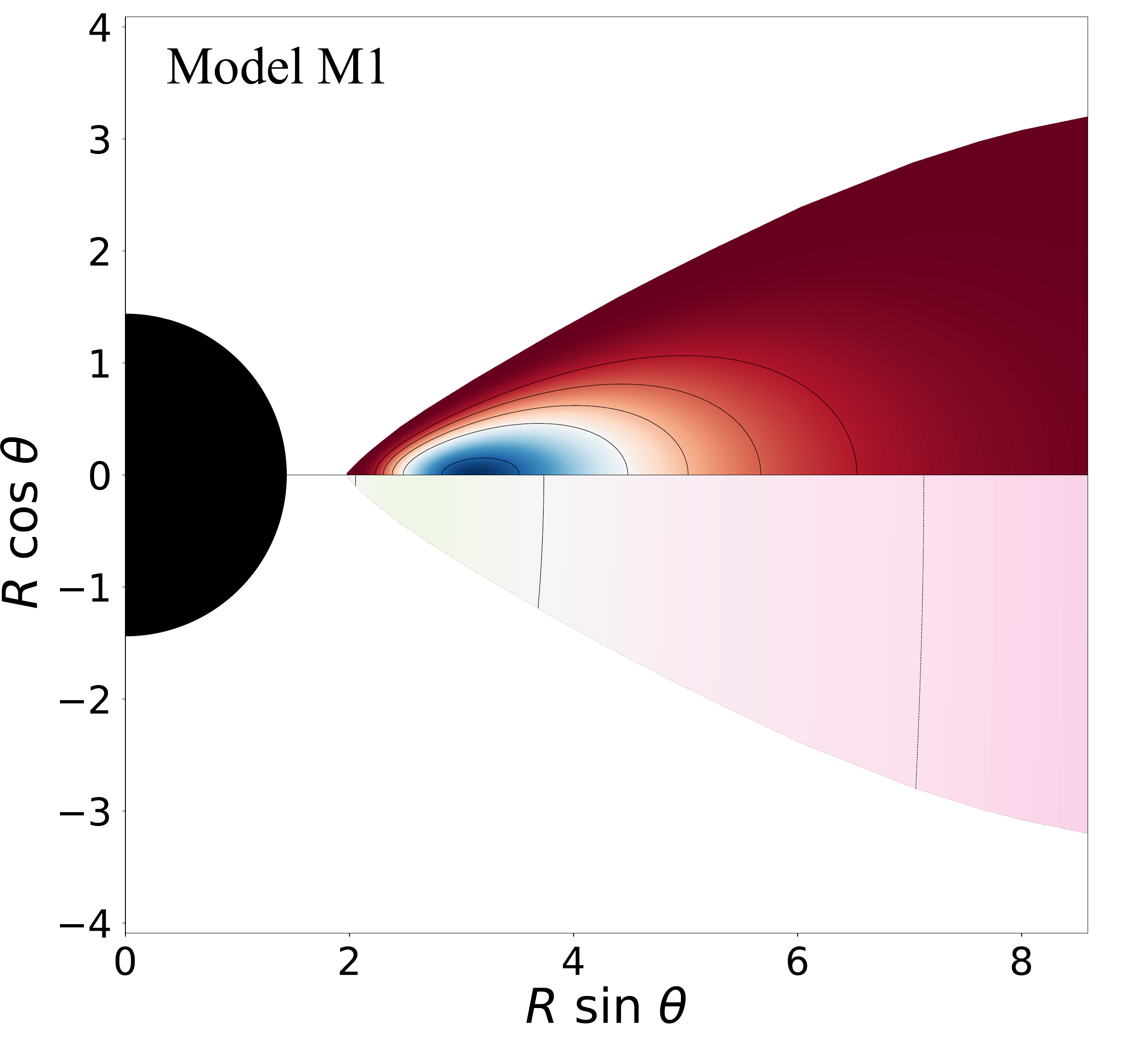}
\hspace{-0.9cm}
\includegraphics[scale=0.1]{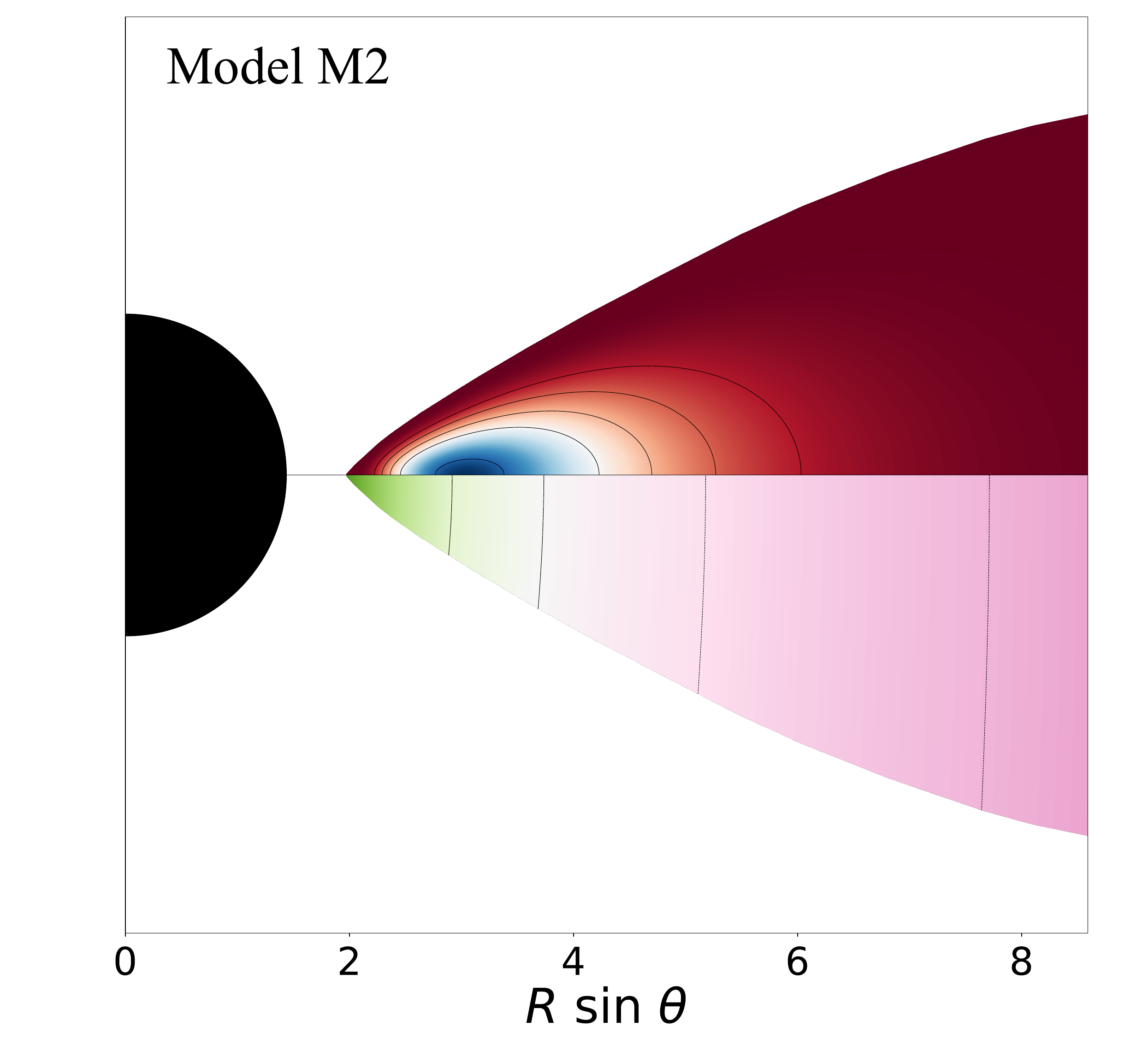}
\hspace{-0.9cm}
\includegraphics[scale=0.1]{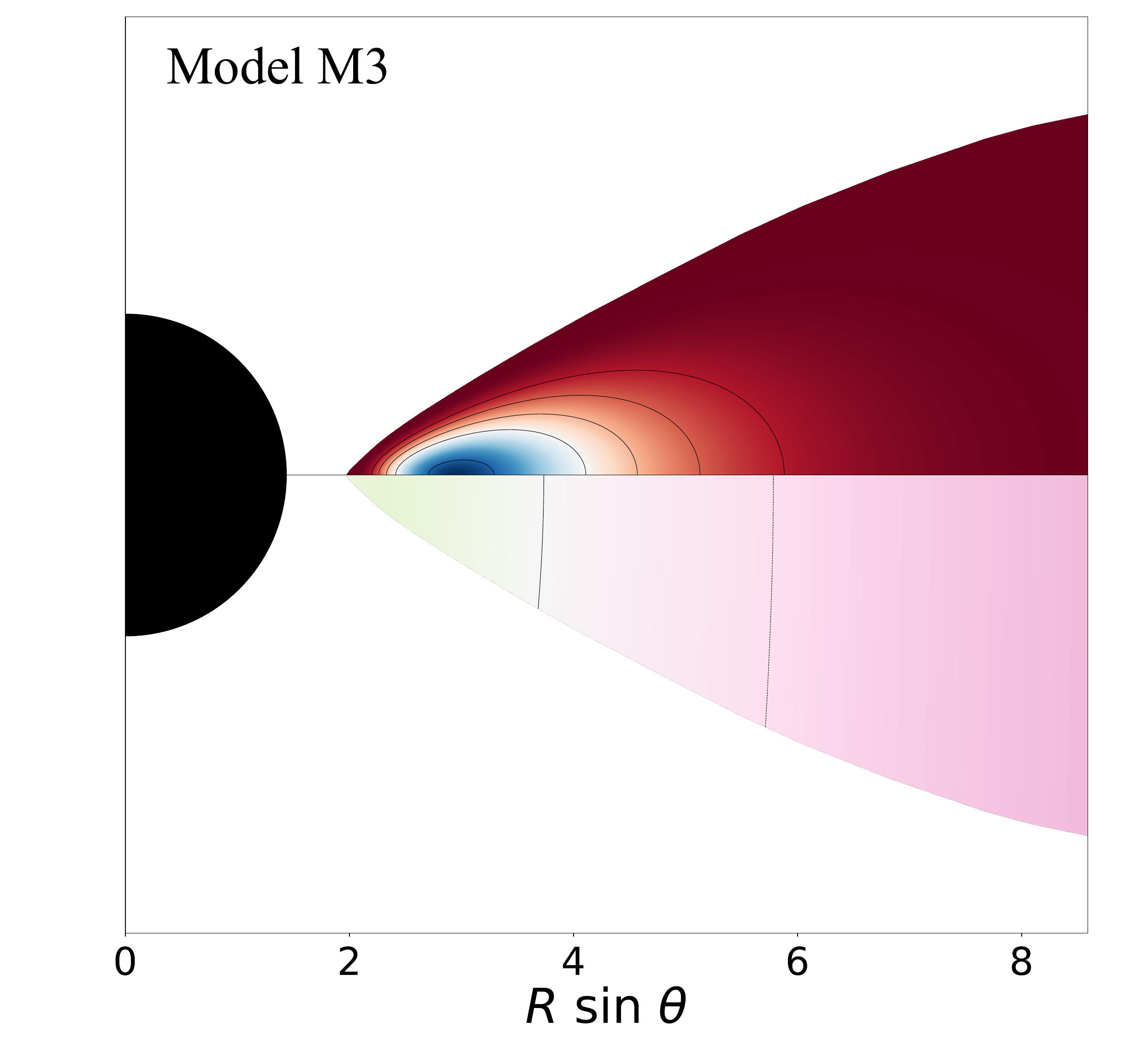}
\hspace{-1.133cm}
\includegraphics[scale=0.1]{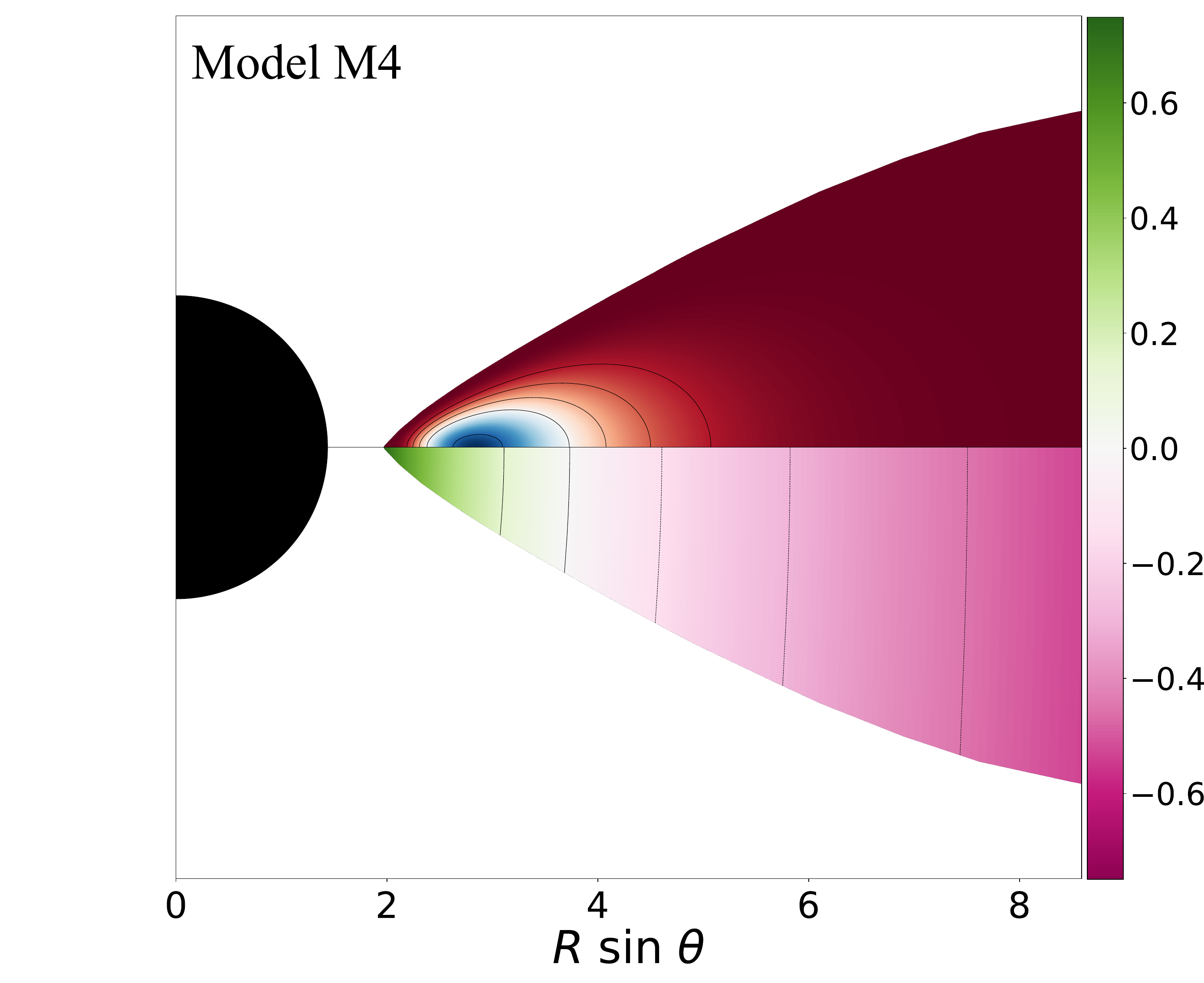}
\caption{Normalized rest-mass density profiles (top part of each panel) and $\log_{10}{\beta_{\mathrm{m}}}$ profiles (bottom half) for our different constant (top row) and non-constant (bottom row) susceptibility models. The columns show, from left to right, models C1 (M1) to C4 (M4) in the top (bottom) row. In this figure, we fix the black hole spin parameter to $a = 0.9$, the exponent of the angular momentum distribution to $\alpha = 0.75$, the constant part of the specific angular momentum distribution to $l_0 = l_{0,2}$ and the magnetization parameter at the center of the disk to $\beta_{\mathrm{m, c}} = 10^{-2}$. The color code is explained in the main text (see Section \ref{results}). The black isocontours represent: i) in the top part of each panel the density isocontours corresponding to $\rho/\rho_{\mathrm{max}} = (0.91, 0.5, 0.33, 0.25, 0.1)$ and ii) in the bottom part the isocontours corresponding to the values of $\log_{10} \beta_{\mathrm{m}} = (-0.45, -0.3, -0.15, 0, 0.15)$. We can observe here, that, from left to right, the rest-mass distribution is more concentrated in the inner region of the disk. It also can be seen that the distribution of the magnetization is almost vertical.}
\label{density_2D_l02}
\end{figure*}

\begin{figure*}
\includegraphics[scale=0.1]{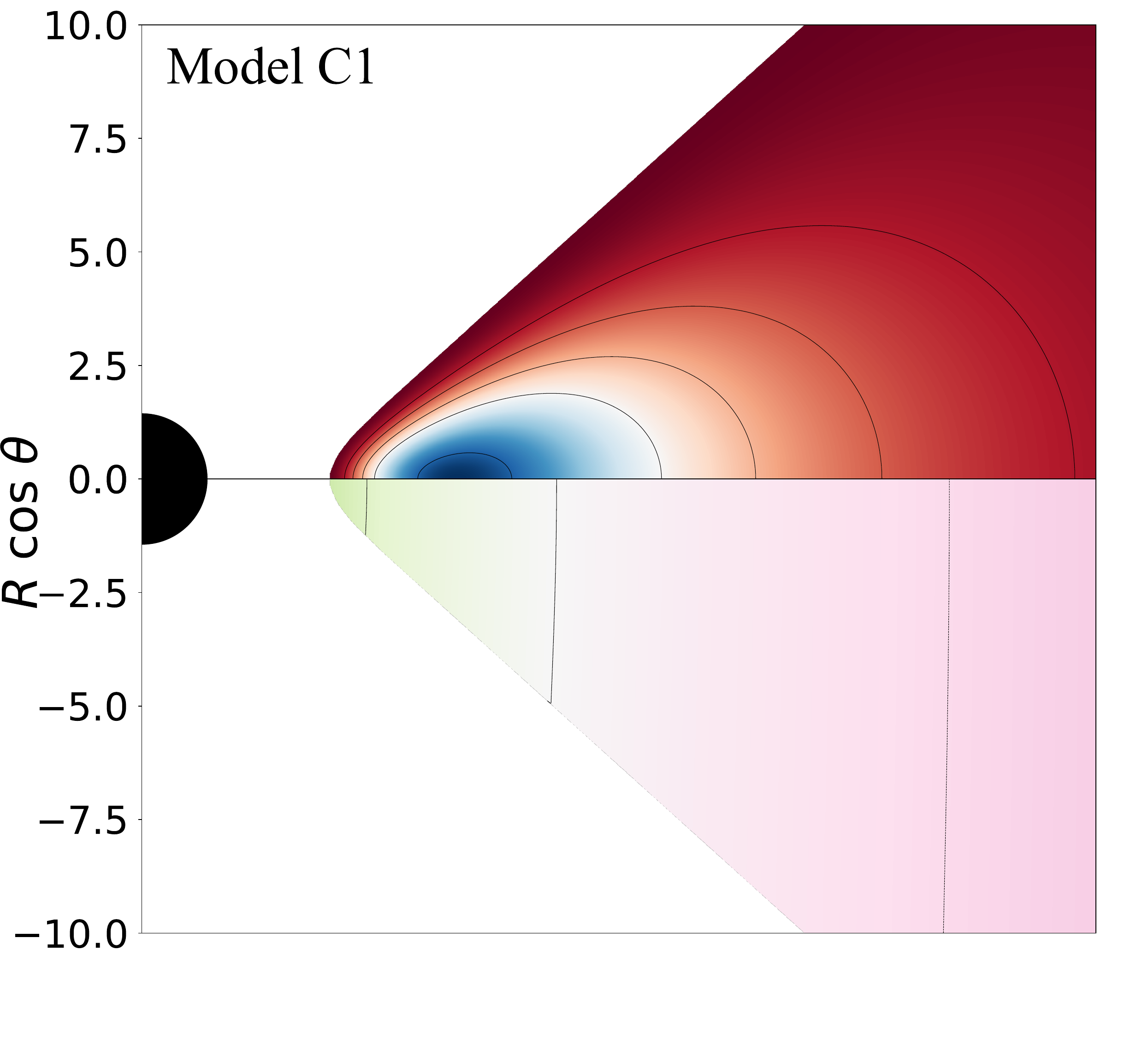}
\hspace{-0.92cm}
\includegraphics[scale=0.1]{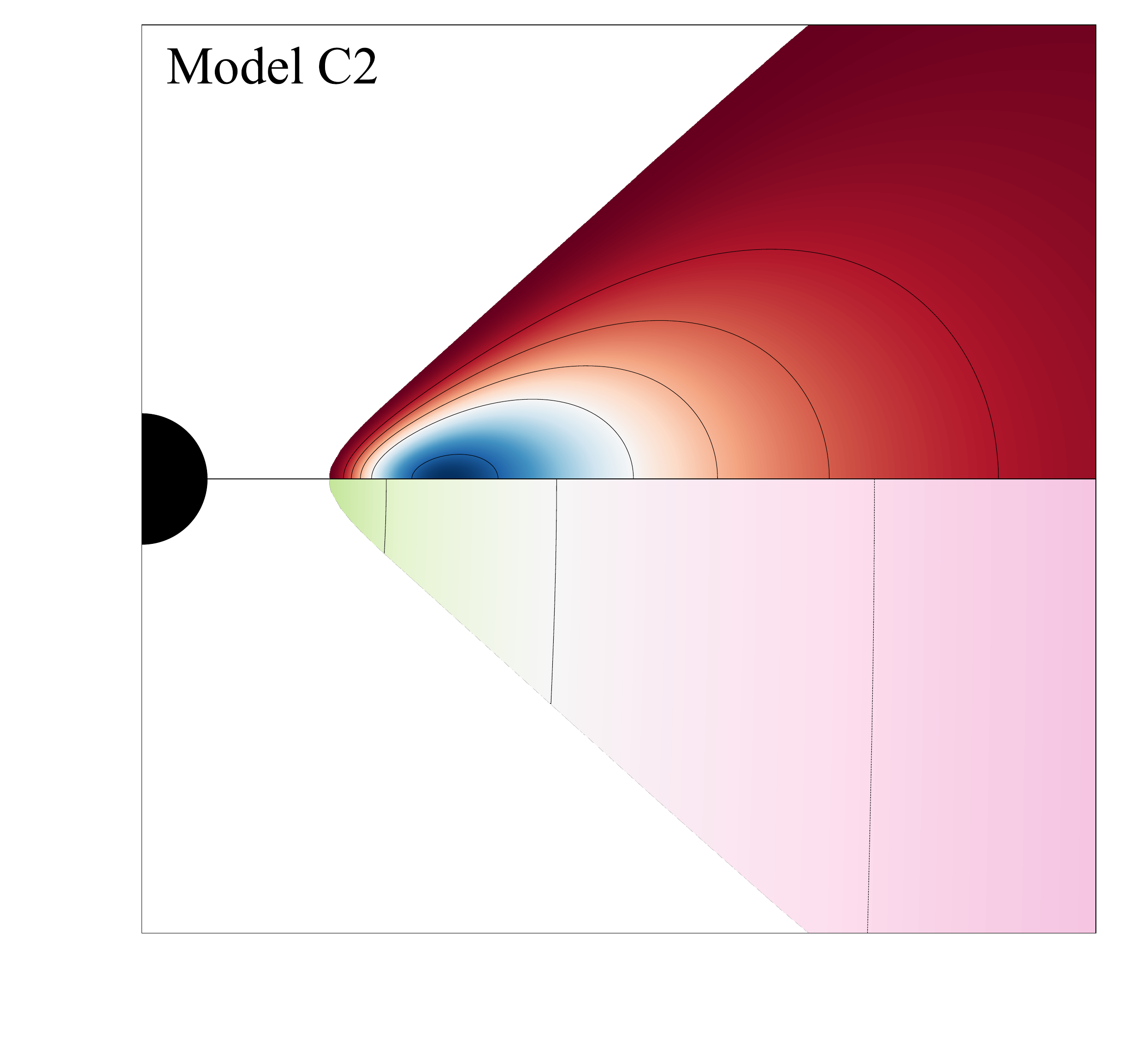}
\hspace{-0.92cm}
\includegraphics[scale=0.1]{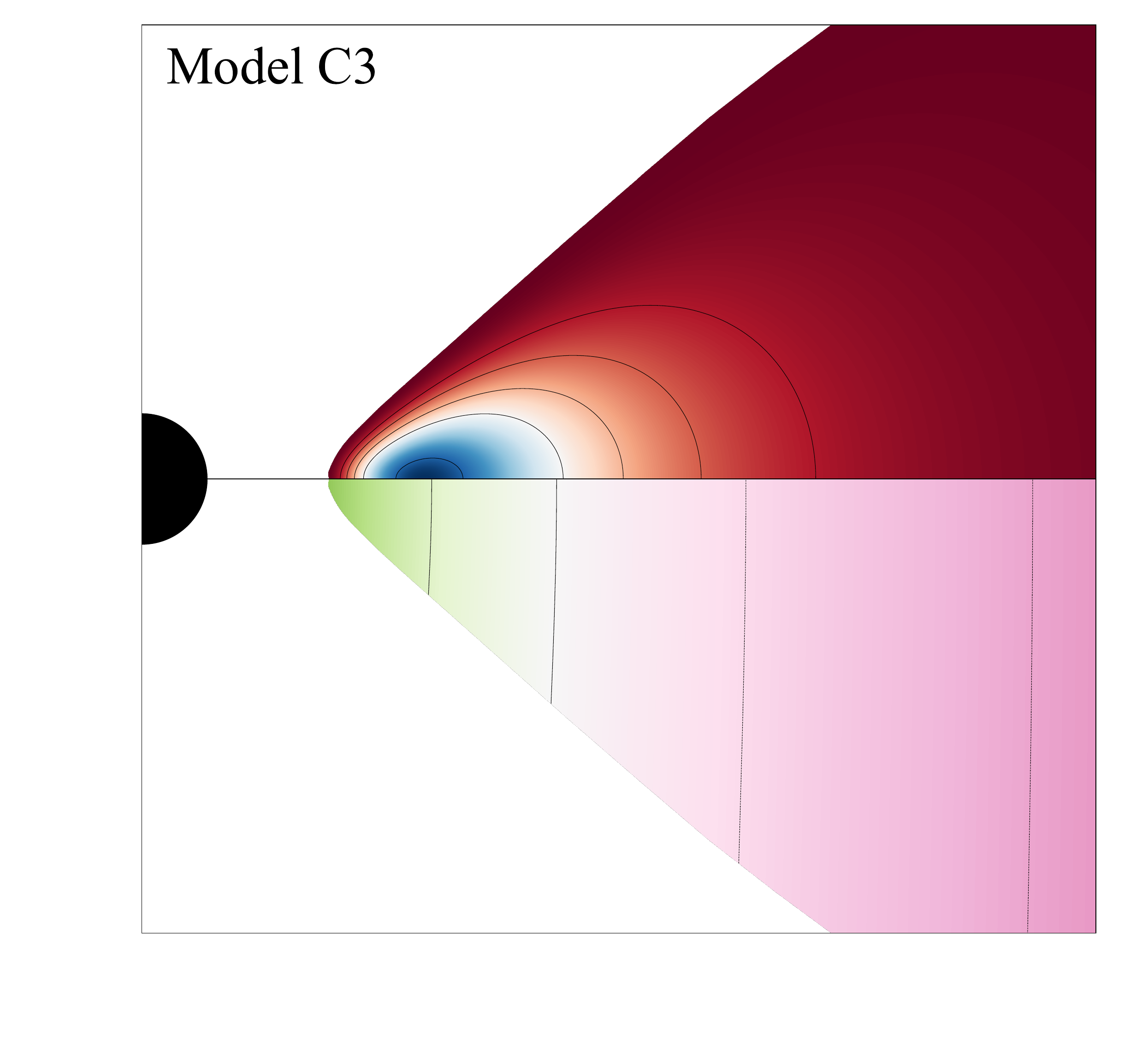}
\hspace{-1.125cm}
\includegraphics[scale=0.1]{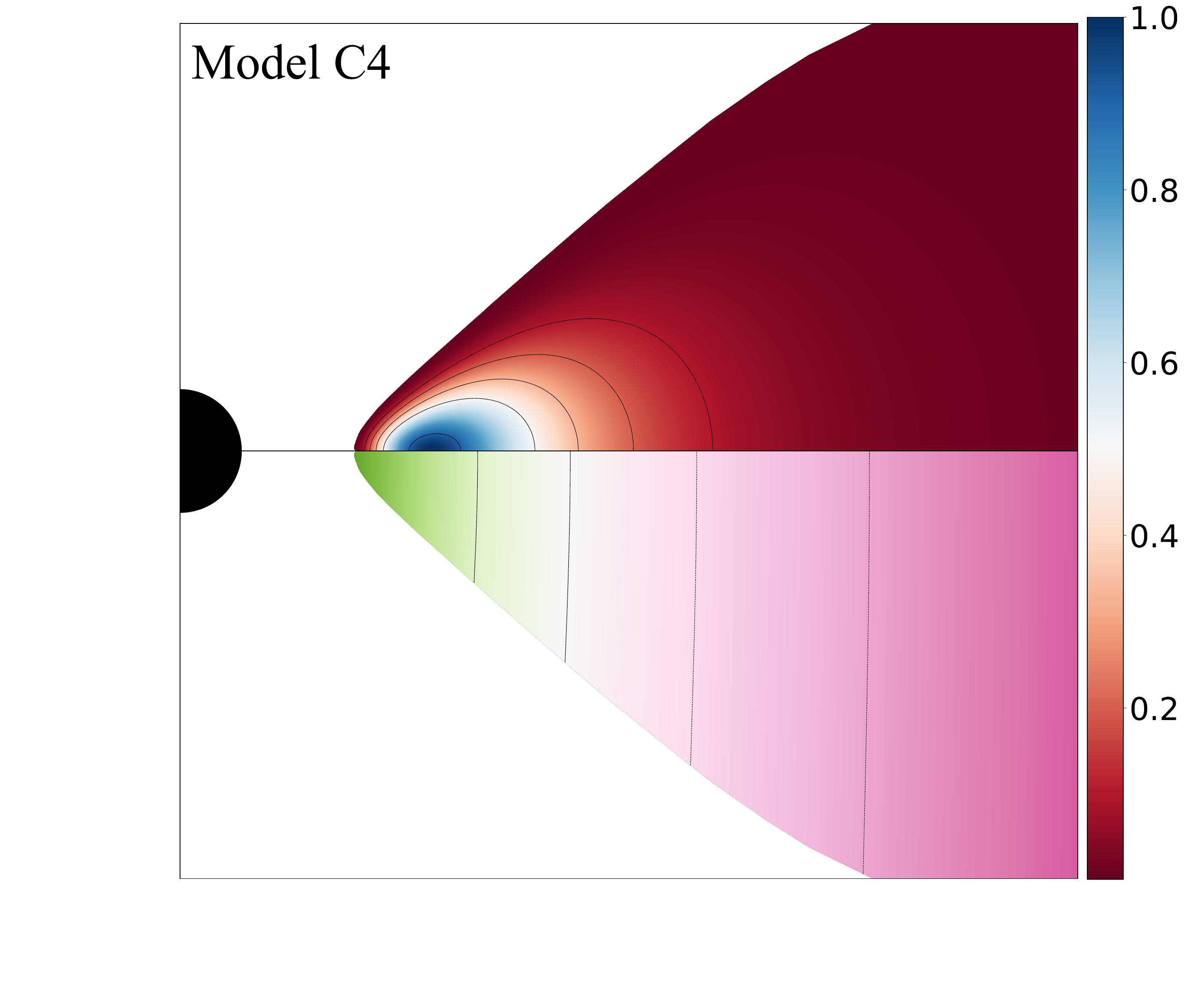}
\vspace{-0.3cm}
\\
\hspace{-0.2cm}
\includegraphics[scale=0.1]{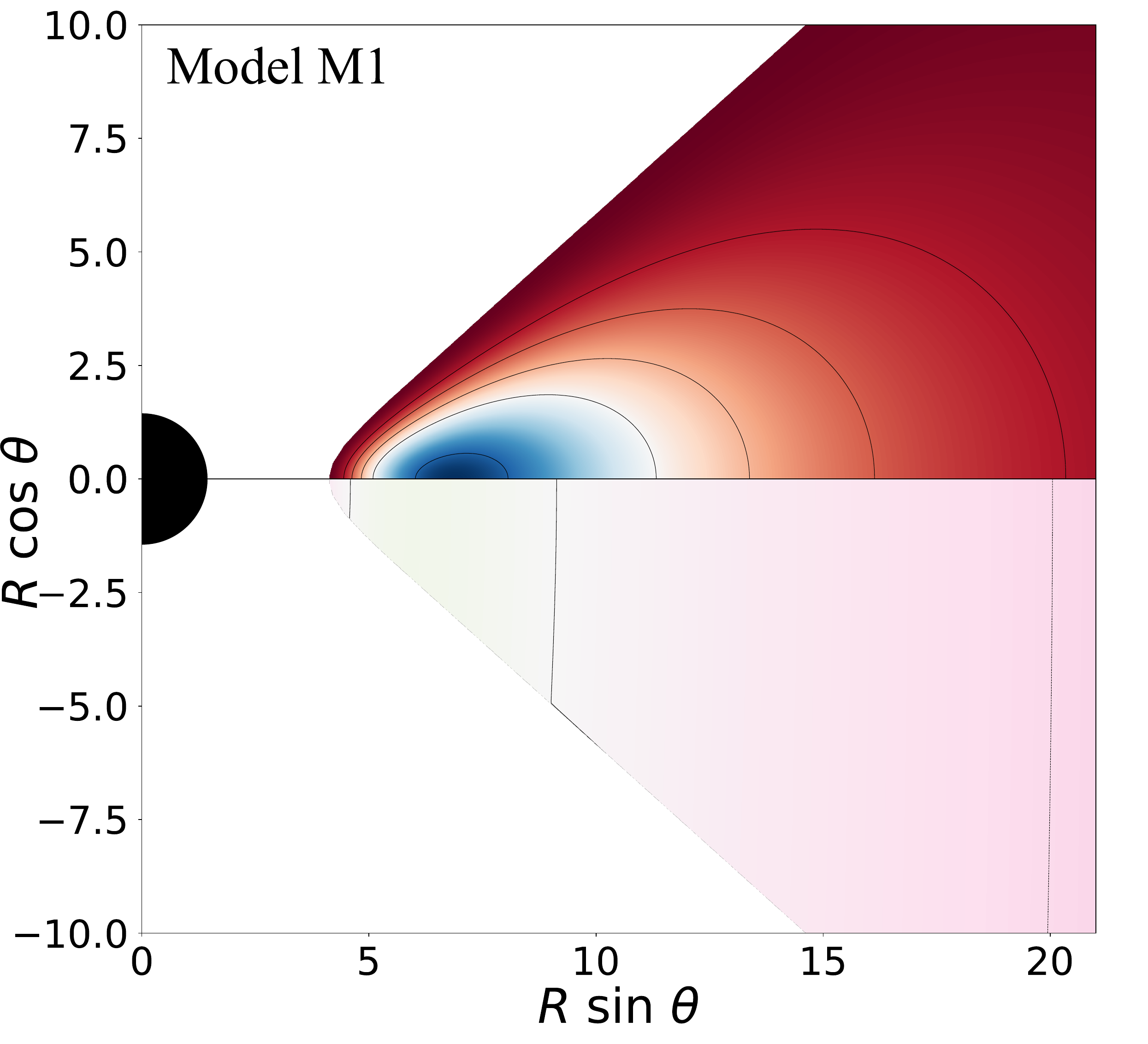}
\hspace{-0.92cm}
\includegraphics[scale=0.1]{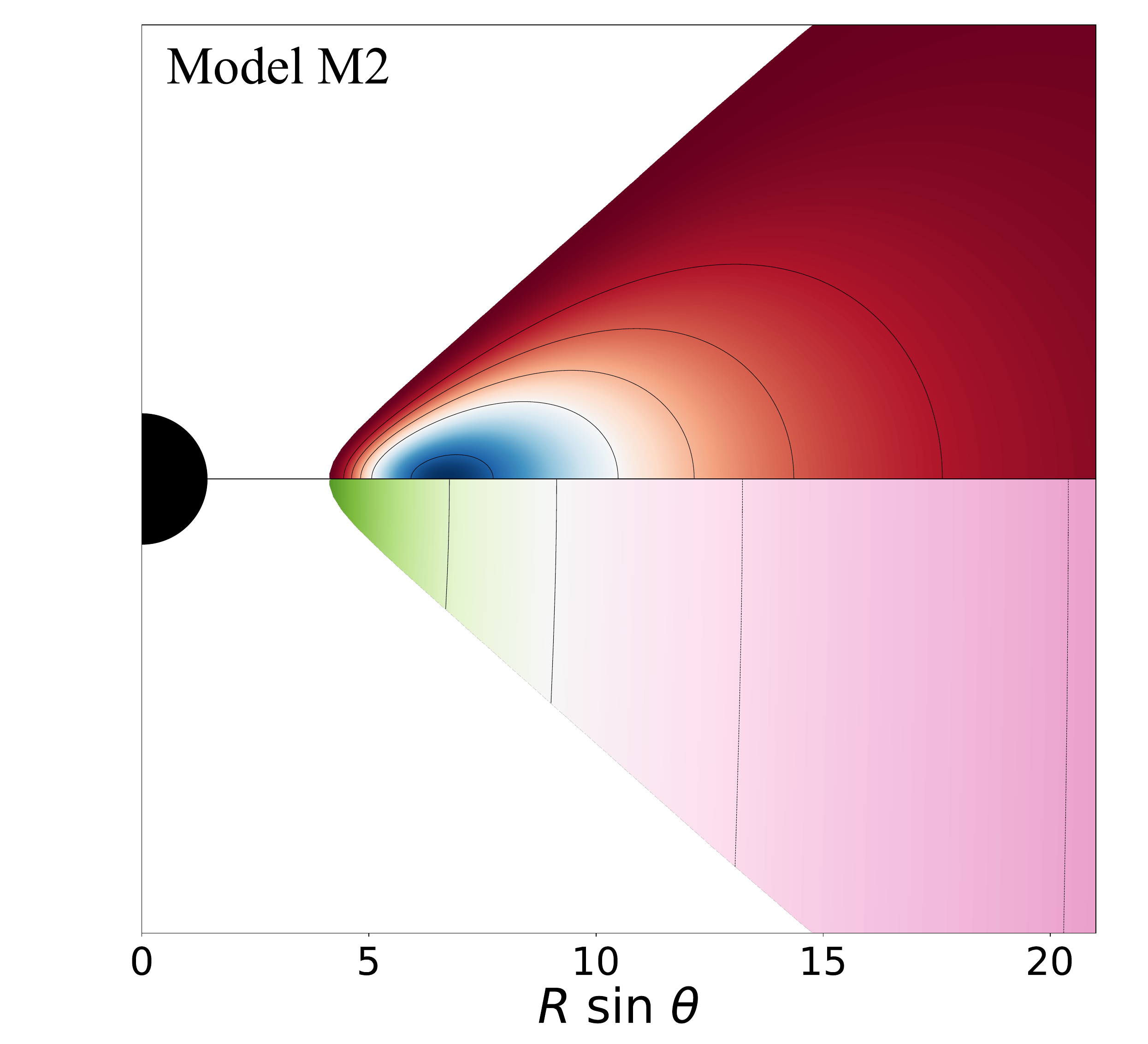}
\hspace{-0.92cm}
\includegraphics[scale=0.1]{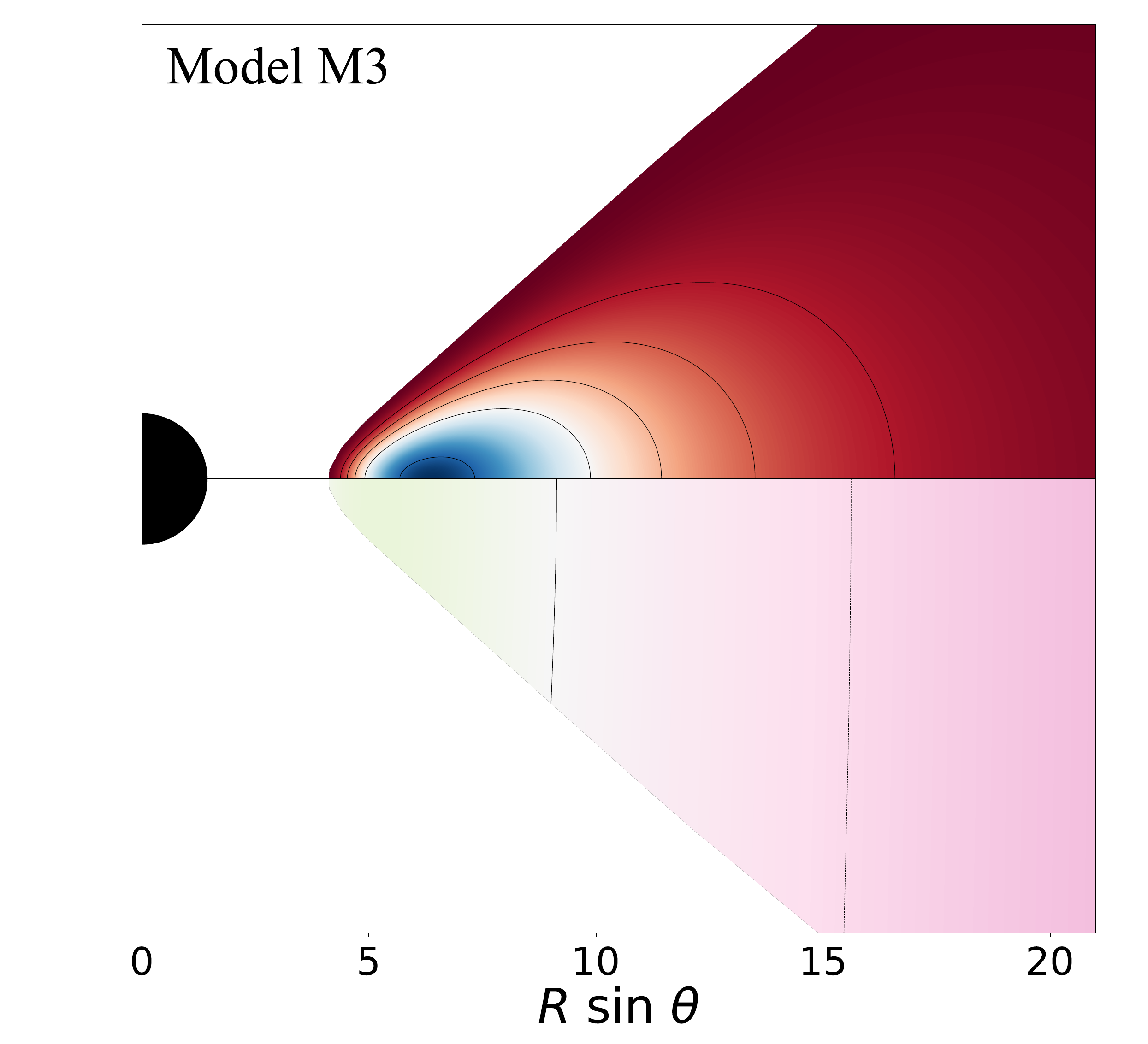}
\hspace{-1.125cm}
\includegraphics[scale=0.1]{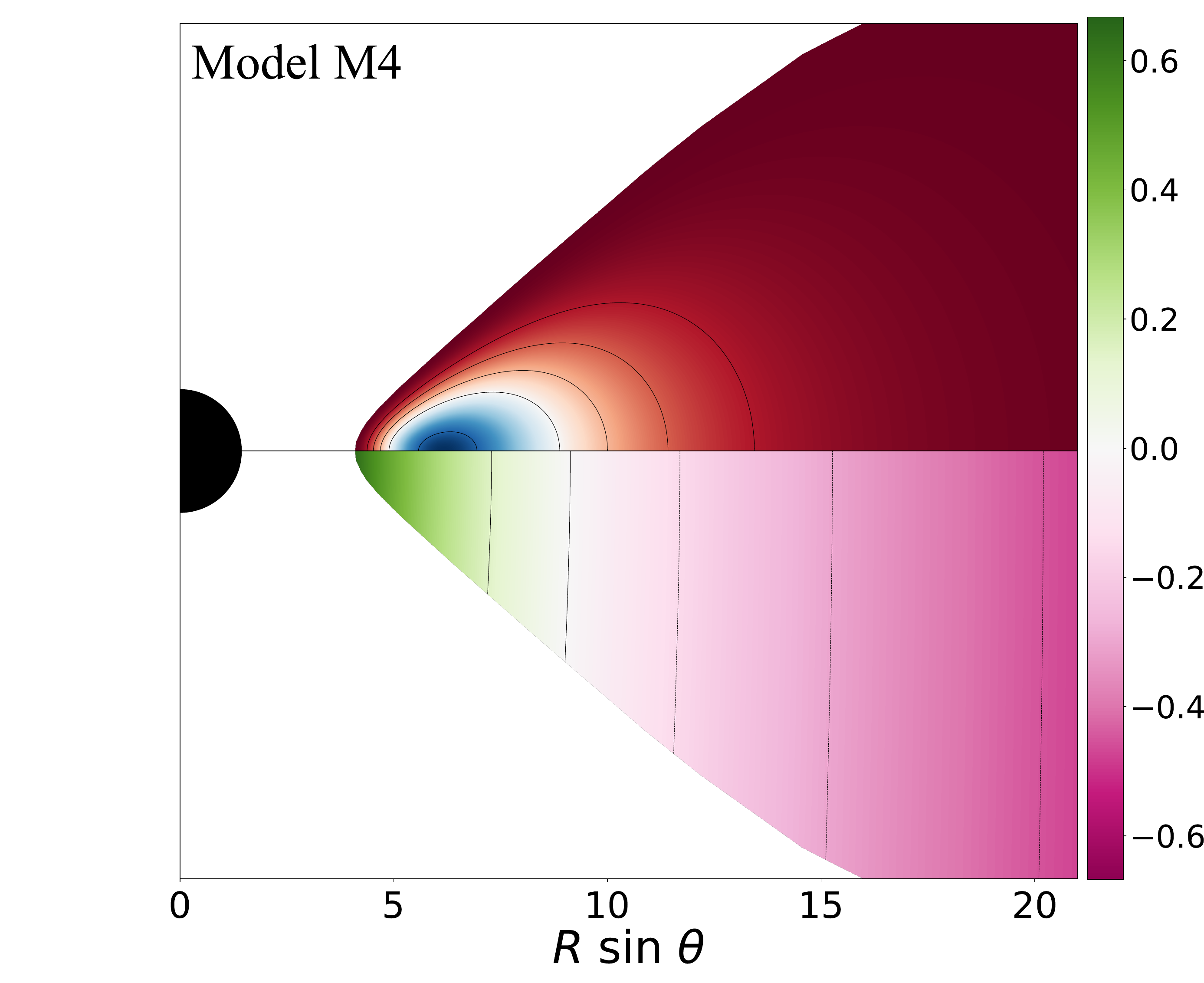}
\caption{Same as in Fig.~\ref{density_2D_l02} but for $l_0 = l_{0,3}$. We also observe here the same trends that appear in Fig.~\ref{density_2D_l02}}
\label{density_2D_l03}
\end{figure*}

\begin{figure*}
\includegraphics[scale=0.08]{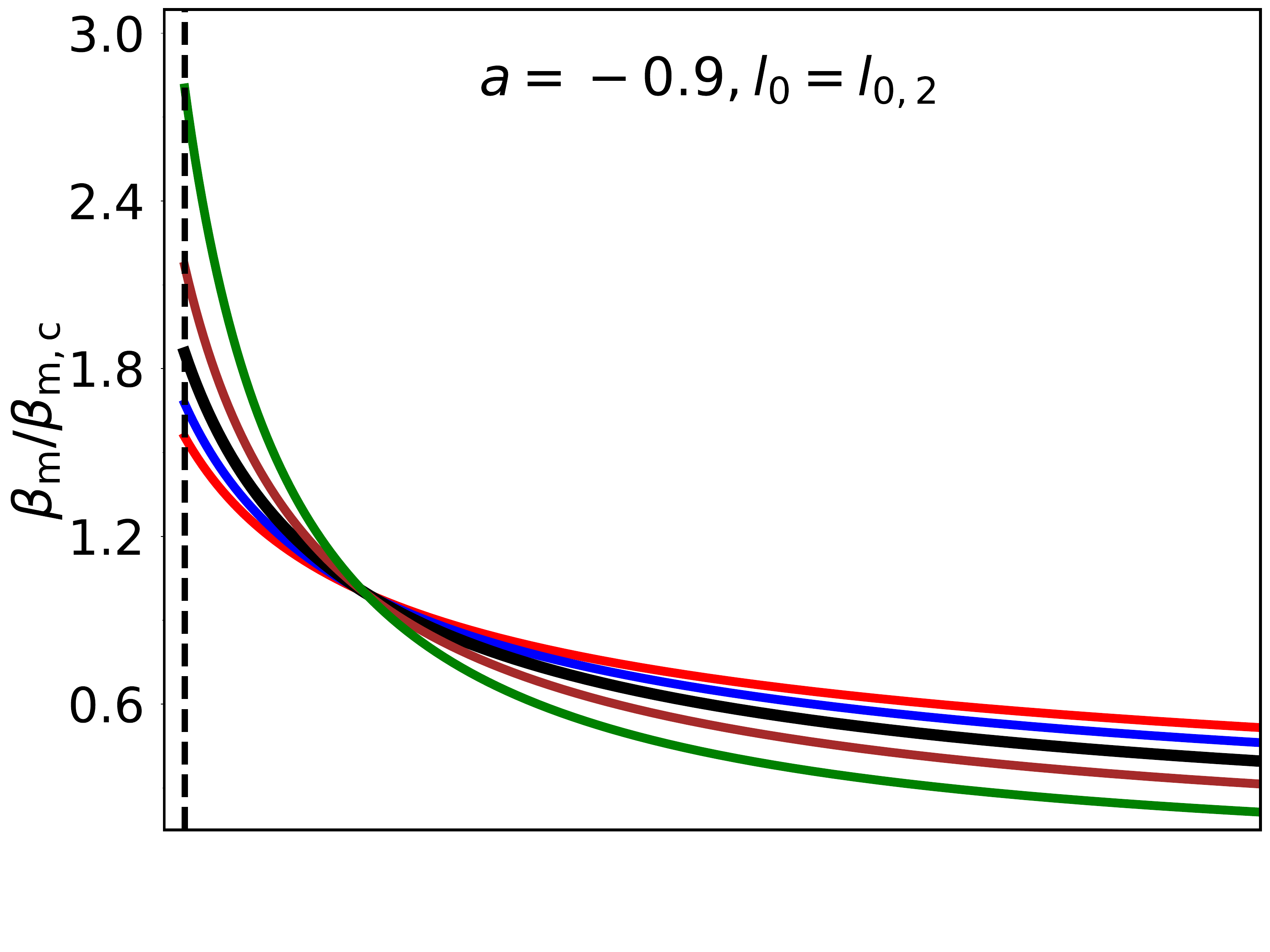}
\hspace{-0.76cm}
\includegraphics[scale=0.08]{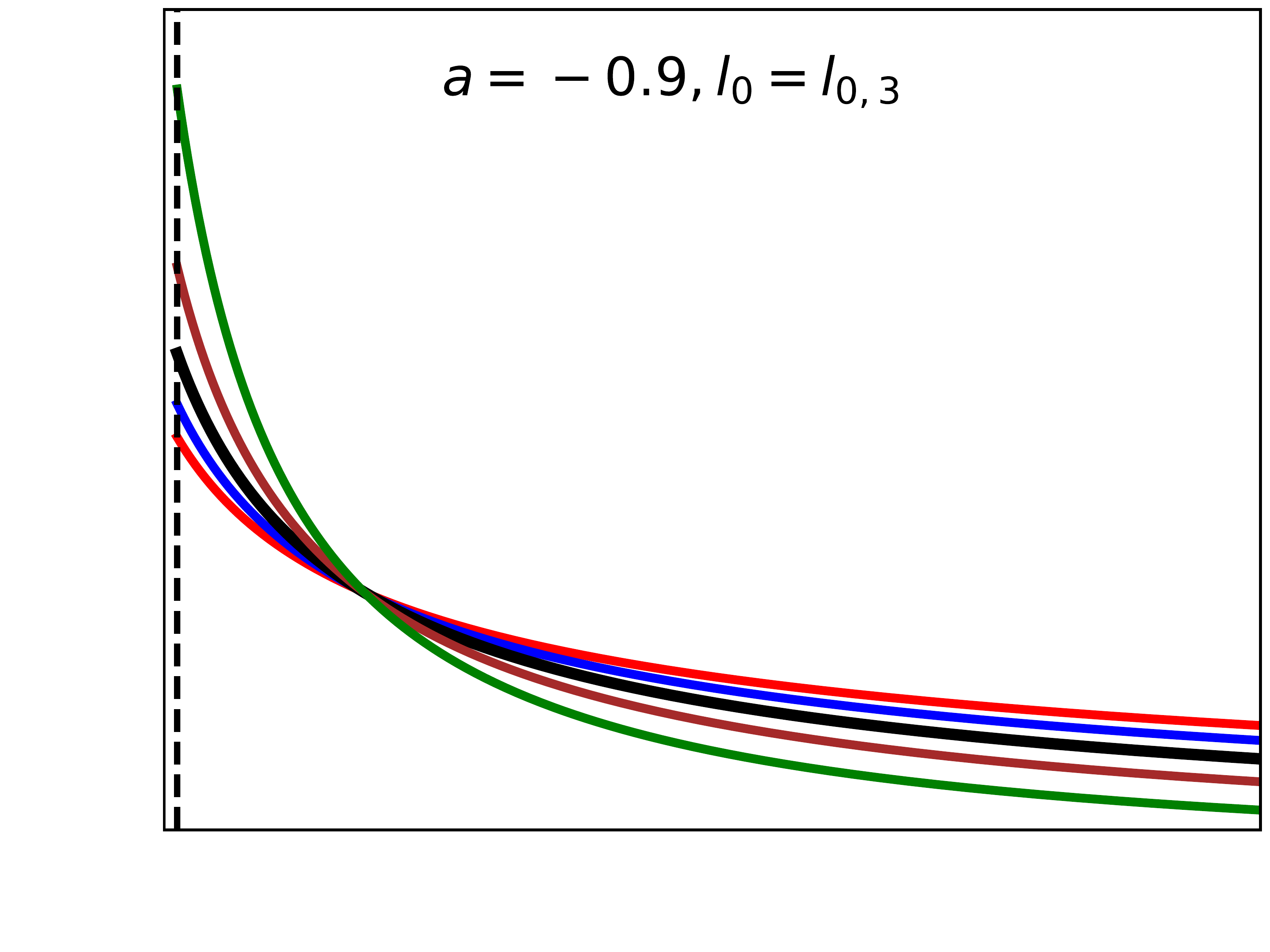}
\hspace{-0.2cm}
\includegraphics[scale=0.08]{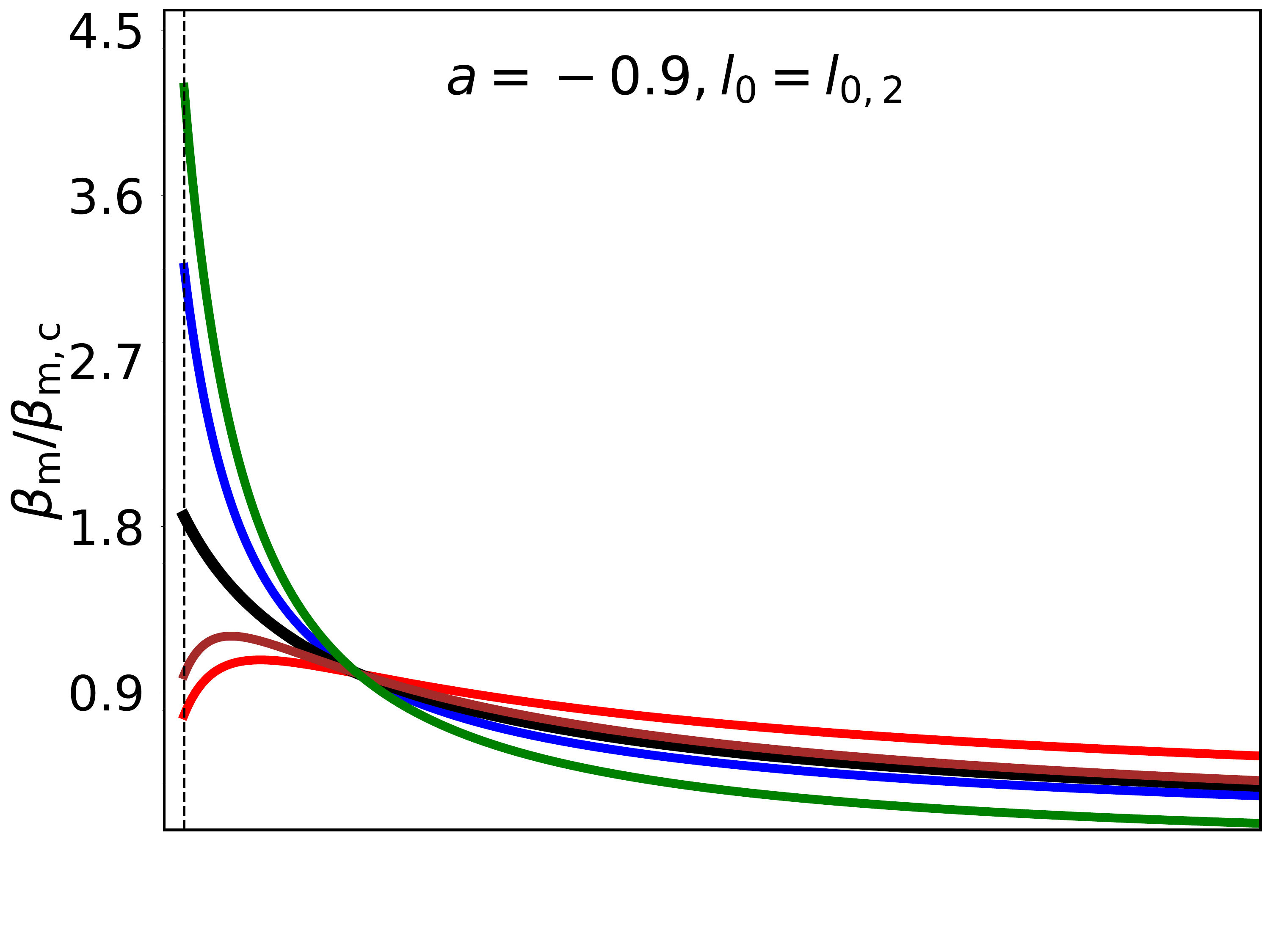}
\hspace{-0.76cm}
\includegraphics[scale=0.08]{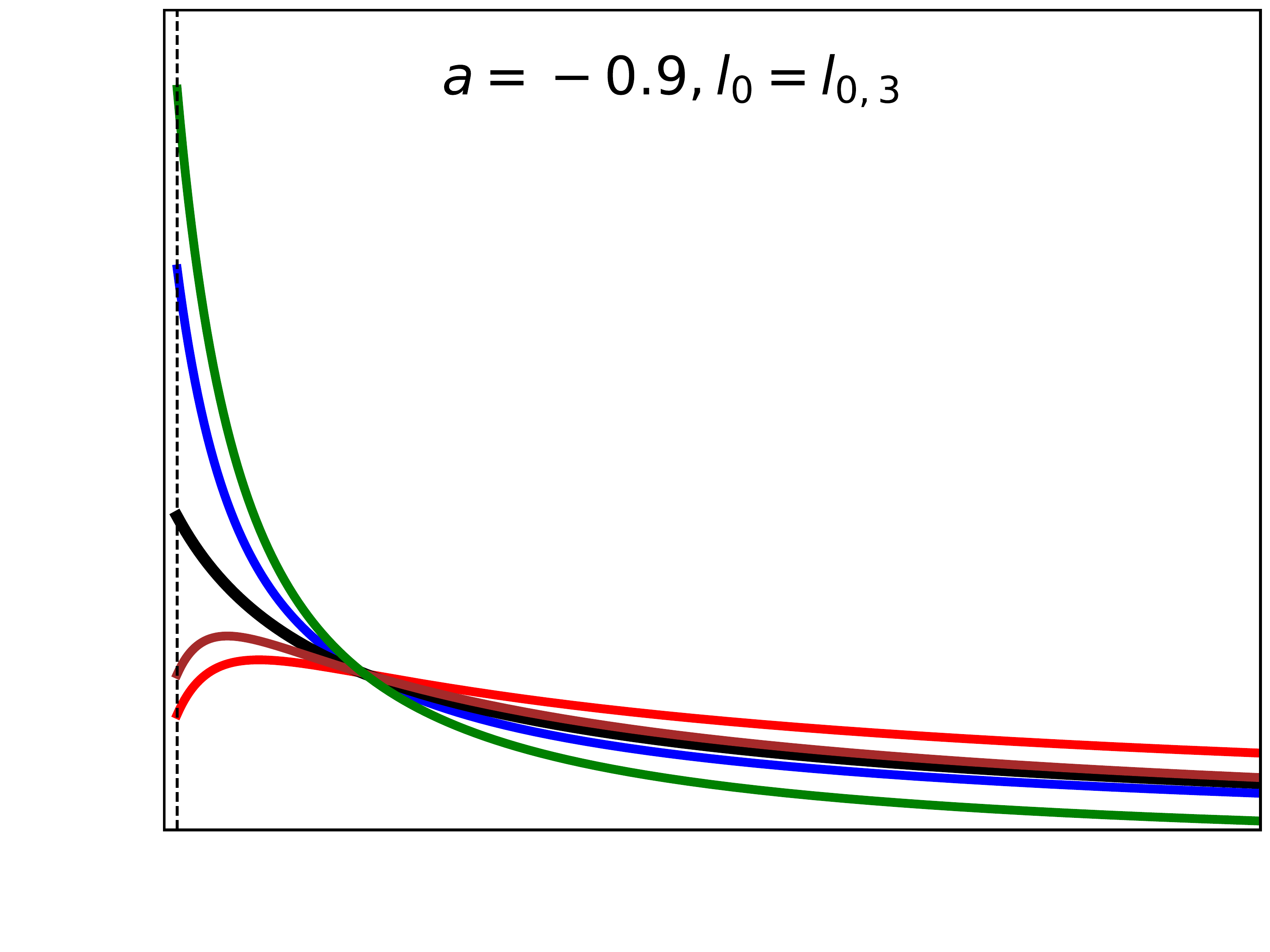}
\vspace{-0.3cm}
\\
\includegraphics[scale=0.08]{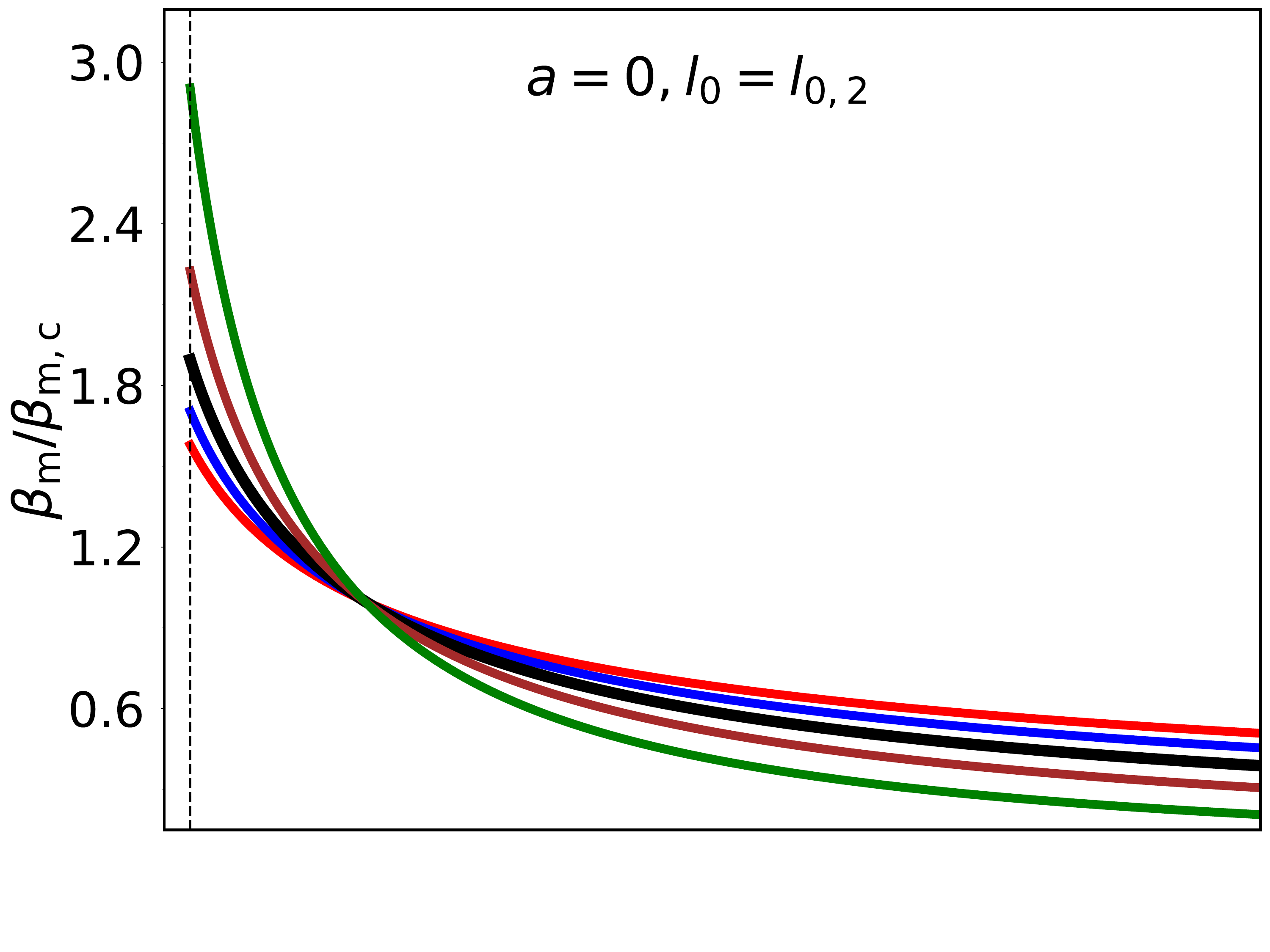}
\hspace{-0.76cm}
\includegraphics[scale=0.08]{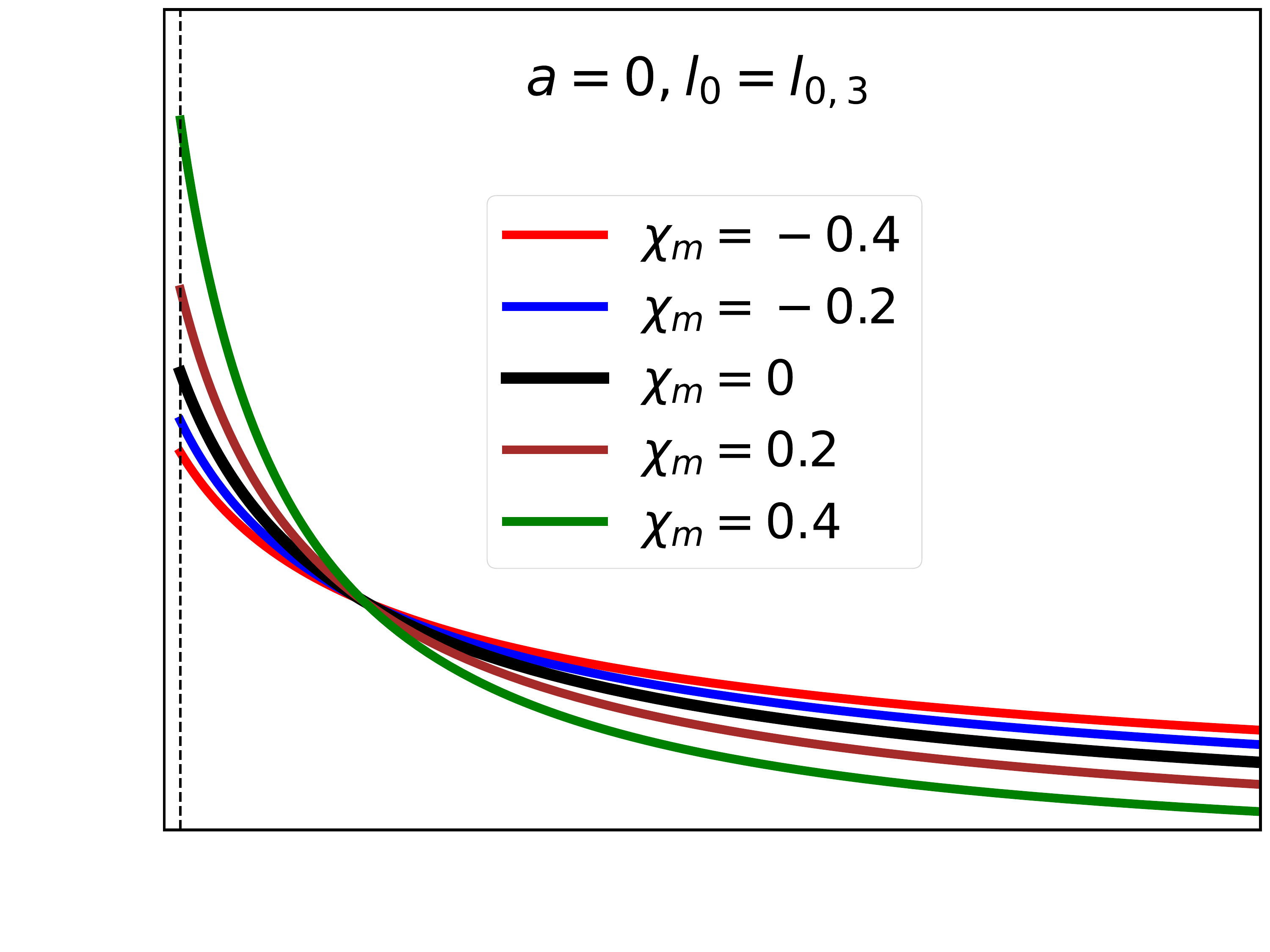}
\hspace{-0.2cm}
\includegraphics[scale=0.08]{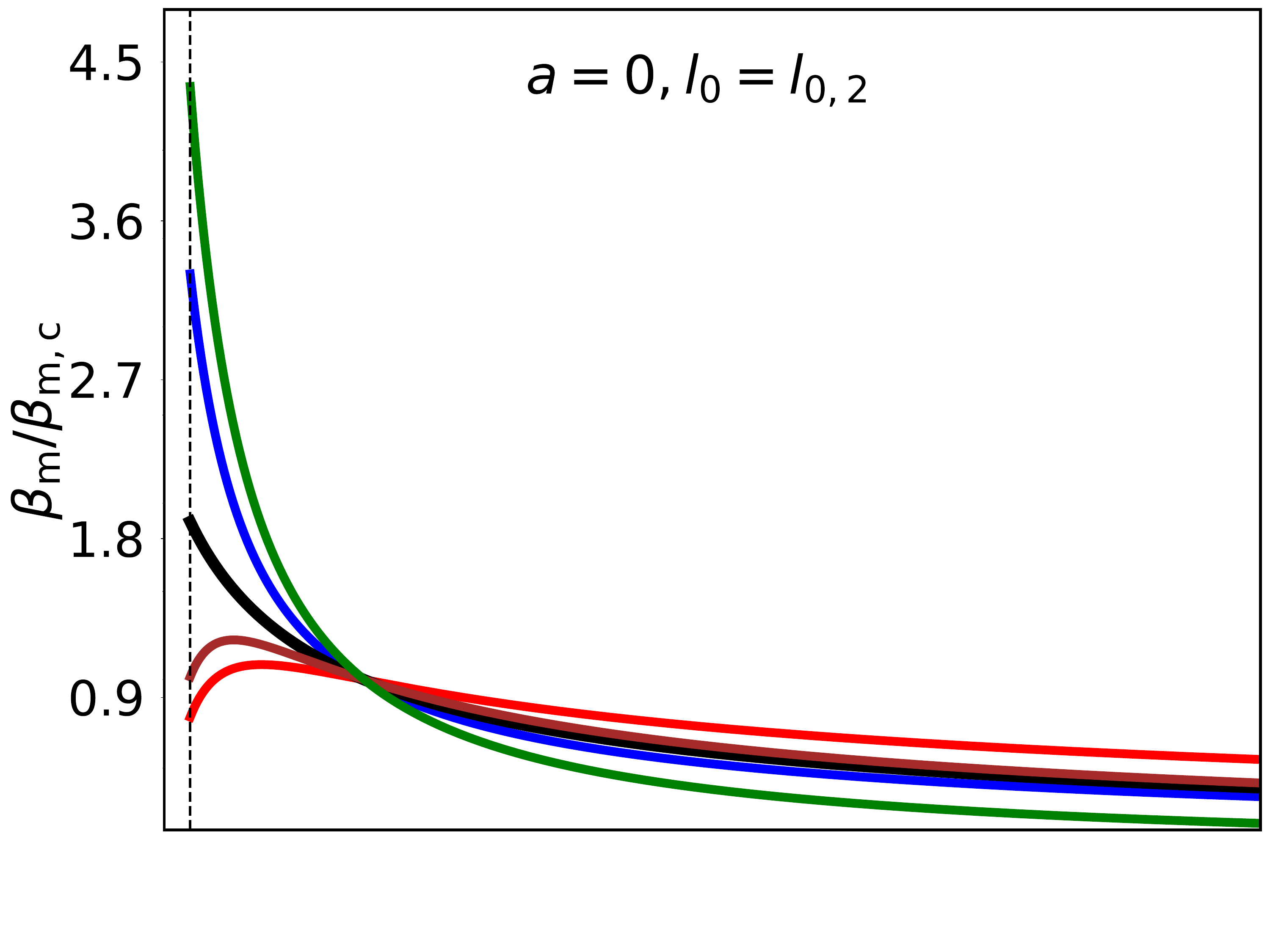}
\hspace{-0.76cm}
\includegraphics[scale=0.08]{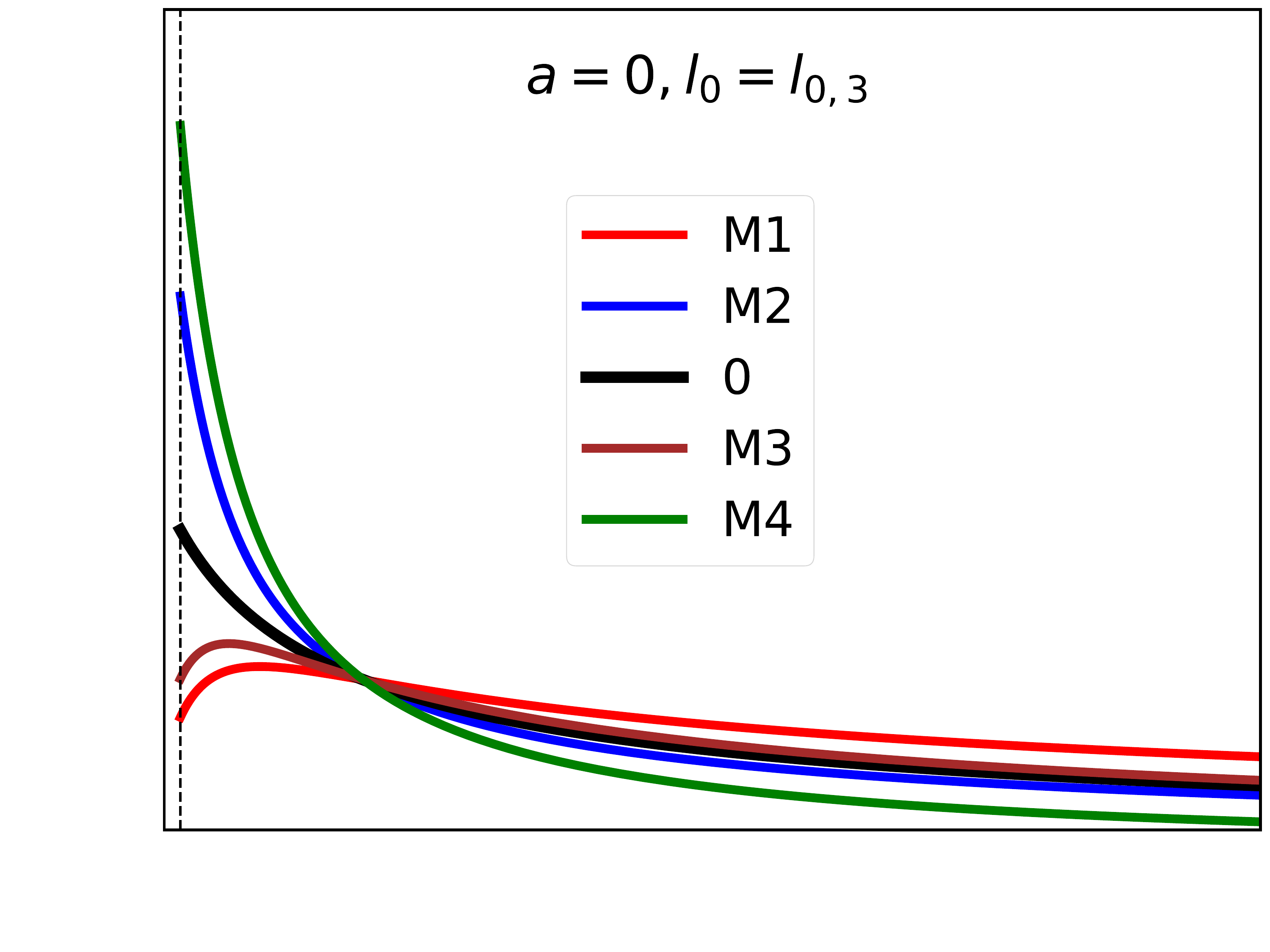}
\vspace{-0.3cm}
\\
\hspace{-0.2cm}
\includegraphics[scale=0.08]{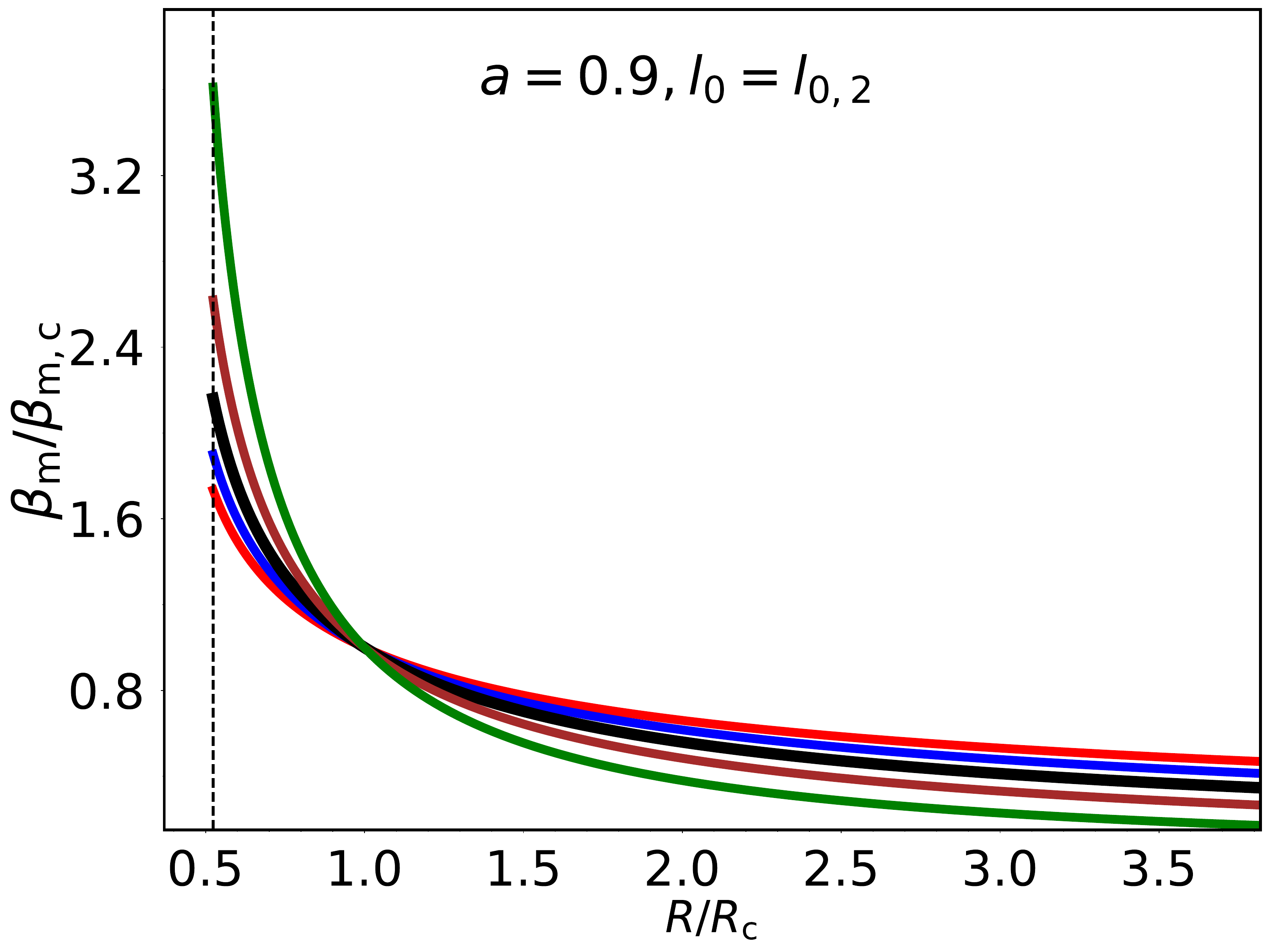}
\hspace{-0.76cm}
\includegraphics[scale=0.08]{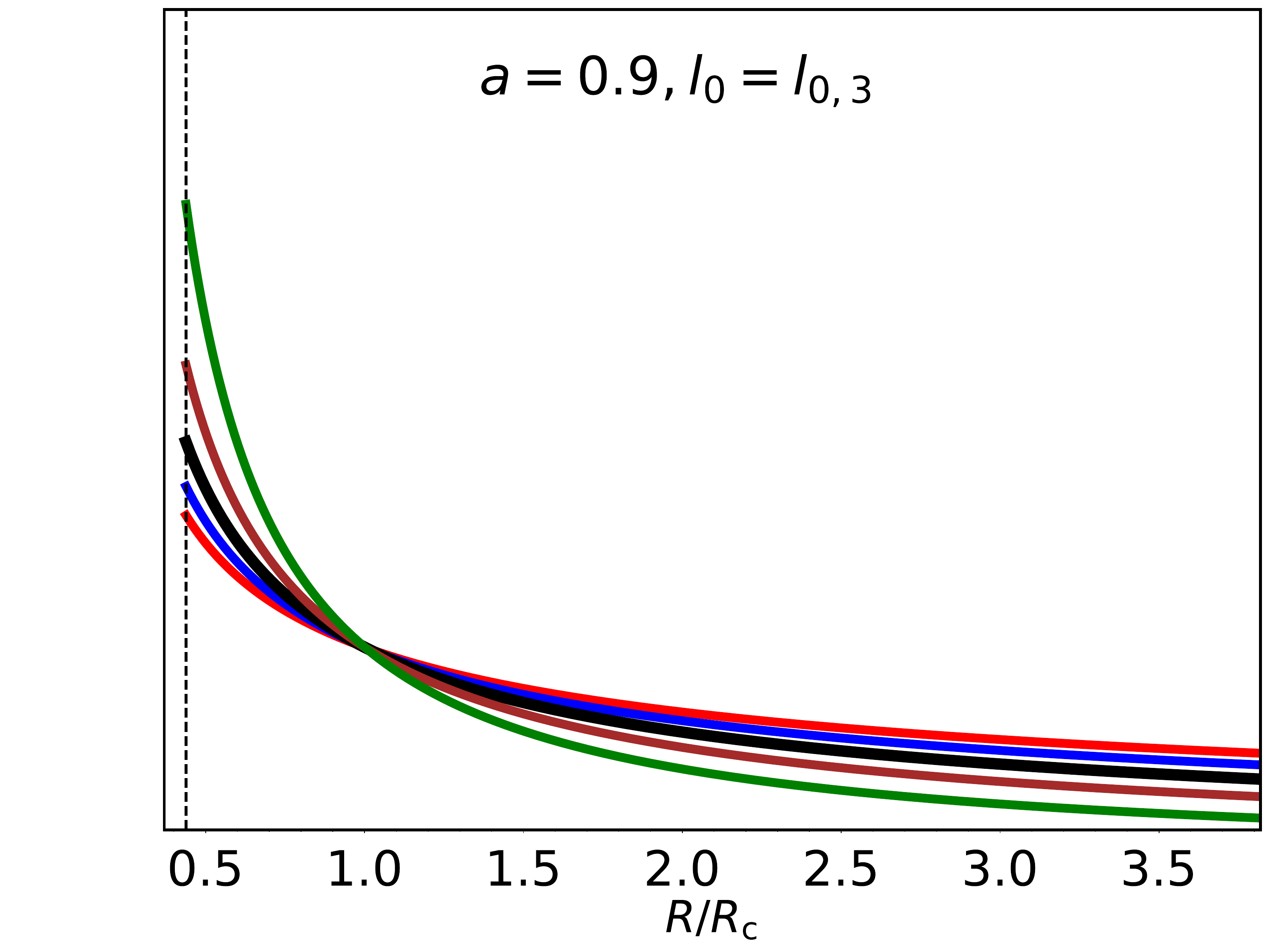}
\hspace{-0.2cm}
\includegraphics[scale=0.08]{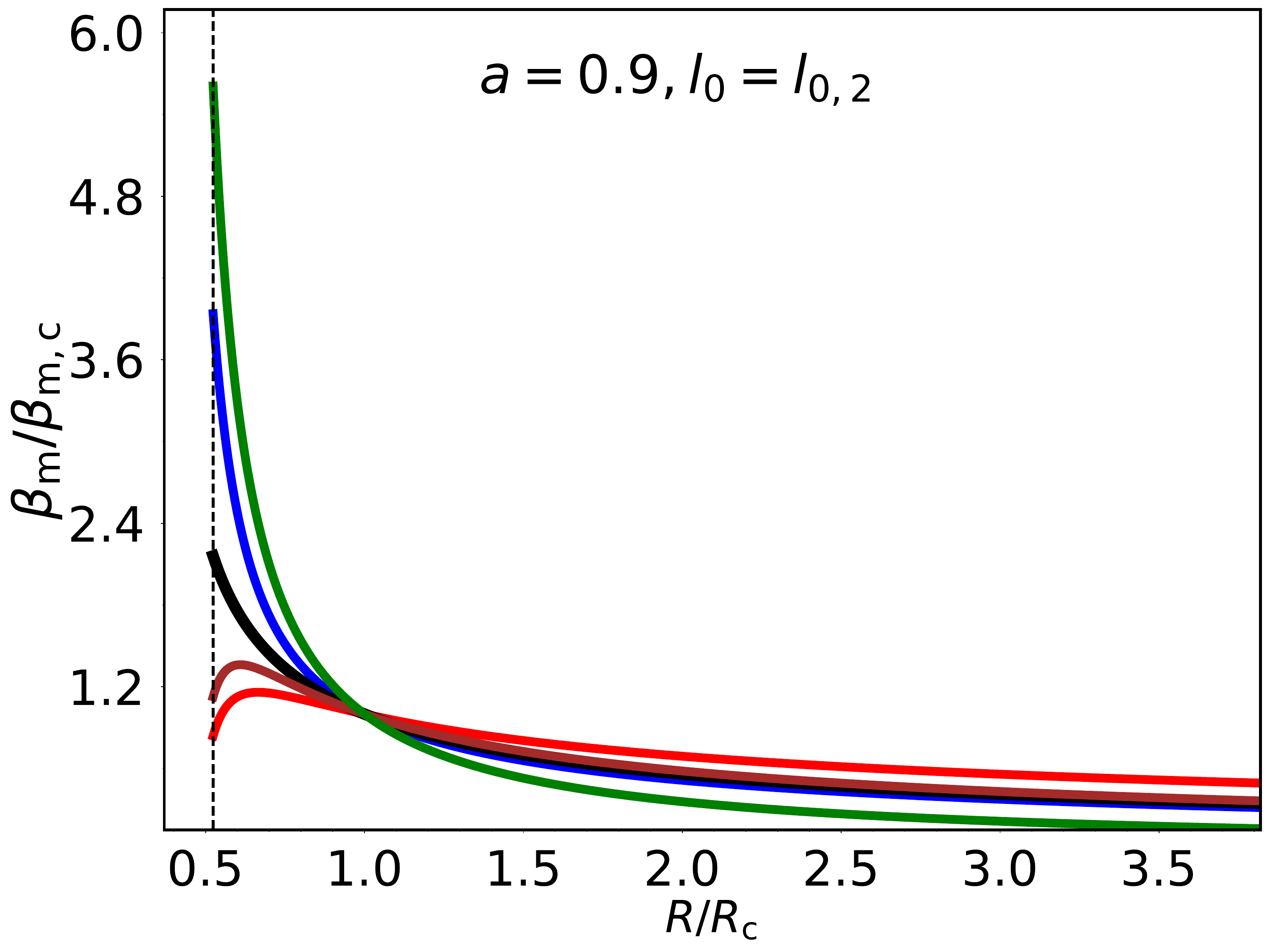}
\hspace{-0.76cm}
\includegraphics[scale=0.08]{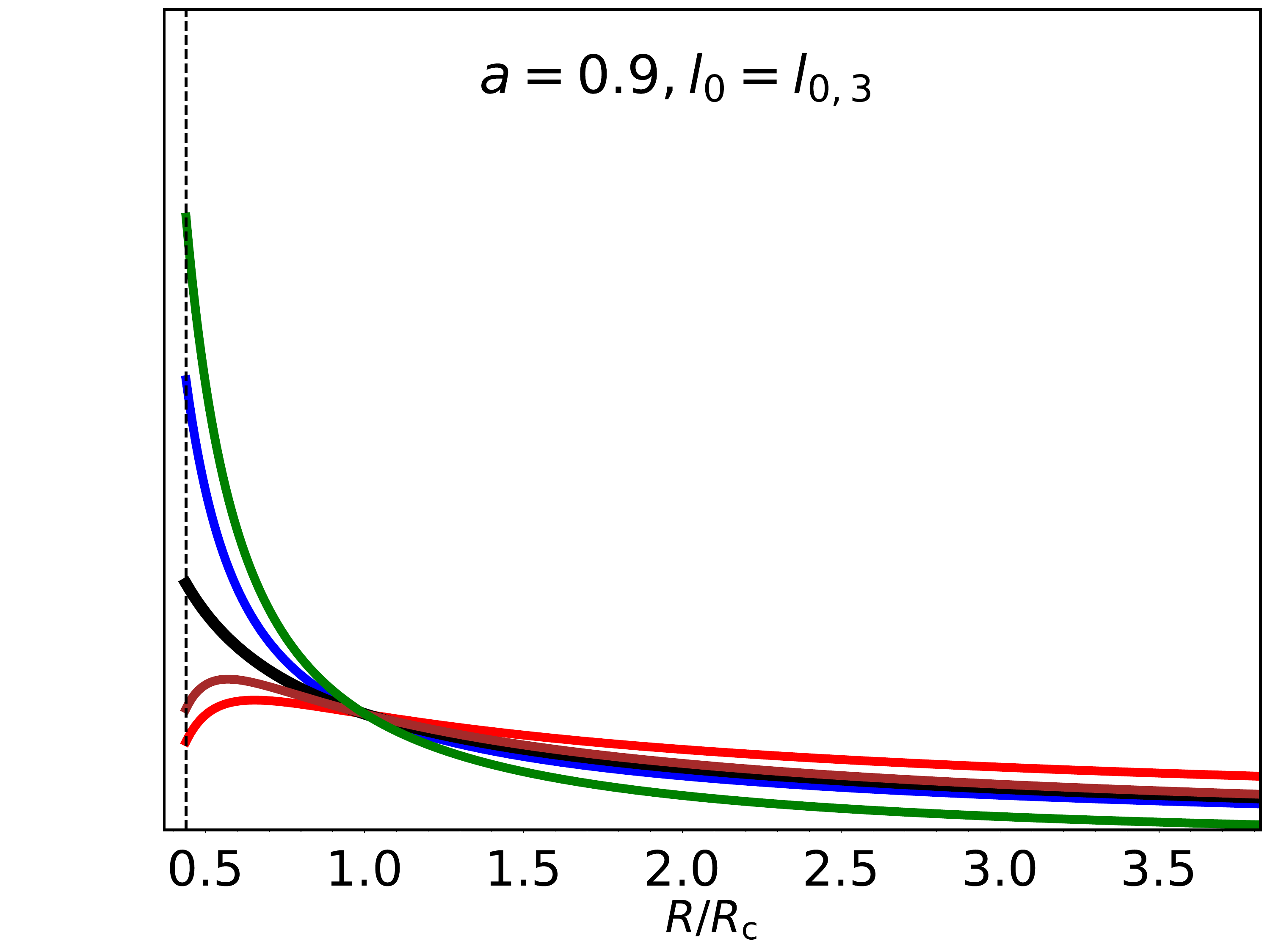}
\caption{Normalized radial profiles of the magnetization function at the equatorial plane, $\beta_{\mathrm{m}}(R)$, for magnetization parameter at the center $\beta_{\mathrm{m, c}} = 10^{-2}$. In the first and second columns we show the constant susceptibility models, while in the third and fourth columns we show the non-constant susceptibility models. Moreover, the first and third columns correspond to models with the constant part of the specific angular momentum distribution $l_0$ equal to $l_{0, 2}$, and the second and fourth columns to models with $l_0 = l_{0, 3}$. From top to bottom the rows depict models with values of the spin parameter of the black hole of $-0.9$, $0$ and $0.9$. The radial coordinate is normalized with respect to the radius of the center of the disk $R_{\mathrm{c}}$ for each model and the exponent of the specific angular momentum law is fixed to $\alpha = 0.75$. The vertical dotted line in each plot indicates the inner edge of the torus.}
\label{beta_r_fig}
\end{figure*}

\begin{figure*}
\includegraphics[scale=0.08]{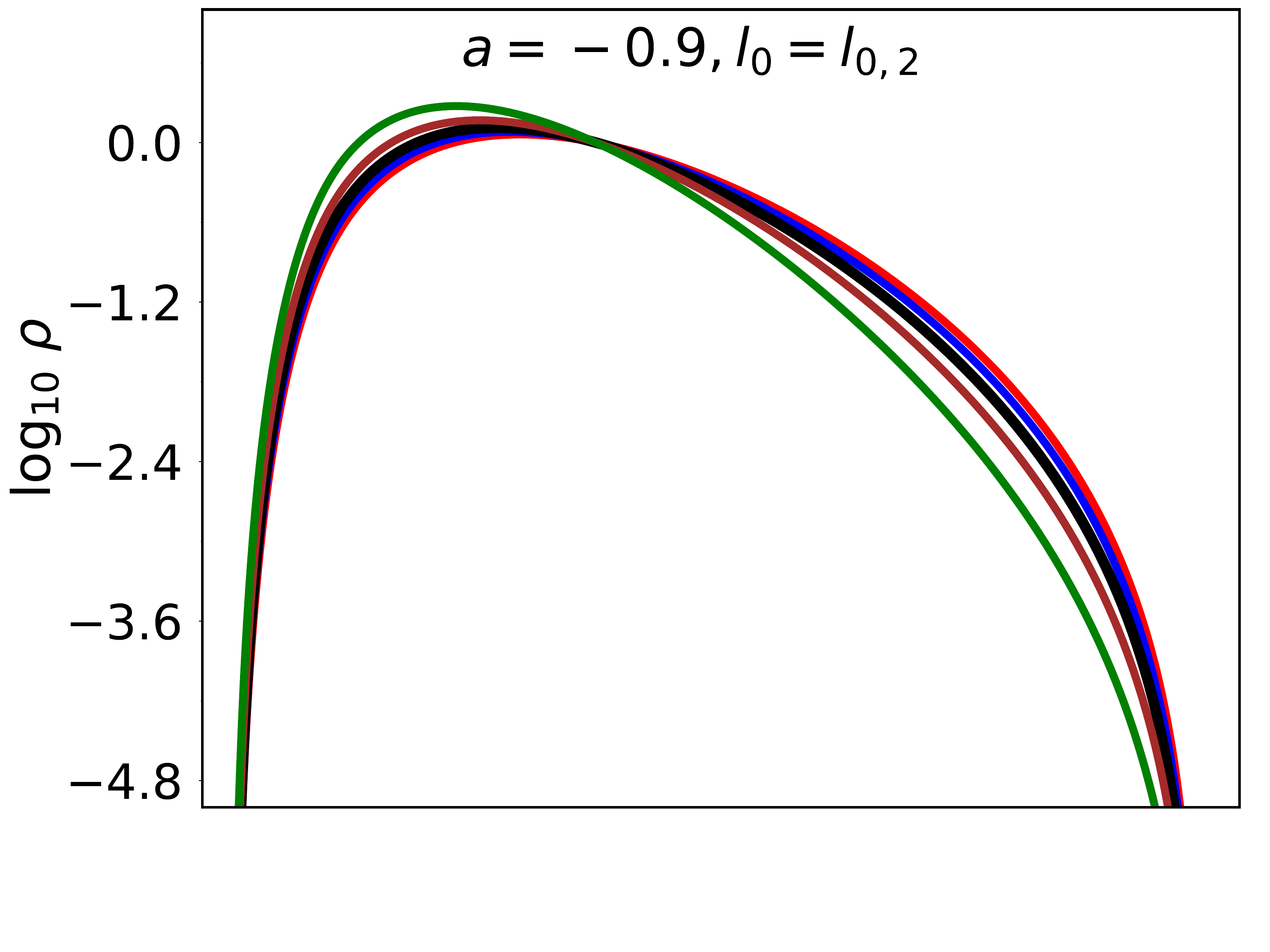}
\hspace{-0.9cm}
\includegraphics[scale=0.08]{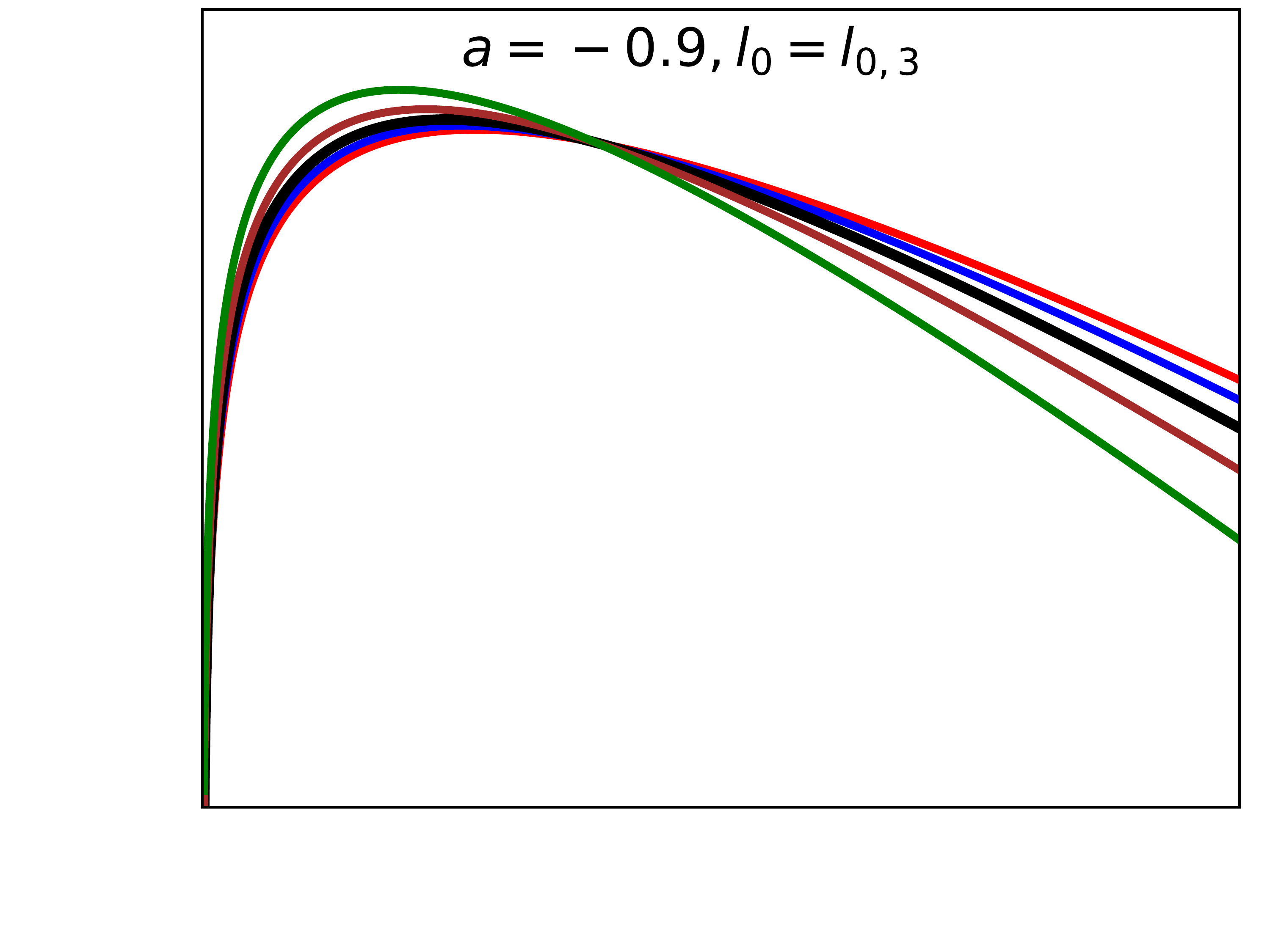}
\hspace{-0.90cm}
\includegraphics[scale=0.08]{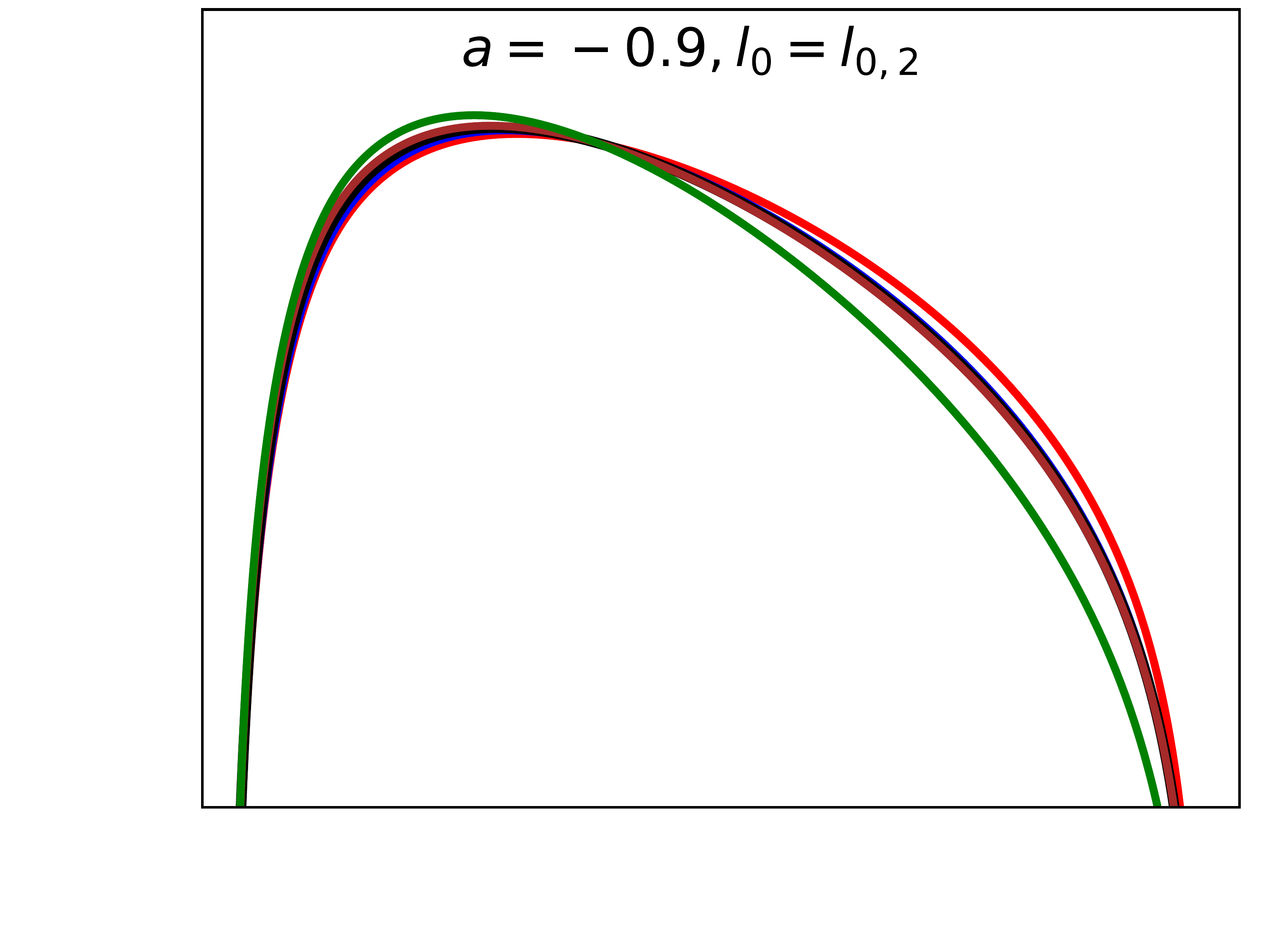}
\hspace{-0.90cm}
\includegraphics[scale=0.08]{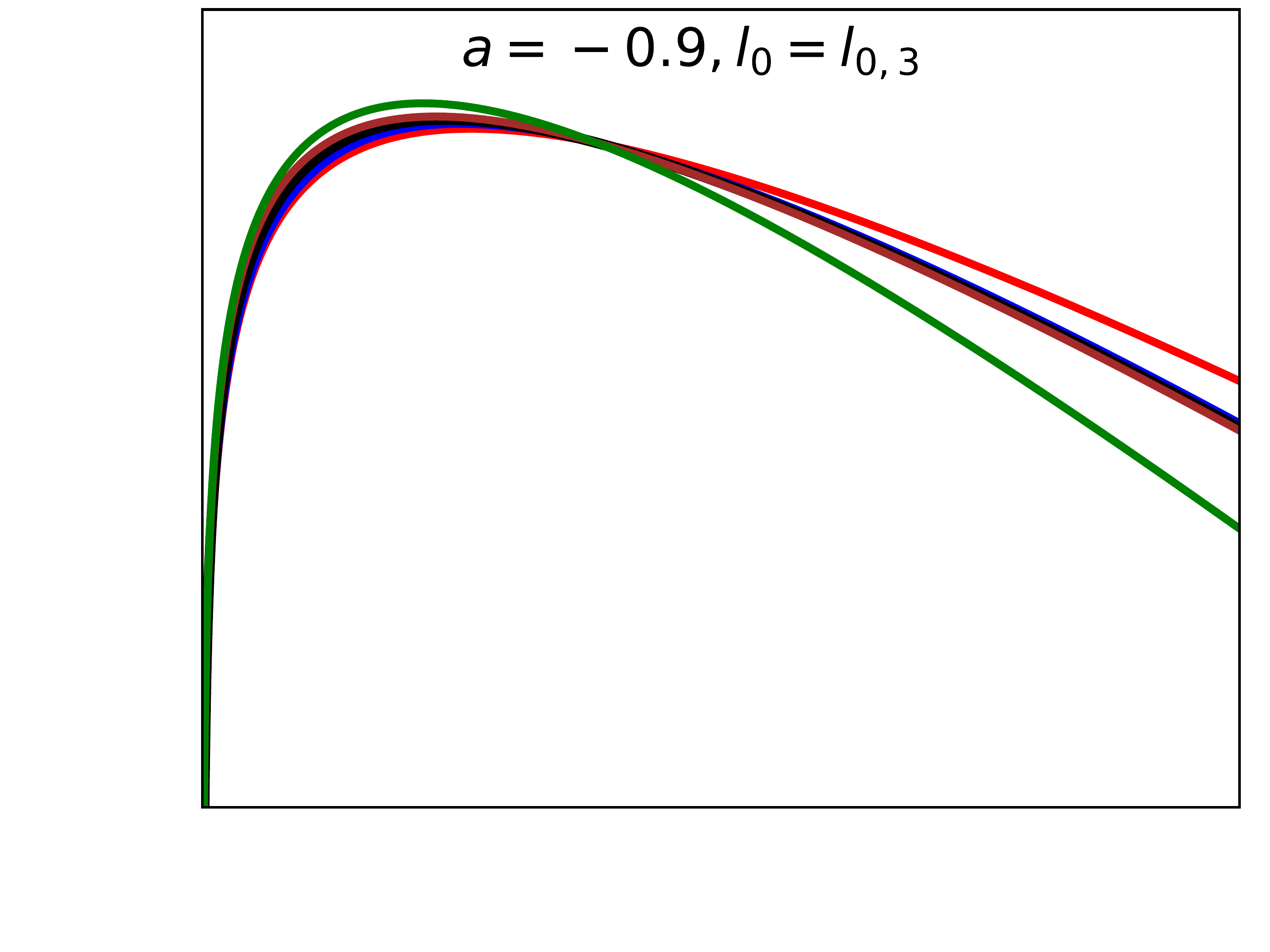}
\vspace{-0.3cm}
\\
\includegraphics[scale=0.08]{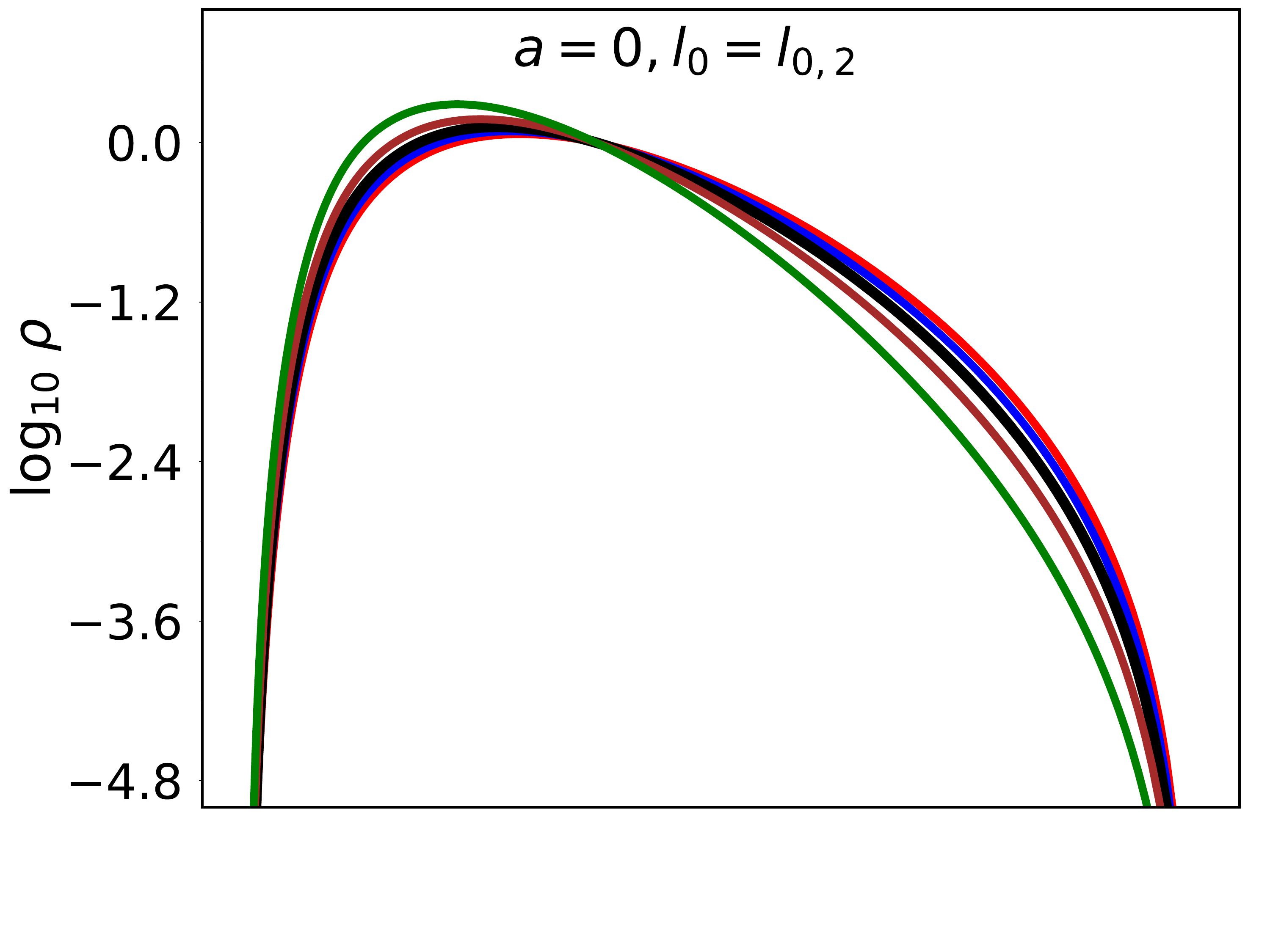}
\hspace{-0.9cm}
\includegraphics[scale=0.08]{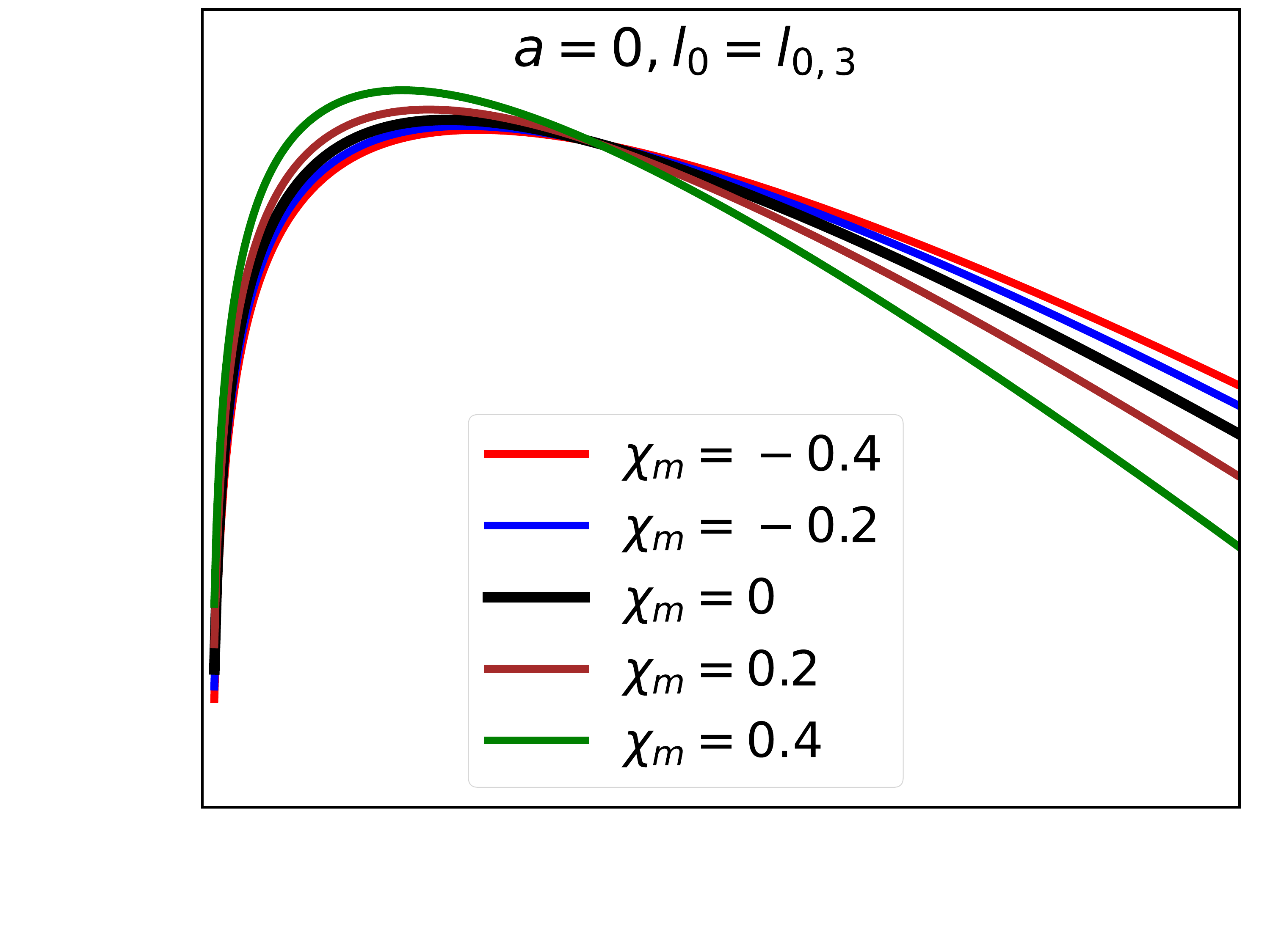}
\hspace{-0.9cm}
\includegraphics[scale=0.08]{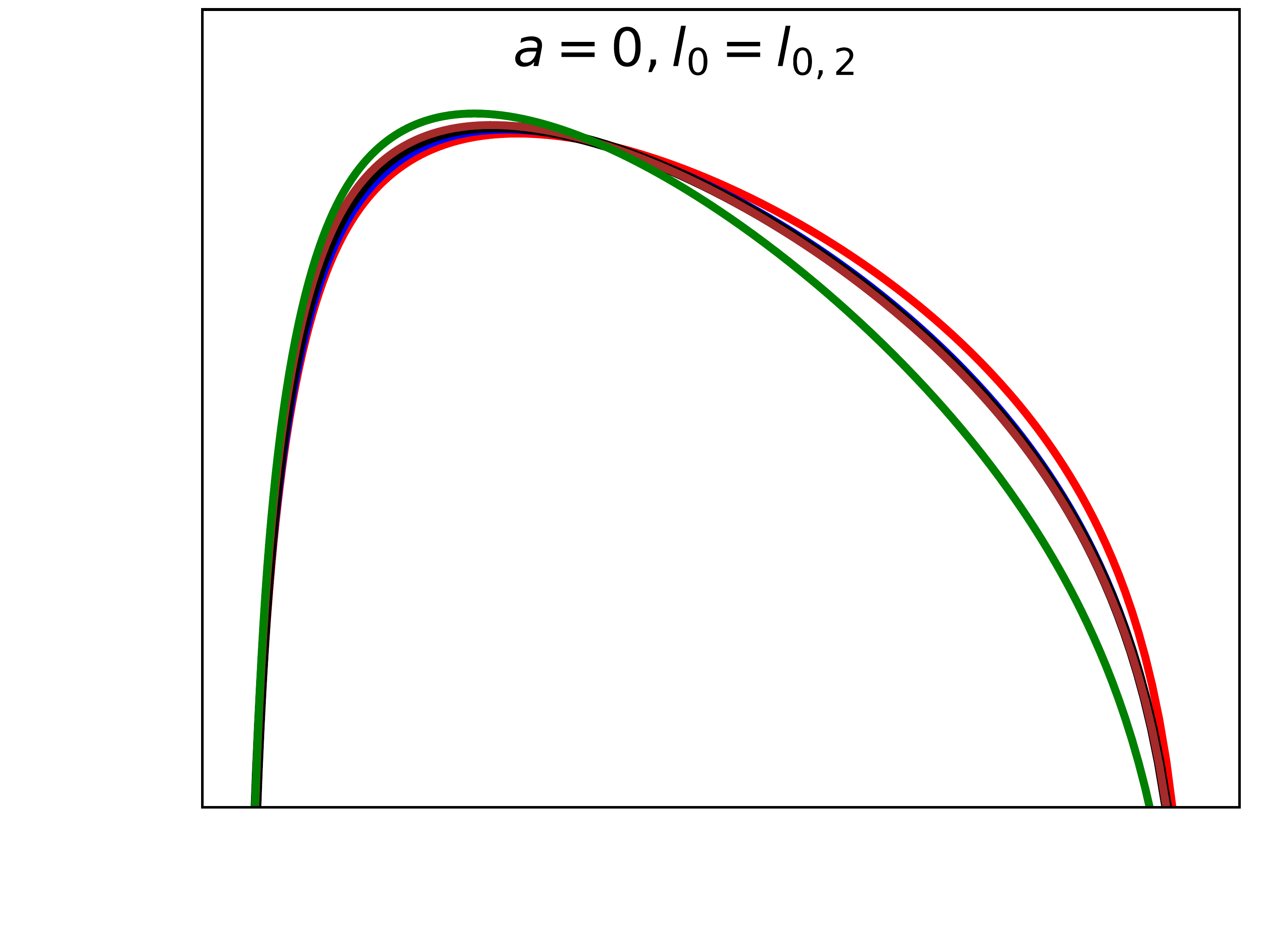}
\hspace{-0.9cm}
\includegraphics[scale=0.08]{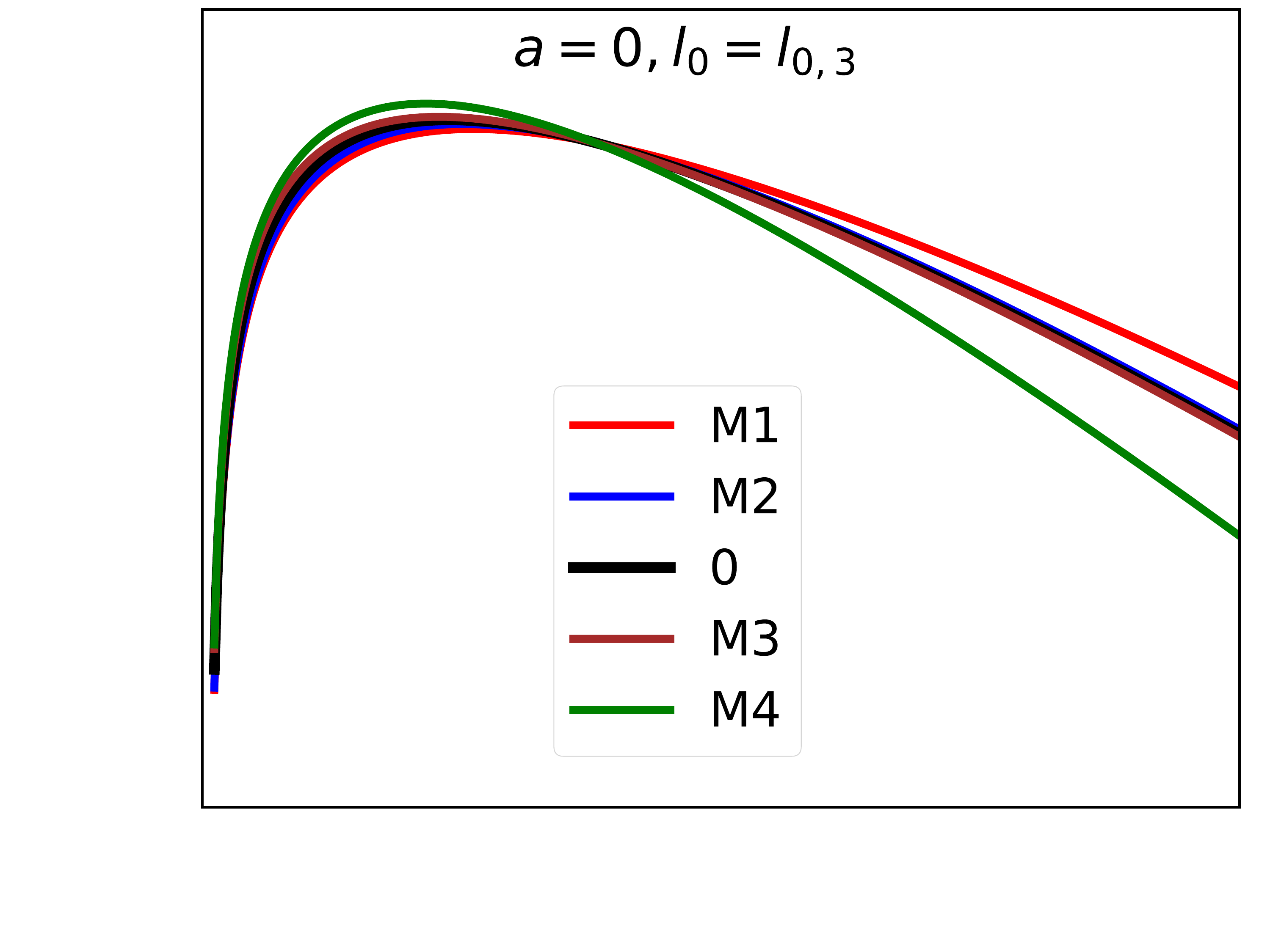}
\vspace{-0.3cm}
\\
\hspace{-0.2cm}
\includegraphics[scale=0.08]{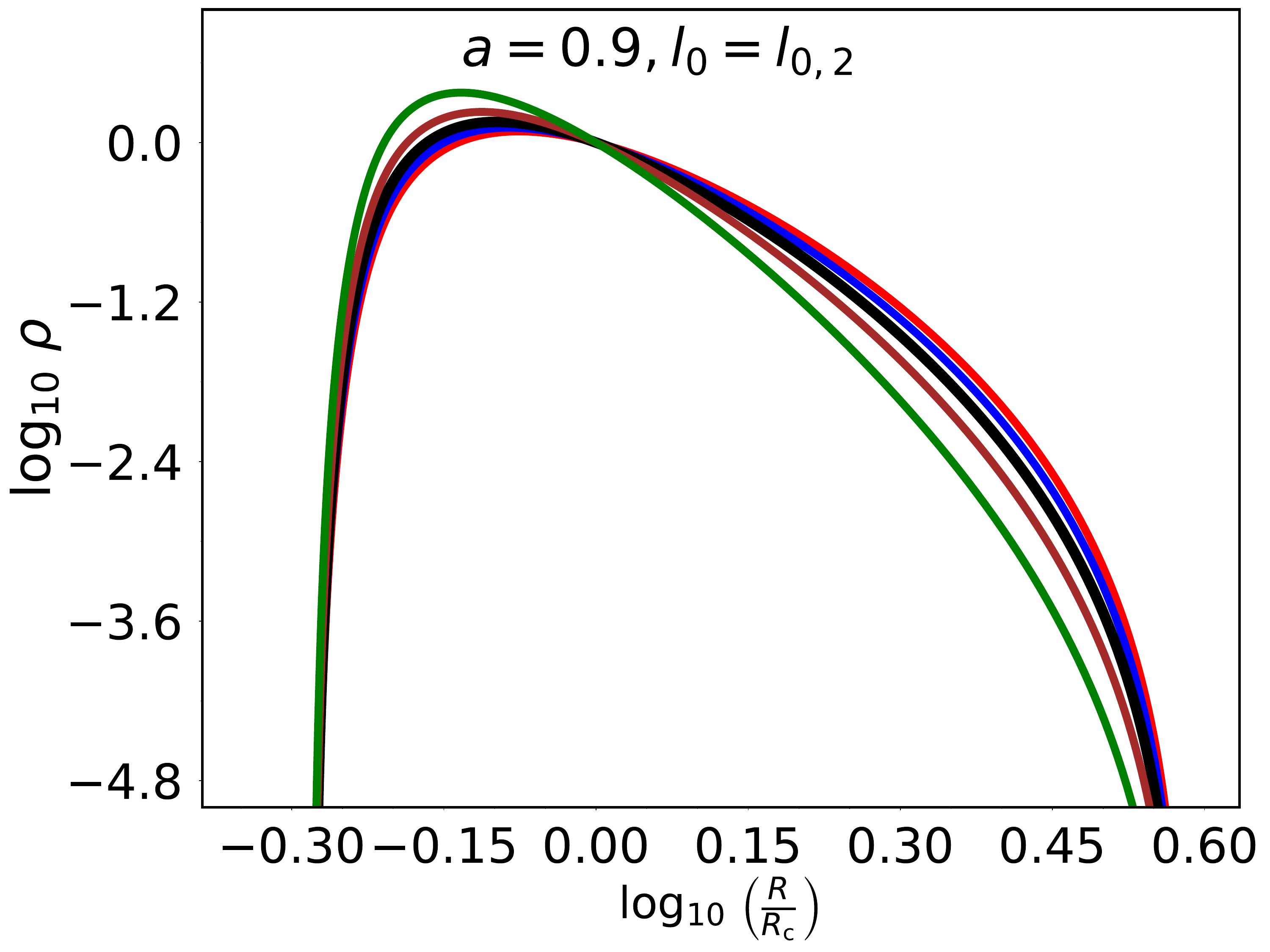}
\hspace{-0.9cm}
\includegraphics[scale=0.08]{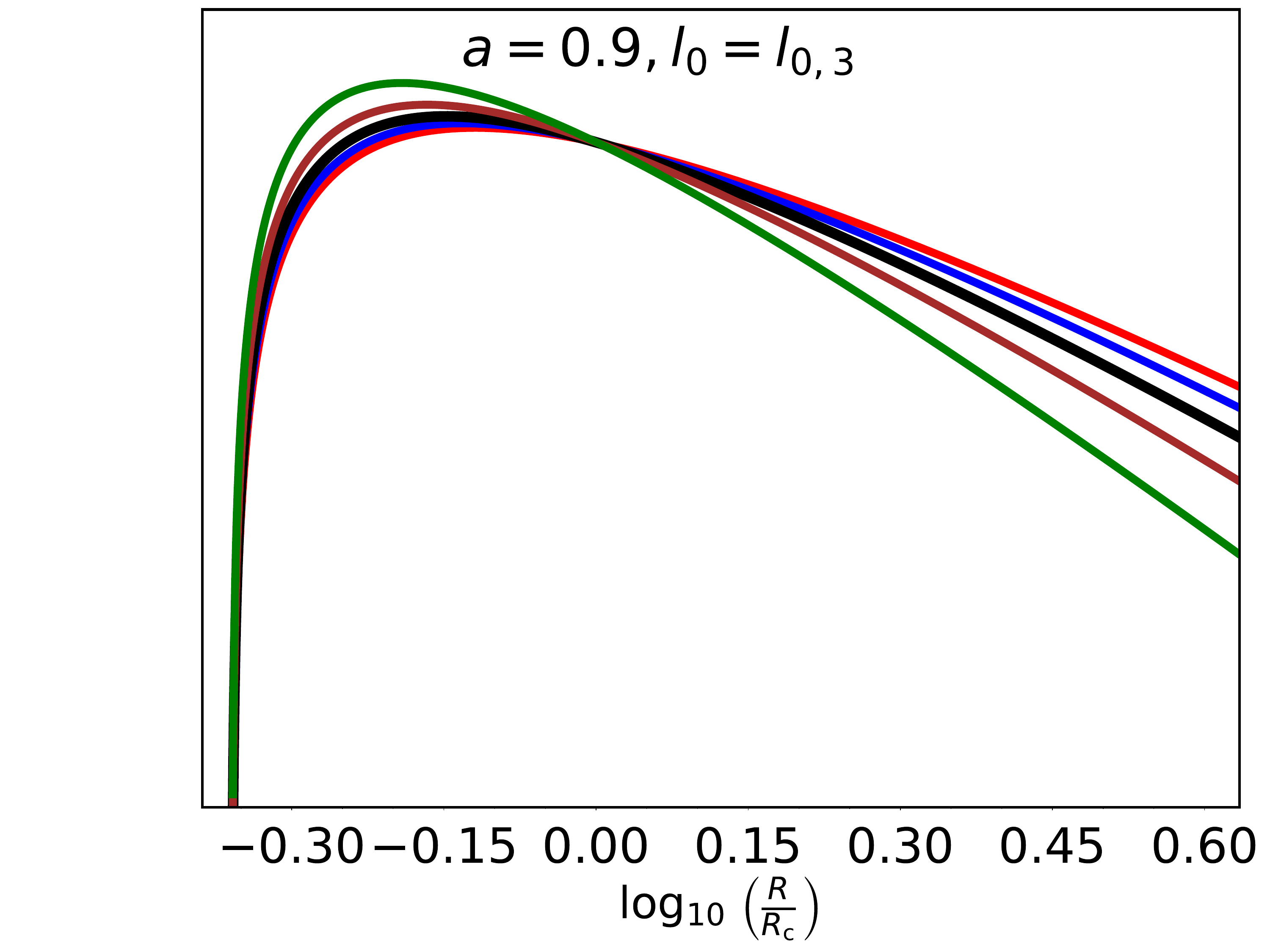}
\hspace{-0.9cm}
\includegraphics[scale=0.08]{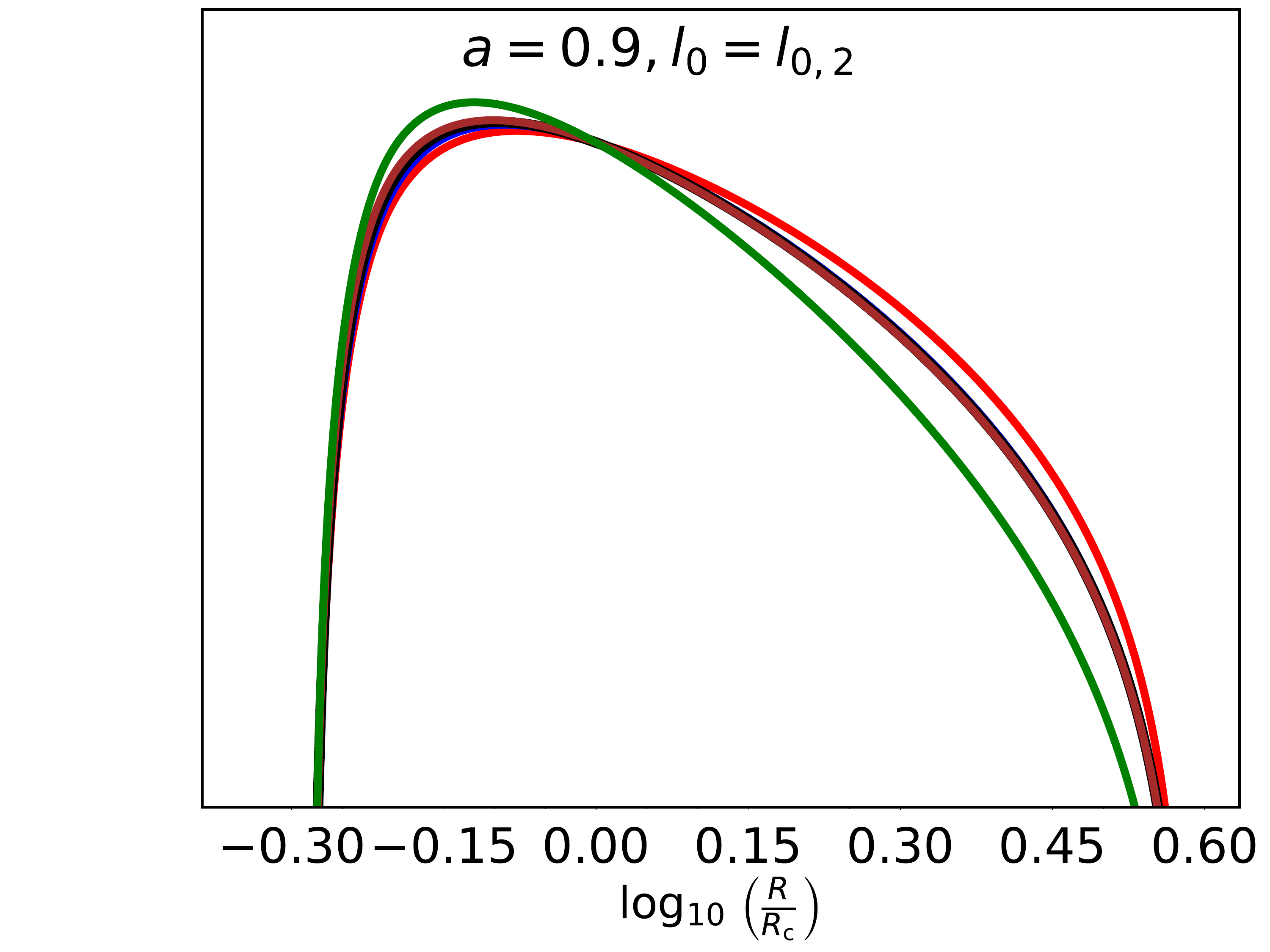}
\hspace{-0.9cm}
\includegraphics[scale=0.08]{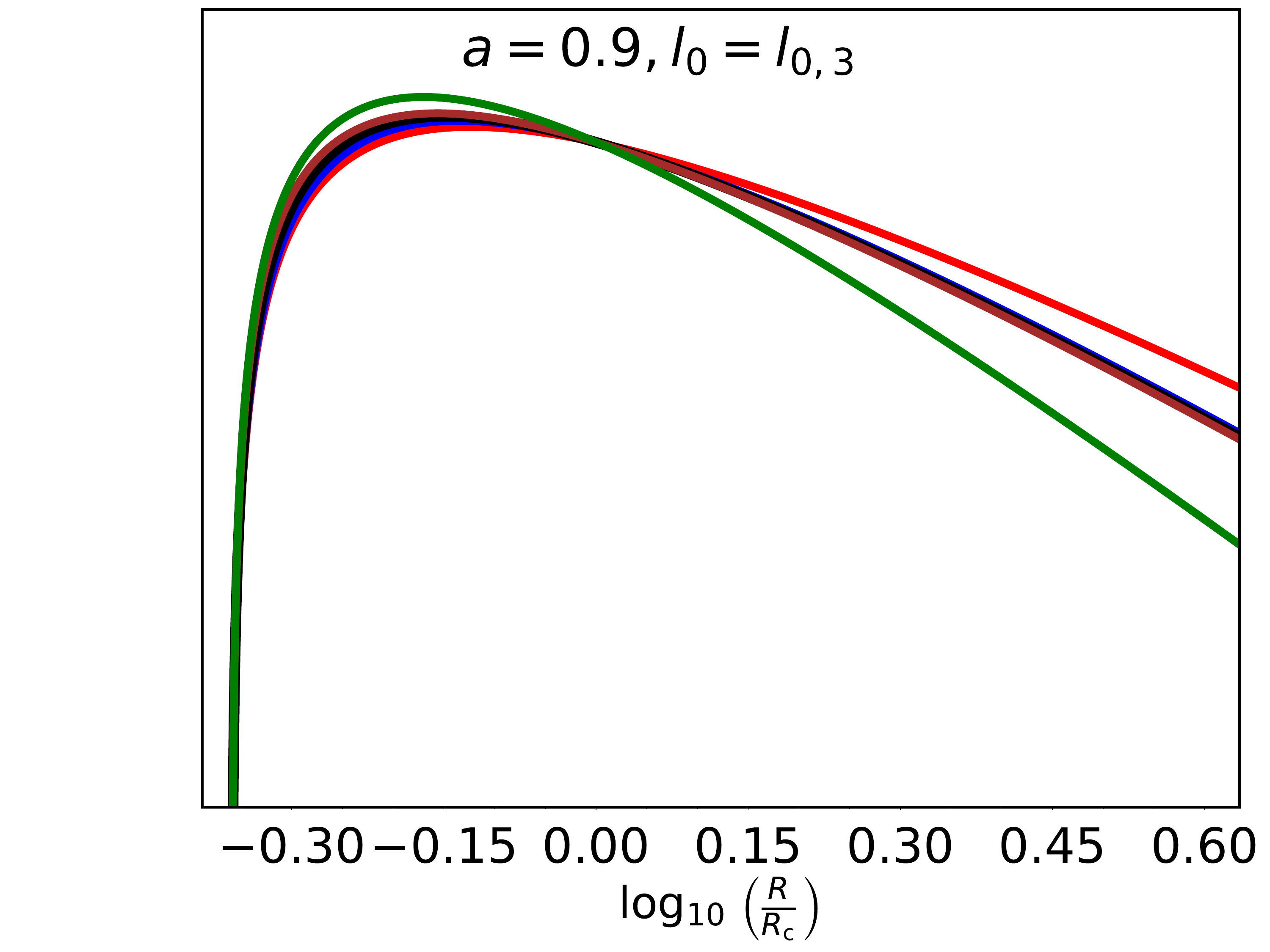}
\caption{Radial profiles of the rest-mass density $\rho$ in double logarithmic scale at the equatorial plane. The layout of this figure is the same as in Fig.~\ref{beta_r_fig}.}
\label{logrho_r_fig}
\end{figure*}

\begin{figure*}[t]
\includegraphics[scale=0.15]{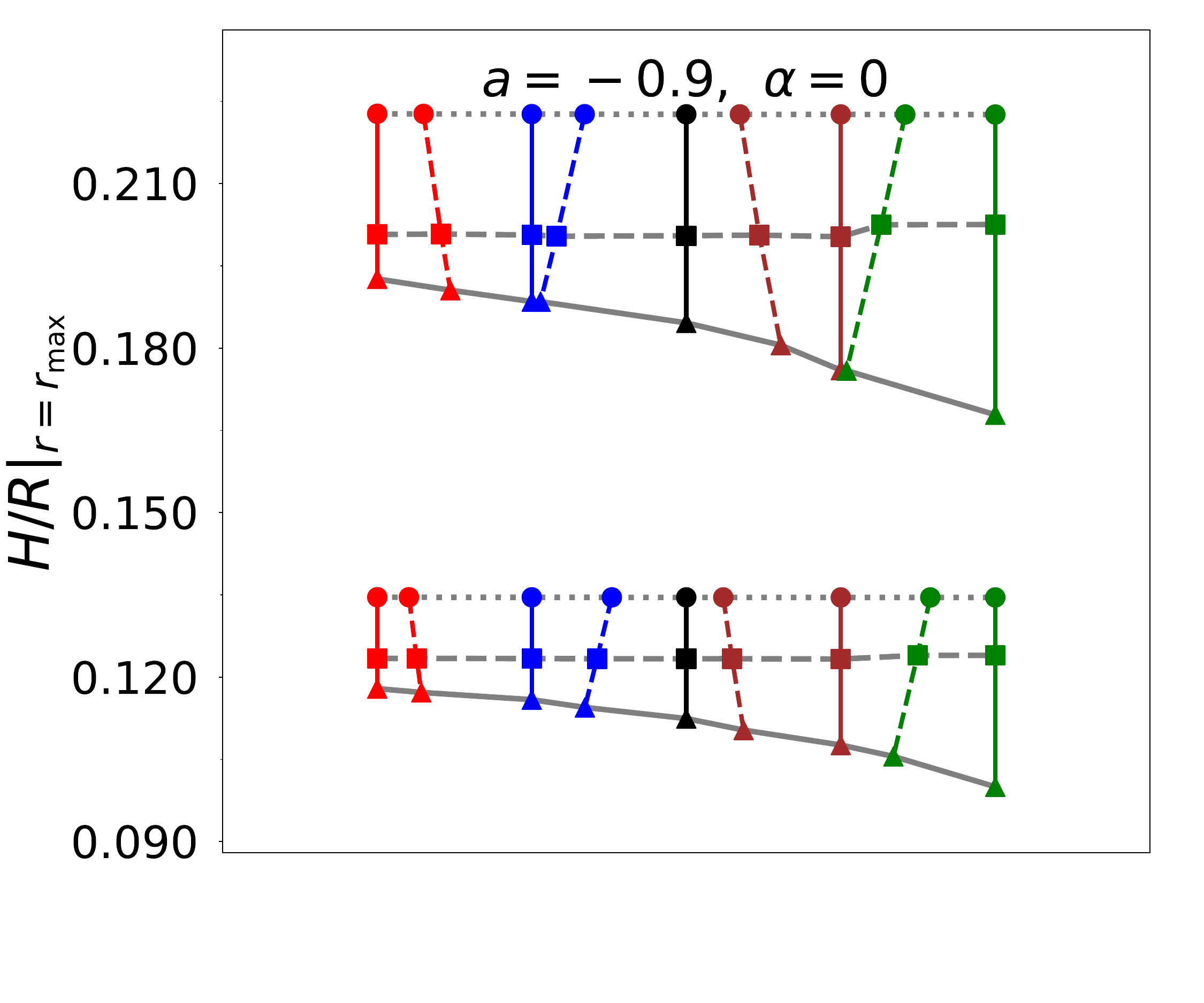}
\hspace{-1.1cm}
\includegraphics[scale=0.15]{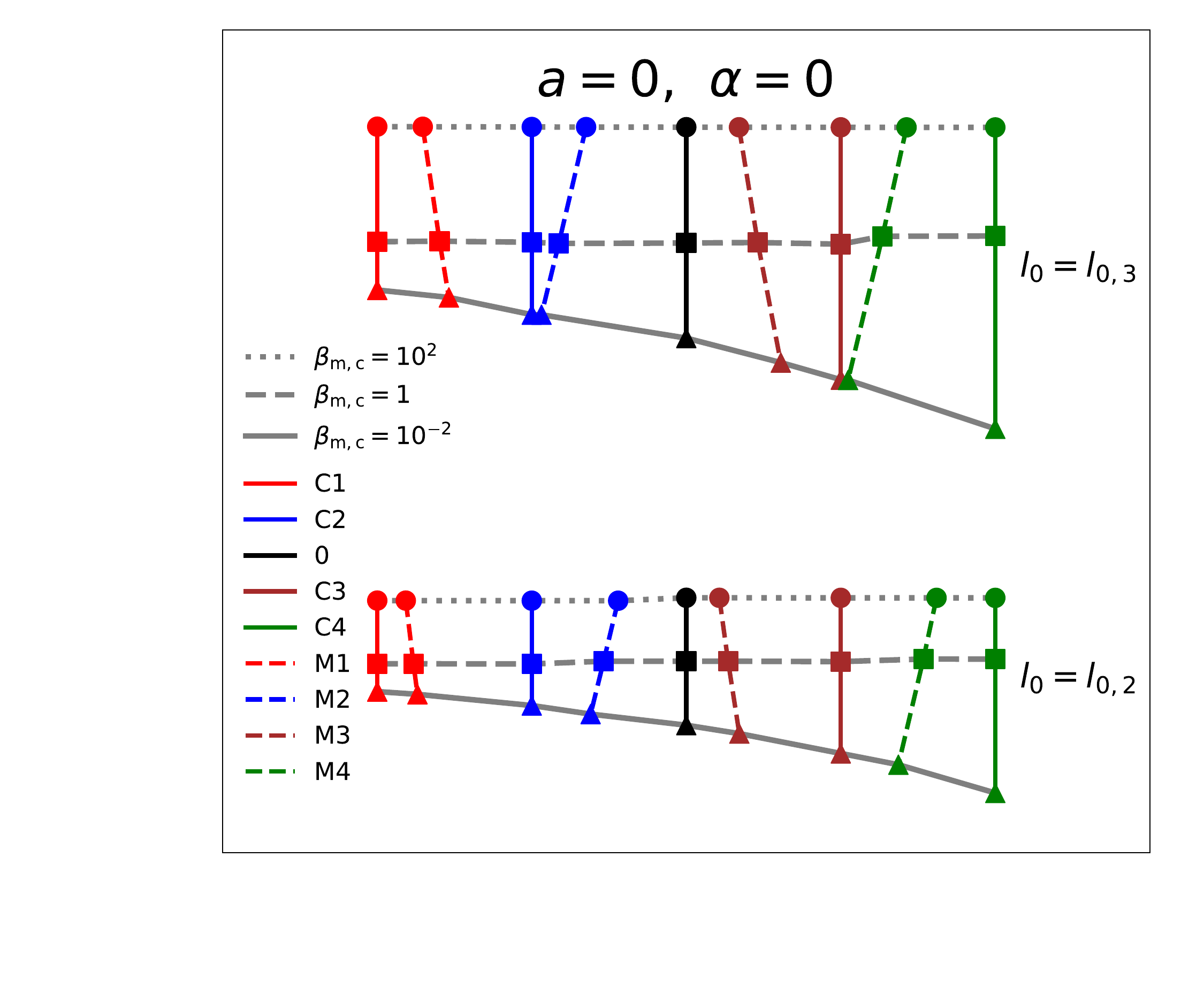}
\hspace{-1.1cm}
\includegraphics[scale=0.15]{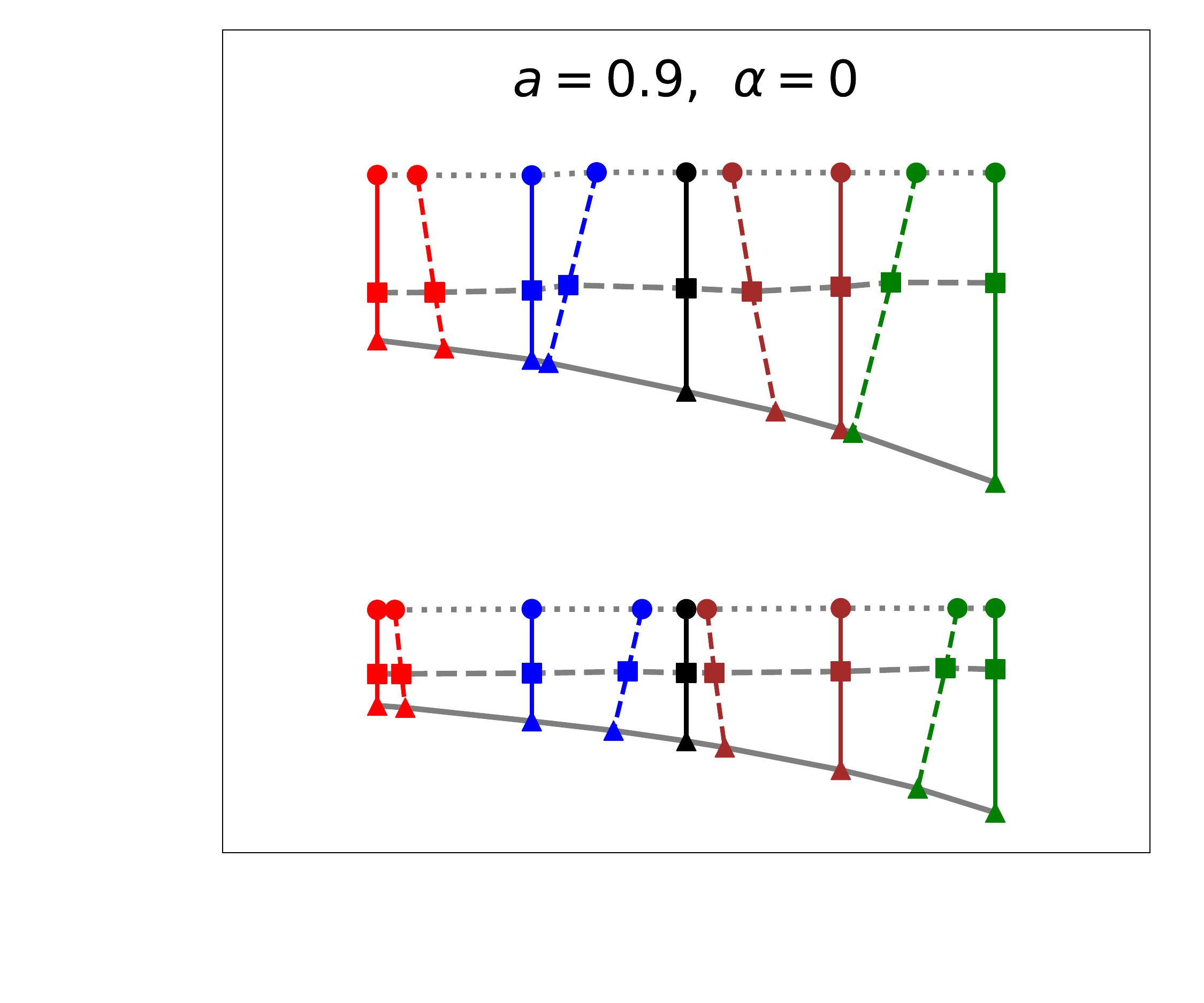}
\vspace{-0.5cm}
\\
\hspace{-0.2cm}
\includegraphics[scale=0.15]{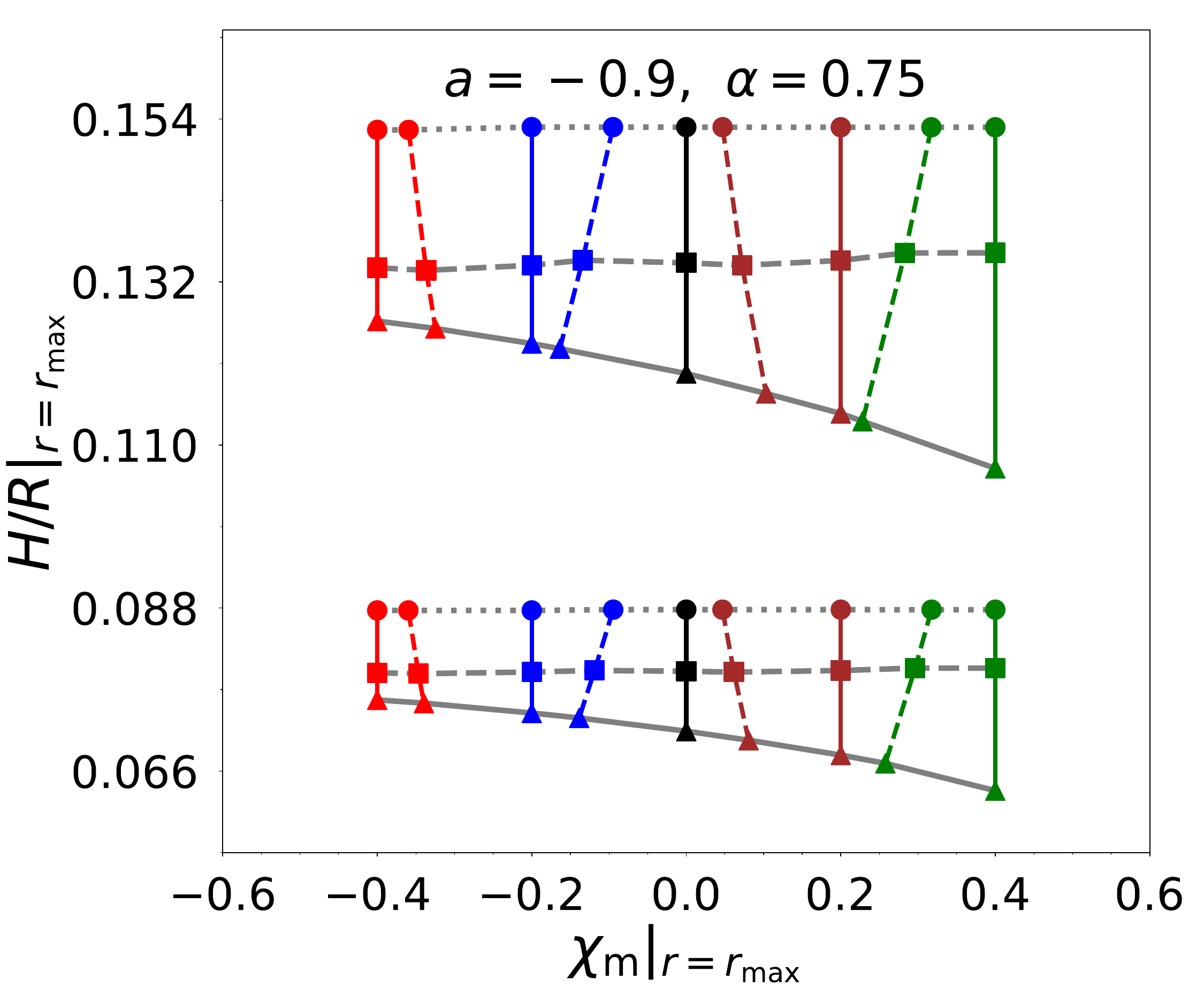}
\hspace{-1.1cm}
\includegraphics[scale=0.15]{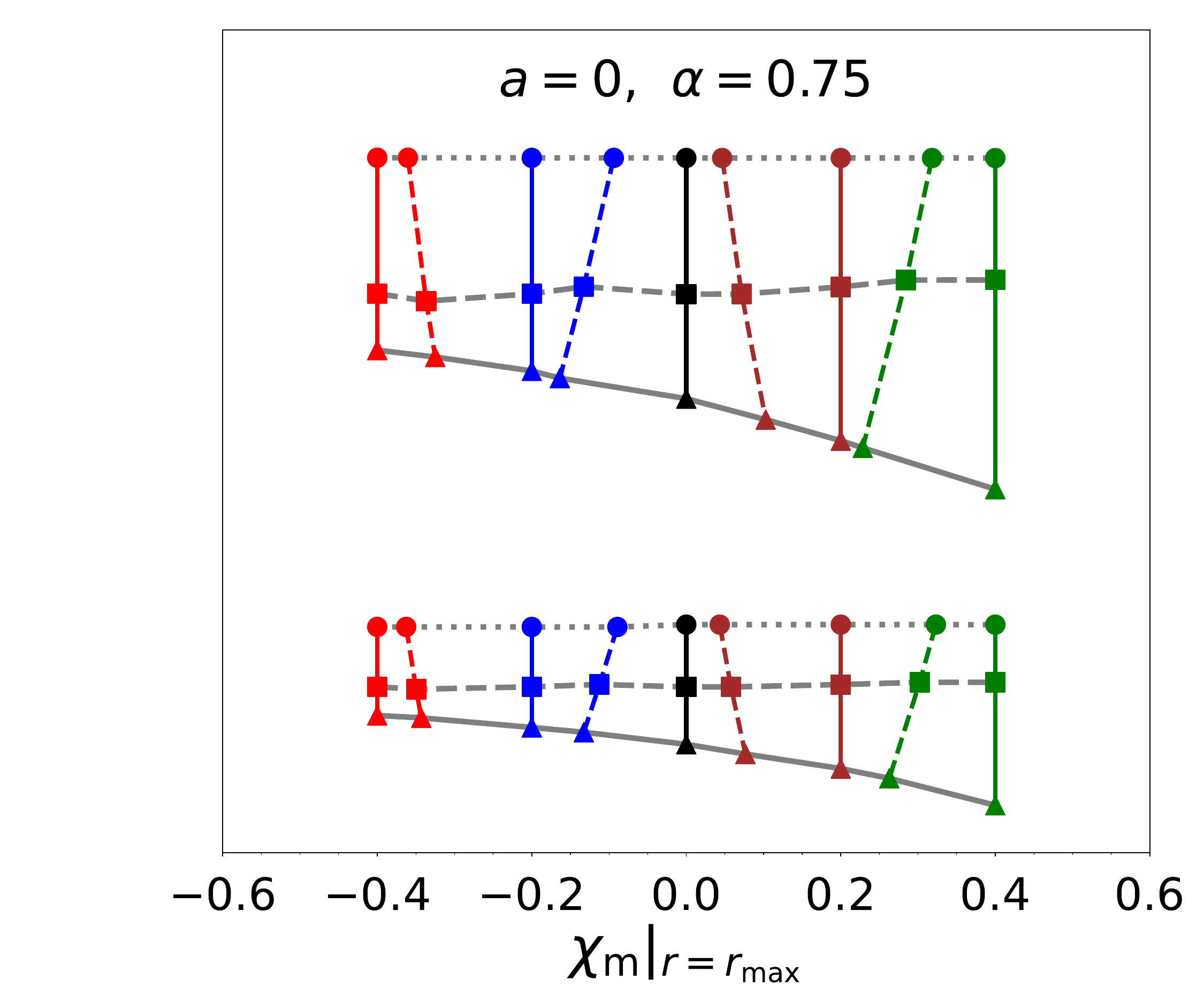}
\hspace{-1.1cm}
\includegraphics[scale=0.15]{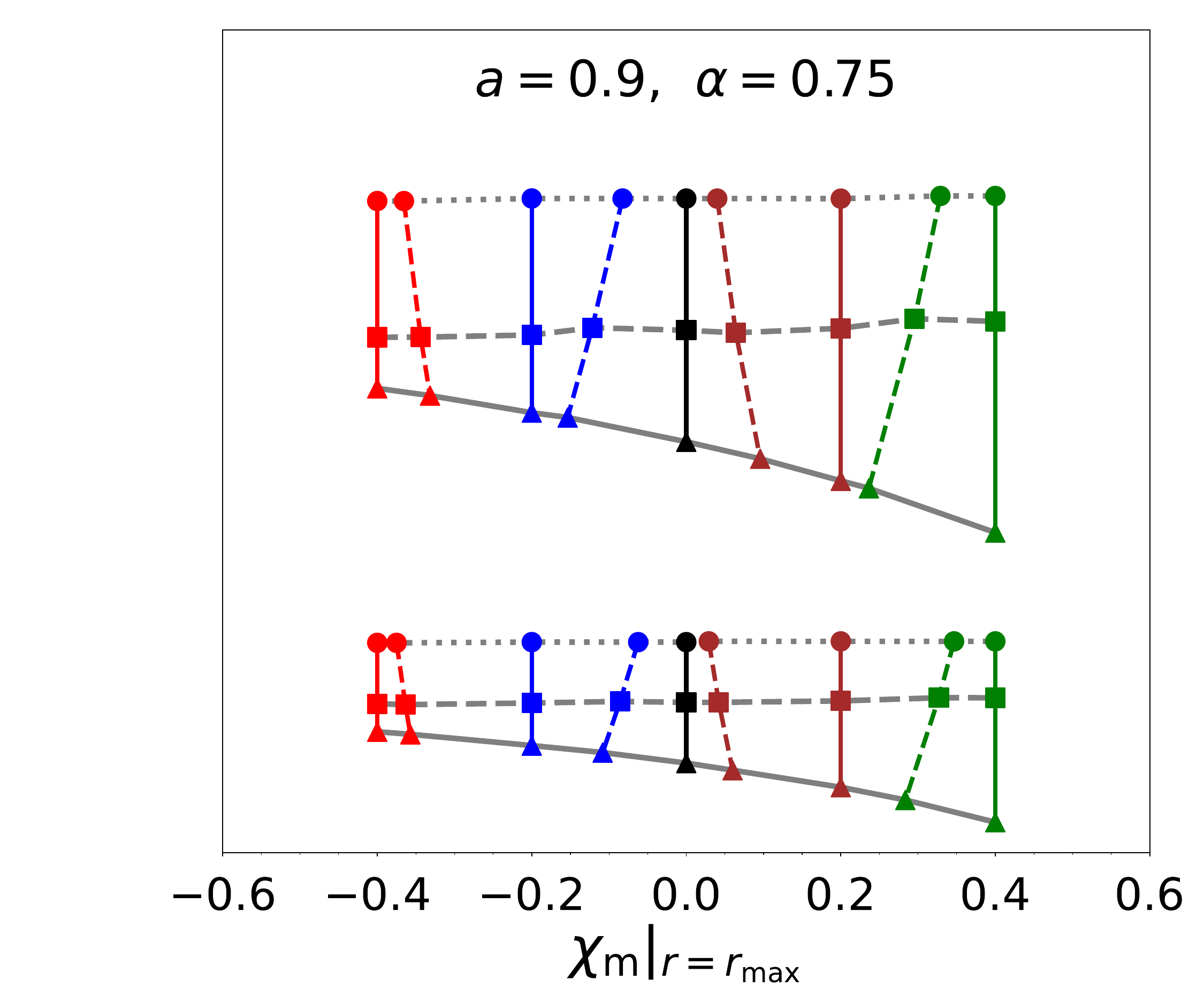}
\caption{Values of the relative thickness $H/R$ evaluated at the point of maximum rest mass density $\rho_{\mathrm{max}}$ as a function of the magnetic susceptibility $\chi_{\mathrm{m}}$ evaluated at the same point. In the rows, we show models with a different value of the exponent of the angular momentum ansatz, namely $\alpha = 0$ (top row) and $\alpha = 0.75$ (bottom row). In the columns, we show models with different values of the spin parameter of the black hole (namely, from left to right $a= -0.9$, $0$ and $0.9$). In each panel we show, at the top part, models with $l_0 = l_{0,3}$ and at the bottom part, the models for $l_0 = l_{0,2}$. Departing from the rest of the paper, here we computed models for three different values of the magnetization parameter at the center of the disk, namely $\beta_{m,c} = 10^2$, $1$ and $10^{-2}$. The points representing models with the same value of $\beta_{m,c}$ are denoted by circular, square and triangular markers respectively and are joined by dotted dashed and solid grey lines. In addition, the models built following the constant magnetic susceptibility prescriptions C1, C2, 0, C3 and C4, are denoted by red, blue, black, brown and green markers respectively (and joined by solid lines of the same colors) and the non-constant magnetic susceptibility models (M1 to M4) are denoted by red, blue, brown and green markers respectively and joined by dashed lines of the same colors.}
\label{h_r_chi_rmax}
\end{figure*}

\begin{table}
\caption{Values of the magnetic susceptibility $\chi_{\mathrm{m}}$ and the parameter $\chi_0$ for the constant models (top). Values of the parameters $\chi_0$ and $\chi_1^*$ for the non-constant susceptibility models (bottom).}        
\label{susceptibility_parameters_values}
\centering          
\begin{tabular}{c c c c c c}
\hline\hline
 & C1 & C2 & 0 & C3 & C4\\
 \hline
$\chi_{\mathrm{m}}$ & $-0.4$ & $-0.2$ & $0$ & $0.2$ & $0.4$\\
$\chi_{0}$ & $-2/3$ & $-1/4$ & $0$ & $1/6$ & $2/7$\\
\hline\hline

 & M1 & M2 & 0 & M3 & M4\\
 \hline
$\chi_{0}$ & $-2/3$ & $0$ & $0$ & $0$ & $2/7$\\
$\chi_{1}^*$ & $2/3$ & $-2/3$ & $0$ & $2/7$ & $-2/7$\\
\hline\hline
\end{tabular}
\end{table}

\begin{table*}
\caption{Values of the relevant physical magnitudes of our results for $a = -0.9$. From left to right, the columns correspond to the particular model of magnetic polarization we are considering, the constant part of the specific angular momentum distribution $l_0$ (for each model, the first and third rows corresponds to $l_0 = l_{0,2}$ and the second and fourth rows to $l_0 = l_{0,3}$), the exponent of the angular momentum distribution $\alpha$, the depth of the potential well at the center $\Delta W_{\mathrm{c}}$, the positions of the inner edge of the disk $R_{\mathrm{in}}$, the center of the disk $R_{\mathrm{c}}$ and the outer edge of the disk $R_{\mathrm{out}}$ \footnote{It must be noted that the mass of the disk is finite even if the disk has an infinite value of $R_{\mathrm{out}}$.}, the maximum value of the magnetization parameter $\beta_{\mathrm{m,max}}$, the value of the magnetization parameter at the inner edge of the disk $\beta_{\mathrm{m,in}}$, the maximum value of the rest-mass density $\rho_{\mathrm{max}}$, the location of the maximum of the rest-mass density $R_{\mathrm{max}}$ and the location of the maximum of the magnetic pressure $R_{\mathrm{m, max}}$.}        
\label{results__09}
\centering          
\begin{tabular}{c c c c  c c c c   c c c c}
\hline\hline       
 $\mathrm{Model}$ & $l_0$ & $\alpha$ & $\Delta W_{\mathrm{c}}$ &  $R_{\mathrm{in}}$ & $R_{\mathrm{c}}$ & $R_{\mathrm{out}}$ & $\beta_{\mathrm{m,max}} (\times 10^{-2})$ & $\beta_{\mathrm{m, in}} (\times 10^{-2})$ &$\rho_{\mathrm{max}}$ & $R_{\mathrm{max}}$ & $R_{\mathrm{m, max}}$\\ 
\hline

C1 & $4.61$ & $0$ & $-1.43 \times 10^{-2}$ & $6.09$ & $13.74$ & $45$ & $1.55$ & $1.55 $ & $1.12$ & $11.96$ & $12.33$ \\

 & $6.21$ & $0$ & $-1.43 \times 10^{-2}$ & $16.27$ & $33.02$ & $\infty$ & $1.42 $ & $1.42 $ & $1.16$ & $27.46$ & $28.42$ \\

 & $4.44$ & $0.75$ & $-4.66 \times 10^{-3}$ & $6.87$ & $15.85$ & $61.9$ & $1.56 $ & $1.56 $ & $1.15$ & $13.33$ & $13.83$ \\

 & $4.85$ & $0.75$ & $-4.66 \times 10^{-3}$ & $16.61$ & $40.60$ & $\infty$ & $1.55 $ & $1.55 $ & $1.25$ & $30.85$ & $32.49$ \\

 C2 & $4.61$ & $0$ & $-1.43 \times 10^{-2}$ & $6.09$ & $13.74$ & $45$ & $1.67$ & $1.67$ & $1.16$ & $11.78$ & $12.14$ \\

& $6.21$ & $0$ & $-1.43 \times 10^{-2}$ & $16.27$ & $33.02$ & $\infty$ & $1.50$ & $1.50$ & $1.22$ & $26.85$ & $28.09$ \\

 & $4.44$ & $0.75$ & $-4.66 \times 10^{-3}$ & $6.87$ & $15.85$ & $61.9$ & $1.67$ & $1.67$ & $1.20$ & $13.03$ & $13.58$ \\

 & $4.85$ & $0.75$ & $-4.66 \times 10^{-3}$ & $16.61$ & $40.60$ & $\infty$ & $1.67 $ & $1.67 $ & $1.34$ & $30.00$ & $31.65$ \\

0 & $4.61$ & $0$ & $-1.43 \times 10^{-2}$ & $6.09$ & $13.74$ & $45$ & $1.84$ & $1.84 $ & $1.23$ & $11.49$ & $11.90$ \\

 & $6.21$ & $0$ & $-1.43 \times 10^{-2}$ & $16.27$ & $33.02$ & $\infty$ & $1.63 $ & $1.63 $ & $1.31$ & $26.27$ & $27.15$ \\

 & $4.44$ & $0.75$ & $-4.66 \times 10^{-3}$ & $6.87$ & $15.85$ & $61.9$ & $1.86 $ & $1.86 $ & $1.29$ & $12.65$ & $13.25$ \\

 & $4.85$ & $0.75$ & $-4.66 \times 10^{-3}$ & $16.61$ & $40.60$ & $\infty$ & $1.85 $ & $1.85 $ & $1.49$ & $28.93$ & $30.66$ \\

 C3 & $4.61$ & $0$ & $-1.43 \times 10^{-2}$ & $6.09$ & $13.74$ & $45$ & $2.15$ & $2.15$ & $1.37$ & $11.10$ & $11.60$ \\

& $6.21$ & $0$ & $-1.43 \times 10^{-2}$ & $16.27$ & $33.02$ & $\infty$ & $1.84$ & $1.84$ & $1.49$ & $25.18$ & $26.56$ \\

 & $4.44$ & $0.75$ & $-4.66 \times 10^{-3}$ & $6.87$ & $15.85$ & $61.9$ & $2.17$ & $2.17 $ & $1.47$ & $12.18$ & $12.82$ \\

& $4.85$ & $0.75$ & $-4.66 \times 10^{-3}$ & $16.61$ & $40.60$ & $\infty$ & $2.16$ & $2.16$ & $1.78$ & $27.62$ & $29.37$ \\

  C4 & $4.61$ & $0$ & $-1.43 \times 10^{-2}$ & $6.09$ & $13.74$ & $45$ & $2.77$ & $2.77$ & $1.69$ & $10.56$ & $11.11$ \\

& $6.21$ & $0$ & $-1.43 \times 10^{-2}$ & $16.27$ & $33.02$ & $\infty$ & $2.25$ & $2.25$ & $1.89$ & $24.18$ & $25.45$ \\

 & $4.44$ & $0.75$ & $-4.66 \times 10^{-3}$ & $6.87$ & $15.85$ & $61.9$ & $2.80$ & $2.80$ & $1.88$ & $11.54$ & $12.22$ \\

& $4.85$ & $0.75$ & $-4.66 \times 10^{-3}$ & $16.61$ & $40.60$ & $\infty$ & $2.80$ & $2.80$ & $2.49$ & $26.00$ & $27.78$ \\
 \hline 
 M1 & $4.61$ & $0$ & $-1.43 \times 10^{-2}$ & $6.09$ & $13.74$ & $45$ & $1.07$ & $0.78$ & $1.13$ & $11.90$ & $12.08$ \\
 & $6.21$ & $0$ & $-1.43 \times 10^{-2}$ & $16.27$ & $33.02$ & $\infty$ & $1.03$ & $0.75$ & $1.18$ & $27.15$ & $27.25$ \\
 & $4.44$ & $0.75$ & $-4.66 \times 10^{-3}$ & $6.87$ & $15.85$ & $61.9$ & $1.07$ & $0.77$ & $1.16$ & $13.25$ & $13.46$ \\
 & $4.85$ & $0.75$ & $-4.66 \times 10^{-3}$ & $16.61$ & $40.60$ & $\infty$ & $1.07$ & $0.78$ & $1.27$ & $30.56$ & $30.95$ \\
 M2 & $4.61$ & $0$ & $-1.43 \times 10^{-2}$ & $6.09$ & $13.74$ & $45$ & $3.17$ & $3.17$ & $1.19$ & $11.66$ & $12.40$ \\
 & $6.21$ & $0$ & $-1.43 \times 10^{-2}$ & $16.27$ & $33.02$ & $\infty$ & $2.62$ & $2.62$ & $1.23$ & $26.85$ & $29.10$ \\
 & $4.44$ & $0.75$ & $-4.66 \times 10^{-3}$ & $6.87$ & $15.85$ & $61.9$ & $3.21$ & $3.21$ & $1.24$ & $12.93$ & $13.93$ \\
 & $4.85$ & $0.75$ & $-4.66 \times 10^{-3}$ & $16.61$ & $40.60$ & $\infty$ & $3.20$ & $3.20$ & $1.38$ & $29.82$ & $33.15$ \\
 M3 & $4.61$ & $0$ & $-1.43 \times 10^{-2}$ & $6.09$ & $13.74$ & $45$ & $1.20$ & $1.00$ & $1.27$ & $11.32$ & $11.60$ \\
 & $6.21$ & $0$ & $-1.43 \times 10^{-2}$ & $16.27$ & $33.02$ & $\infty$ & $1.10$ & $0.92$ & $1.38$ & $25.71$ & $25.99$ \\
 & $4.44$ & $0.75$ & $-4.66 \times 10^{-3}$ & $6.87$ & $15.85$ & $61.9$ & $1.20$ & $0.99$ & $1.34$ & $12.47$ & $12.80$ \\
 & $4.85$ & $0.75$ & $-4.66 \times 10^{-3}$ & $16.61$ & $40.60$ & $\infty$ & $1.20$ & $1.00$ & $1.57$ & $28.26$ & $28.93$ \\
  M4 & $4.61$ & $0$ & $-1.43 \times 10^{-2}$ & $6.09$ & $13.74$ & $45$ & $4.13$ & $4.13$ & $1.48$ & $10.95$ & $11.72$ \\
 & $6.21$ & $0$ & $-1.43 \times 10^{-2}$ & $16.27$ & $33.02$ & $\infty$ & $3.19$ & $3.19$ & $1.55$ & $25.18$ & $27.46$ \\
 & $4.44$ & $0.75$ & $-4.66 \times 10^{-3}$ & $6.87$ & $15.85$ & $61.9$ & $4.19$ & $4.19$ & $1.60$ & $12.02$ & $13.03$ \\
 & $4.85$ & $0.75$ & $-4.66 \times 10^{-3}$ & $16.61$ & $40.60$ & $\infty$ & $4.18$ & $4.18$ & $1.97$ & $27.39$ & $30.37$ \\
\hline\hline
\end{tabular}
\end{table*}

\begin{table*}
\caption{Same as in Table~\ref{results__09} but for $a = 0$.}        
\label{results_0}
\centering          
\begin{tabular}{c c c c  c c c c   c c c c}
\hline\hline       
 $\mathrm{Model}$ & $l_0$ & $\alpha$ & $\Delta W_{\mathrm{c}}$ &  $R_{\mathrm{in}}$ & $R_{\mathrm{c}}$ & $R_{\mathrm{out}}$ & $\beta_{\mathrm{m,max}} (\times 10^{-2})$ & $\beta_{\mathrm{m, in}} (\times 10^{-2})$ &$\rho_{\mathrm{max}}$ & $R_{\mathrm{max}}$ & $R_{\mathrm{m, max}}$\\ 
\hline 
C1 & $3.88$ & $0$ & $-2.14 \times 10^{-2}$ & $4.28$ & $9.30$ & $30$ & $1.58$ & $1.58$ & $1.13$ & $8.09$ & $8.33$ \\
 & $5.18$ & $0$ & $-2.14 \times 10^{-2}$ & $11.02$ & $22.21$ & $\infty$ & $1.42$ & $1.42$ & $1.16$ & $18.41$ & $19.18$ \\
 & $3.76$ & $0.75$ & $-6.84 \times 10^{-3}$ & $4.79$ & $10.65$ & $41$ & $1.58$ & $1.58$ & $1.16$ & $8.95$ & $9.29$ \\
 & $4.08$ & $0.75$ & $-6.84 \times 10^{-3}$ & $11.15$ & $26.74$ & $\infty$ & $1.55$ & $1.55$ & $1.24$ & $20.36$ & $21.50$ \\
 C2 & $3.88$ & $0$ & $-2.14 \times 10^{-2}$ & $4.28$ & $9.30$ & $30$ & $1.71$ & $1.71$ & $1.17$ & $7.94$ & $8.21$ \\
 & $5.18$ & $0$ & $-2.14 \times 10^{-2}$ & $11.02$ & $22.21$ & $\infty$ & $1.51$ & $1.51$ & $1.22$ & $17.98$ & $18.71$ \\
 & $3.76$ & $0.75$ & $-6.84 \times 10^{-3}$ & $4.79$ & $10.65$ & $41$ & $1.70$ & $1.70$ & $1.21$ & $8.77$ & $9.14$ \\
 & $4.08$ & $0.75$ & $-6.84 \times 10^{-3}$ & $11.15$ & $26.74$ & $\infty$ & $1.70$ & $1.70$ & $1.33$ & $19.84$ & $20.91$ \\
0 & $3.88$ & $0$ & $-2.14 \times 10^{-2}$ & $4.28$ & $9.30$ & $30$ & $1.90$ & $1.90$ & $1.25$ & $7.74$ & $8.06$ \\
 & $5.18$ & $0$ & $-2.14 \times 10^{-2}$ & $11.02$ & $22.21$ & $\infty$ & $1.64$ & $1.64$ & $1.31$ & $17.56$ & $18.26$ \\
 & $3.76$ & $0.75$ & $-6.84 \times 10^{-3}$ & $4.79$ & $10.65$ & $41$ & $1.90$ & $1.90$ & $1.31$ & $8.53$ & $8.91$ \\
 & $4.08$ & $0.75$ & $-6.84 \times 10^{-3}$ & $11.15$ & $26.74$ & $\infty$ & $1.85$ & $1.85$ & $1.48$ & $19.18$ & $20.36$ \\
 C3 & $3.88$ & $0$ & $-2.14 \times 10^{-2}$ & $4.28$ & $9.30$ & $30$ & $2.24$ & $2.24$ & $1.40$ & $7.47$ & $7.82$ \\
 & $5.18$ & $0$ & $-2.14 \times 10^{-2}$ & $11.02$ & $22.21$ & $\infty$ & $1.85$ & $1.85$ & $1.49$ & $16.91$ & $17.70$ \\
 & $3.76$ & $0.75$ & $-6.84 \times 10^{-3}$ & $4.79$ & $10.65$ & $41$ & $2.23$ & $2.23$ & $1.50$ & $8.21$ & $8.63$ \\
 & $4.08$ & $0.75$ & $-6.84 \times 10^{-3}$ & $11.15$ & $26.74$ & $\infty$ & $2.16$ & $2.16$ & $1.77$ & $18.26$ & $19.50$ \\
  C4 & $3.88$ & $0$ & $-2.14 \times 10^{-2}$ & $4.28$ & $9.30$ & $30$ & $2.92$ & $2.92$ & $1.74$ & $7.13$ & $7.50$ \\
 & $5.18$ & $0$ & $-2.14 \times 10^{-2}$ & $11.02$ & $22.21$ & $\infty$ & $2.27$ & $2.27$ & $1.89$ & $16.19$ & $17.04$ \\
 & $3.76$ & $0.75$ & $-6.84 \times 10^{-3}$ & $4.79$ & $10.65$ & $41$ & $2.91$ & $2.91$ & $1.94$ & $7.77$ & $8.24$ \\
 & $4.08$ & $0.75$ & $-6.84 \times 10^{-3}$ & $11.15$ & $26.74$ & $\infty$ & $2.79$ & $2.79$ & $2.47$ & $17.30$ & $18.41$ \\
 \hline 
 M1 & $3.88$ & $0$ & $-2.14 \times 10^{-2}$ & $4.28$ & $9.30$ & $30$ & $1.09$ & $0.79$ & $1.13$ & $8.06$ & $8.18$ \\
 & $5.18$ & $0$ & $-2.14 \times 10^{-2}$ & $11.02$ & $22.21$ & $\infty$ & $1.03$ & $0.75$ & $1.18$ & $18.26$ & $18.34$ \\
 & $3.76$ & $0.75$ & $-6.84 \times 10^{-3}$ & $4.79$ & $10.65$ & $41$ & $1.09$ & $0.79$ & $1.17$ & $8.91$ & $9.06$ \\
 & $4.08$ & $0.75$ & $-6.84 \times 10^{-3}$ & $11.15$ & $26.74$ & $\infty$ & $1.08$ & $0.79$ & $1.26$ & $20.18$ & $20.54$ \\
 M2 & $3.88$ & $0$ & $-2.14 \times 10^{-2}$ & $4.28$ & $9.30$ & $30$ & $3.32$ & $3.32$ & $1.20$ & $7.85$ & $8.37$ \\
 &  $5.18$ & $0$ & $-2.14 \times 10^{-2}$ & $11.02$ & $22.21$ & $\infty$ & $2.64$ & $2.64$ & $1.23$ & $17.98$ & $19.50$ \\
 & $3.76$ & $0.75$ & $-6.84 \times 10^{-3}$ & $4.79$ & $10.65$ & $41$ & $3.30$ & $3.30$ & $1.25$ & $8.70$ & $9.33$ \\
 & $4.08$ & $0.75$ & $-6.84 \times 10^{-3}$ & $11.15$ & $26.74$ & $\infty$ & $3.17$ & $3.17$ & $1.38$ & $19.67$ & $21.90$ \\
 M3 & $3.88$ & $0$ & $-2.14 \times 10^{-2}$ & $4.28$ & $9.30$ & $30$ & $1.23$ & $1.02$ & $1.28$ & $7.66$ & $7.85$ \\
 & $5.18$ & $0$ & $-2.14 \times 10^{-2}$ & $11.02$ & $22.21$ & $\infty$ & $1.11$ & $0.93$ & $1.38$ & $17.17$ & $17.43$ \\
 & $3.76$ & $0.75$ & $-6.84 \times 10^{-3}$ & $4.79$ & $10.65$ & $41$ & $1.23$ & $1.02$ & $1.35$ & $8.40$ & $8.63$ \\
 & $4.08$ & $0.75$ & $-6.84 \times 10^{-3}$ & $11.15$ & $26.74$ & $\infty$ & $1.21$ & $1.01$ & $1.56$ & $18.71$ & $19.18$ \\
 M4 & $3.88$ & $0$ & $-2.14 \times 10^{-2}$ & $4.28$ & $9.30$ & $30$ & $4.39$ & $4.39$ & $1.53$ & $7.37$ & $7.88$ \\
 & $5.18$ & $0$ & $-2.14 \times 10^{-2}$ & $11.02$ & $22.21$ & $\infty$ & $3.23$ & $3.23$ & $1.55$ & $16.91$ & $18.41$ \\
 & $3.76$ & $0.75$ & $-6.84 \times 10^{-3}$ & $4.79$ & $10.65$ & $41$ & $4.36$ & $4.36$ & $1.65$ & $8.09$ & $8.74$ \\
 & $4.08$ & $0.75$ & $-6.84 \times 10^{-3}$ & $11.15$ & $26.74$ & $\infty$ & $4.14$ & $4.14$ & $1.96$ & $18.12$ & $20.18$ \\
\hline\hline
\end{tabular}
\end{table*}

\begin{table*}
\caption{Same as in Table~\ref{results__09} but for $a = 0.9$.}        
\label{results_09}
\centering          
\begin{tabular}{c c c c  c c c c   c c c c}
\hline\hline       
 $\mathrm{Model}$ & $l_0$ & $\alpha$ & $\Delta W_{\mathrm{c}}$ &  $R_{\mathrm{in}}$ & $R_{\mathrm{c}}$ & $R_{\mathrm{out}}$ & $\beta_{\mathrm{m,max}} (\times 10^{-2})$ & $\beta_{\mathrm{m, in}} (\times 10^{-2})$ &$\rho_{\mathrm{max}}$ & $R_{\mathrm{max}}$ & $R_{\mathrm{m, max}}$\\ 
\hline 
C1 & $2.58$ & $0$ & $-6.44 \times 10^{-2}$ & $1.81$ & $3.30$ & $10.83$ & $1.77$ & $1.77$ & $1.18$ & $2.86$ & $2.95$ \\
 & $3.27$ & $0$ & $-6.44 \times 10^{-2}$ & $3.90$ & $7.51$ & $\infty$ & $1.47$ & $1.47$ & $1.20$ & $6.18$ & $6.43$ \\
 & $2.53$ & $0.75$ & $-2.03 \times 10^{-2}$ & $1.95$ & $3.73$ & $14.18$ & $1.73$ & $1.73$ & $1.21$ & $3.13$ & $3.25$ \\
 & $2.71$ & $0.75$ & $-2.03 \times 10^{-3}$ & $3.99$ & $9.13$ & $\infty$ & $1.62$ & $1.62$ & $1.29$ & $6.94$ & $7.29$ \\
 C2 & $2.58$ & $0$ & $-6.44 \times 10^{-2}$ & $1.81$ & $3.30$ & $10.83$ & $1.95$ & $1.95$ & $1.25$ & $2.80$ & $2.90$ \\
 & $3.27$ & $0$ & $-6.44 \times 10^{-2}$ & $3.90$ & $7.51$ & $\infty$ & $1.57$ & $1.57$ & $1.27$ & $6.07$ & $6.30$ \\
 & $2.53$ & $0.75$ & $-2.03 \times 10^{-2}$ & $1.95$ & $3.73$ & $14.18$ & $1.90$ & $1.90$ & $1.29$ & $3.07$ & $3.19$ \\
 & $2.71$ & $0.75$ & $-2.03 \times 10^{-3}$ & $3.99$ & $9.13$ & $\infty$ & $1.75$ & $1.75$ & $1.39$ & $6.73$ & $7.11$ \\
0 & $2.58$ & $0$ & $-6.44 \times 10^{-2}$ & $1.81$ & $3.30$ & $10.83$ & $2.23$ & $2.23$ & $1.37$ & $2.74$ & $2.85$ \\
 & $3.27$ & $0$ & $-6.44 \times 10^{-2}$ & $3.90$ & $7.51$ & $\infty$ & $1.72$ & $1.72$ & $1.38$ & $5.89$ & $6.17$ \\
 & $2.53$ & $0.75$ & $-2.03 \times 10^{-2}$ & $1.95$ & $3.73$ & $14.18$ & $2.16$ & $2.16$ & $1.43$ & $2.98$ & $3.12$ \\
 & $2.71$ & $0.75$ & $-2.03 \times 10^{-3}$ & $3.99$ & $9.13$ & $\infty$ & $1.96$ & $1.96$ & $1.57$ & $6.50$ & $6.90$ \\
 C3 & $2.58$ & $0$ & $-6.44 \times 10^{-2}$ & $1.81$ & $3.30$ & $10.83$ & $2.72$ & $2.72$ & $1.59$ & $2.66$ & $2.77$ \\
 & $3.27$ & $0$ & $-6.44 \times 10^{-2}$ & $3.90$ & $7.51$ & $\infty$ & $1.97$ & $1.97$ & $1.60$ & $5.70$ & $5.97$ \\
 & $2.53$ & $0.75$ & $-2.03 \times 10^{-2}$ & $1.95$ & $3.73$ & $14.18$ & $2.62$ & $2.62$ & $1.70$ & $2.88$ & $3.02$ \\
 & $2.71$ & $0.75$ & $-2.03 \times 10^{-3}$ & $3.99$ & $9.13$ & $\infty$ & $2.32$ & $2.32$ & $1.92$ & $6.22$ & $6.62$ \\
  C4 & $2.58$ & $0$ & $-6.44 \times 10^{-2}$ & $1.81$ & $3.30$ & $10.83$ & $3.80$ & $3.80$ & $2.16$ & $2.54$ & $2.66$ \\
 & $3.27$ & $0$ & $-6.44 \times 10^{-2}$ & $3.90$ & $7.51$ & $\infty$ & $2.47$ & $2.47$ & $2.11$ & $5.45$ & $5.73$ \\
 & $2.53$ & $0.75$ & $-2.03 \times 10^{-2}$ & $1.95$ & $3.73$ & $14.18$ & $3.61$ & $3.61$ & $2.37$ & $2.75$ & $2.89$ \\
 & $2.71$ & $0.75$ & $-2.03 \times 10^{-3}$ & $3.99$ & $9.13$ & $\infty$ & $3.07$ & $3.07$ & $2.80$ & $5.88$ & $6.25$ \\
 \hline 
 M1 & $2.58$ & $0$ & $-6.44 \times 10^{-2}$ & $1.81$ & $3.30$ & $10.83$ & $1.18$ & $0.85$ & $1.19$ & $2.85$ & $2.91$ \\
 & $3.27$ & $0$ & $-6.44 \times 10^{-2}$ & $3.90$ & $7.51$ & $\infty$ & $1.04$ & $0.76$ & $1.22$ & $6.13$ & $6.17$ \\
 & $2.53$ & $0.75$ & $-2.03 \times 10^{-2}$ & $1.95$ & $3.73$ & $14.18$ & $1.16$ & $0.84$ & $1.22$ & $3.12$ & $3.19$ \\
 & $2.71$ & $0.75$ & $-2.03 \times 10^{-3}$ & $3.99$ & $9.13$ & $\infty$ & $1.10$ & $0.80$ & $1.31$ & $6.88$ & $7.00$ \\
 M2 & $2.58$ & $0$ & $-6.44 \times 10^{-2}$ & $1.81$ & $3.30$ & $10.83$ & $4.09$ & $4.09$ & $1.31$ & $2.77$ & $2.93$ \\
 & $3.27$ & $0$ & $-6.44 \times 10^{-2}$ & $3.90$ & $7.51$ & $\infty$ & $2.86$ & $2.86$ & $1.29$ & $6.05$ & $6.56$ \\
 & $2.53$ & $0.75$ & $-2.03 \times 10^{-2}$ & $1.95$ & $3.73$ & $14.18$ & $3.94$ & $3.94$ & $1.36$ & $3.03$ & $3.23$ \\
 & $2.71$ & $0.75$ & $-2.03 \times 10^{-3}$ & $3.99$ & $9.13$ & $\infty$ & $3.45$ & $3.45$ & $1.46$ & $6.69$ & $7.41$ \\
 M3 & $2.58$ & $0$ & $-6.44 \times 10^{-2}$ & $1.81$ & $3.30$ & $10.83$ & $1.40$ & $1.16$ & $1.40$ & $2.72$ & $2.80$ \\
 & $3.27$ & $0$ & $-6.44 \times 10^{-2}$ & $3.90$ & $7.51$ & $\infty$ & $1.14$ & $0.95$ & $1.46$ & $5.79$ & $5.88$ \\
 & $2.53$ & $0.75$ & $-2.03 \times 10^{-2}$ & $1.95$ & $3.73$ & $14.18$ & $1.36$ & $1.13$ & $1.47$ & $2.95$ & $3.05$ \\
 & $2.71$ & $0.75$ & $-2.03 \times 10^{-3}$ & $3.99$ & $9.13$ & $\infty$ & $1.26$ & $1.04$ & $1.66$ & $6.37$ & $6.54$ \\
 M4 & $2.58$ & $0$ & $-6.44 \times 10^{-2}$ & $1.81$ & $3.30$ & $10.83$ & $5.93$ & $5.93$ & $1.88$ & $2.60$ & $2.76$ \\
 & $3.27$ & $0$ & $-6.44 \times 10^{-2}$ & $3.90$ & $7.51$ & $\infty$ & $3.58$ & $3.58$ & $1.71$ & $5.68$ & $6.15$ \\
 & $2.53$ & $0.75$ & $-2.03 \times 10^{-2}$ & $1.95$ & $3.73$ & $14.18$ & $5.61$ & $5.61$ & $2.01$ & $2.83$ & $3.03$ \\
 & $2.71$ & $0.75$ & $-2.03 \times 10^{-3}$ & $3.99$ & $9.13$ & $\infty$ & $4.65$ & $4.65$ & $2.20$ & $6.17$ & $6.79$ \\
\hline\hline
\end{tabular}
\end{table*}

\subsection{Parameter space}
\label{parameters}

The parameters defining our tori can be classified into three groups: on the one hand, we have the parameters related to the magnetic susceptibility such as the susceptibility itself $\chi_{\mathrm{m}}$, and parameters $\chi_0$ and $\chi_1^*$, which we explain below. The values of those parameters are reported in Table~\ref{susceptibility_parameters_values}. A second group comprises  parameters $l_0$ and $\alpha$ characterising the angular momentum distribution (which, in our approach is closely related to the potential well depth $\Delta W_{\mathrm{c}} = W_{\mathrm{c}}-W_{\mathrm{in}}$). Finally, a third group of parameters is the one related to the black hole properties (mass $M$, spin $a$, and sense of rotation) and disk properties (magnetization $\beta_{\mathrm{m}} (=\hspace{-0.1cm}p/p_{\mathrm{m}})$, density $\rho$ at the center of the disk\footnote{The center of the disk is defined as the location of the minimum of the potential $W$.} (i.e. $\beta_{\mathrm{m,c}}$ and $\rho_{\mathrm{c}}$), and the exponents $\Gamma$ and $\eta$.

%It is worth mentioning that the susceptibility values used in this work are motivated by the theory of Langevin, in which, for an electron gas, the paramagnetism is associated with the intrinsic magnetic moment of electrons and the alignment of the spins due to magnetic torques, while the diamagnetism is associated with the electron orbital motion around the magnetic field lines. For a detailed description in the context of magnetized relativistic disks, see \cite{velasquez2022synchrotron}.

\subsubsection{Magnetic susceptibility}\label{magsus_subsubsection}

In the first place, regarding the magnetic susceptibility of the disk, we follow an approach similar to that of~\citet{Pimentel:2018}. Therefore, we select 5 models with a constant distribution of the magnetic susceptibility (2 diamagnetic, 1 non-magnetic and 2 paramagnetic; see the top rows of Table~\ref{susceptibility_parameters_values} for the exact values of the parameters) and 4 models with non-constant magnetic susceptibility. The parameters $\chi_0$ and $\chi_1$ for the latter 4 models are chosen in the following way: i) 2 diamagnetic models, the first one going from $\chi_{\mathrm{m}} = 0$ at $R = R_{\mathrm{in}}$ to $\chi_{\mathrm{m}} = -0.4$ for $R \rightarrow \infty$, and the second one going from $\chi_{\mathrm{m}} = -0.4$ at $R = R_{\mathrm{in}}$ to $\chi_{\mathrm{m}} = 0$ for $R \rightarrow \infty$; ii) 2 paramagnetic models chosen in a similar way, i.e.~the first one going from $\chi_{\mathrm{m}} = 0.4$ at $R = R_{\mathrm{in}}$ to $\chi_{\mathrm{m}} = 0$ for $R \rightarrow \infty$ and the second one going from $\chi_{\mathrm{m}} = 0$ at $R = R_{\mathrm{in}}$ to $\chi_{\mathrm{m}} = 0.4$ for $R \rightarrow \infty$.
The specific values of $\chi_0$ and $\chi_1$ are obtained by using the following procedure (note that the parameter $\sigma$ is fixed to $\sigma = -1$ for the non-constant magnetic susceptibility cases\footnote{It is important to note that, even though it seems that $\sigma$ is not relevant for the constant $\chi$ cases, the parameter $\sigma$ is present through the definition of $\tilde{f}$ (see Eq.~\eqref{eq:functions_case2}) in the constant cases. To arrive at the same results as in~\cite{Pimentel:2018} we need to set $\sigma \rightarrow -\infty$.}): First, for the cases with $\chi_m(R_{\mathrm{in}},\pi/2) = \pm0.4$ and $\lim_{R\rightarrow \infty} \chi_{\mathrm{m}}(R,\pi/2) = 0$, we can use Eq.~\eqref{eq:chi} and, employing the relation between $\chi$ and $\chi_{\mathrm{m}}$,
\begin{equation}
\chi(\chi_{\mathrm{m}}) = \frac{\chi_{\mathrm{m}}}{1+\chi_{\mathrm{m}}},
\end{equation}
we can see that, at $R = R_{\mathrm{in}}$
\begin{equation}
\chi(\chi_{\mathrm{m}} = \pm0.4) = \chi_0 + \chi_1 \mathcal{L}^{\sigma}(R_{\mathrm{in}}, \pi/2),
\end{equation}
and for $R \rightarrow \infty$
\begin{equation}
0 = \chi_0 +\lim_{R \rightarrow \infty} (\chi_1 \mathcal{L}^{\sigma}(R,\pi/2)) = \chi_0,
\end{equation}
where we have used that $\sigma = -1$ and $\lim_{R \rightarrow \infty} \mathcal{L}(R,\pi/2) = \infty$.
Then, it can be seen that for these models:
\begin{equation}
\chi_0 = 0, \hspace{5mm} \chi_1 = \frac{\chi(\chi_{\mathrm{m}} = \pm0.4)}{\mathcal{L}^{\sigma}(R_{\mathrm{in}}, \pi/2)},
\end{equation}
so the function $\chi$ can be written as
\begin{equation}
\chi = \chi(\chi_{\mathrm{m}} = \pm0.4) \left(\frac{\mathcal{L}(R, \theta)}{\mathcal{L}_{\mathrm{in}}}\right)^{\sigma},
\end{equation}
where $\mathcal{L}_{\mathrm{in}} = \mathcal{L}(R_{\mathrm{in}}, \pi/2)$.
Following a similar procedure we can derive the values of the parameters for the cases $\chi_m(R = R_{\mathrm{in}}) = 0$ and $\lim_{R \rightarrow \infty }\chi_{\mathrm{m}}(R,\pi/2) = \pm0.4$. At $R = R_{\mathrm{in}}$ we can write
\begin{equation}
\chi_0 = -\chi_1 \mathcal{L}^{\sigma}(R_{\mathrm{in}}, \pi/2),
\end{equation}
and at $R \rightarrow \infty$
\begin{equation}
\chi(\chi_{\mathrm{m}} = \pm0.4) = \chi_0,
\end{equation}
so the values of the $\chi_0$ and $\chi_1$ parameters are
\begin{equation}
\chi_0 = \chi(\chi_{\mathrm{m}} = \pm0.4), \hspace{5mm} \chi_1 = -\frac{\chi(\chi_{\mathrm{m}} = \pm0.4)}{\mathcal{L}^{\sigma}(R_{\mathrm{in}}, \pi/2)},
\end{equation}
and the general form of the $\chi$ function is
\begin{equation}
\chi = \chi(\chi_{\mathrm{m}} = \pm0.4) \left(1- \left(\frac{\mathcal{L}(R, \theta)}{\mathcal{L}_{\mathrm{in}}}\right)^{\sigma}\right).
\end{equation}

We can see that the expression for the parameter $\chi_1$ depends on the function $\mathcal{L}$, so its particular value for different values of the parameters $(R_{\mathrm{in}}, M, a)$ will be different. However, we can define a new dimensionless parameter $\chi_1^*$ as
\begin{equation}\label{eq:chi1star_def}
    \chi_1^* = \chi_1\mathcal{L}^{\sigma}_{\mathrm{in}},
\end{equation}
and therefore we write $\chi$ as
\begin{equation}
    \chi(R, \theta) = \chi_0 + \chi_1^*\left(\frac{\mathcal{L}(R, \theta)}{\mathcal{L}_{\mathrm{in}}}\right)^{\sigma}.
\end{equation}
In this way, the values of the parameters $\chi_0$ and $\chi_1^*$ are independent of the parameters $(R_{\mathrm{in}}, M, a)$ and, therefore, are the same for all the different Kerr spacetimes and disk specific angular momentum distributions we are considering, and only depend on the kind of magnetic susceptibility model we are studying. The particular values of $(\chi_0, \chi_1^*)$ for all models we compute are reported in the bottom rows of Table~\ref{susceptibility_parameters_values}. The actual values of the parameter $\chi_1$ can be computed from the table using the definition~\eqref{eq:chi1star_def} and the values of $\mathcal{L}_{\mathrm{in}}$ (which can be computed from the data in Tables~\ref{results__09}, ~\ref{results_0} and ~\ref{results_09}).

Additionally, it is worth discussing the allowed values for our magnetic susceptibility parameters $(\chi_0, \chi_1^*)$. In the left panel of Figure~\ref{existence}, we show a diagram that depicts the location of our set of values for $(\chi_0, \chi_1^*)$ on the parameter space. The symbols flag the specific models of our sample reported in Table~\ref{susceptibility_parameters_values}. Some of these symbols are joined by dotted lines which correspond to constant values of the magnetic susceptibility $\chi$ (and thus, $\chi_{\mathrm{m}}$ as well) evaluated at the inner edge of the disk $R = R_{\mathrm{in}}$. The grey shaded regions in Fig.~\ref{existence} correspond to forbidden regions of the parameter space while the blue shaded regions designate values of $(\chi_0, \chi_1^*)$ such that the corresponding disks attain a maximum of the magnetization function $\beta_{\mathrm{m}}$ for some value of the radial coordinate. The condition for a maximum of $\beta_{\mathrm{m}}$ to appear is discussed in detail in Appendix~\ref{partial_beta_0}. In addition, in the right panel of Fig.~\ref{existence} we depict the radial profiles of $\chi_{\mathrm{m}}(R)$ for our choice of non-constant susceptibility models. This is only shown for an illustrative combination of the $(a, \alpha, l_0)$ parameters. We note that the radial profiles are similar for all the values of the parameters $(a, \alpha, l_0)$ we have considered in our study.

To better understand what regions of the parameter space $(\chi_0, \chi_1^*)$ are available to build our models it is useful to discuss first the correspondence between values of the function $\chi$ and the magnetic susceptibility $\chi_{\mathrm{m}}$. This is plotted in Figure~\ref{chim_chi}. 
In this figure we can identify the following regions: i) $\chi \in (-\infty,1)$ corresponds to $\chi_{\mathrm{m}} \in (-1, \infty)$ and ii) $\chi \in (1, \infty)$ corresponds to $\chi_{\mathrm{m}} \in (-\infty, -1)$. Then, it is apparent that $\chi = 1$ is an ill-defined value of the function $\chi$, so it must be excluded from our parameter space. Further pathological regions come from the definitions of $\tilde{\lambda}$ and $\tilde{f}$ (Eqs.~\eqref{eq:functions_case2}), namely  $\chi_0\neq 1/2$ to avoid the denominator in the definition of $\tilde{\lambda}$ from being zero, and $\chi< 1/2$, since if $(1-2\chi) <0$, then $\tilde{f}$ does not exist for fractional values of the exponent. We can also exclude the region $\chi_0 \geq 1/2$ taking into account that, in the constant case $\chi = \chi_0$ and in the non-constant case, if $\chi_1^* >0$ and $\chi_0 \geq 1/2$, then $\chi > 1/2$ and if $\chi_1^* < 0$, $\chi$ will approach $1/2$ asymptotically for $R \rightarrow \infty$. For that reason, it is apparent that the $\chi > 1/2$ restriction must be enforced at the inner edge of the disk $R = R_{\mathrm{in}}$. Physically speaking, we can relate this limit to the fact that $\chi_{\mathrm{m}} \geq 1$ ($1/2 \leq \chi < 1$) no longer represents a paramagnetic fluid. Using the same argument, we can exclude the region for which $\chi_{\mathrm{m}} \leq 1$ ($\chi \in (1, \infty)$) as well. 
In the left panel of Fig.~\ref{existence}, the lines $\chi_0 = 1/2$ and $\chi = 1/2$ (i.e.~$\chi_1^* = -\chi_0 + 1/2$) are represented by black solid lines and the regions $\chi_0 \geq 1/2$ and $\chi \geq 1/2$ are shaded in grey. Moreover, the region of the parameter space in which there is a maximum of the function $\beta_{\mathrm{m}}$ in the disk is shaded in blue. The critical line corresponds to the values of $(\chi_0, \chi_1^*)$ for which the maximum of the magnetization function is located at the inner edge of the disk, and it is displayed as a blue line.
The constant $\chi$ (and $\chi_{\mathrm m}$) lines follow the equation $\chi_1^* = (\mathcal{L}/\mathcal{L}_{\mathrm{in}})*(-\chi_0 + \chi)$ which corresponds to parallel oblique lines ($\chi_1^* = -\chi_0 + \chi$) for $R = R_{\mathrm{in}}$, and to parallel vertical lines ($\chi = \chi_0$) when $R \rightarrow \infty$.

\subsubsection{Angular momentum}

We turn next to discuss the second group of model parameters, that is, those related to the specific angular momentum distribution at the equatorial plane, Eq.~\eqref{eq:ansatz}, namely the constant part $l_0$ and the exponent $\alpha$. Following~\cite{Gimeno-Soler_etal:2021}, we focus on two values for the exponent $\alpha$, namely $\alpha = 0$ (constant angular momentum) and $\alpha = 0.75$ (nearly Keplerian rotation). The values of the parameter $l_0$ are chosen in the following way: First, we compute the value of the depth of the potential well, $\Delta W_{\mathrm{c}}$, which is defined as the difference between the potential at the center and the potential at the inner edge of the disk, $\Delta W_{\mathrm{c}} \equiv W_{\mathrm{c}} - W_{\mathrm{in}}$. This quantity achieves its maximum for a particular value of $l_0$, given a set of parameters $(M, a, \alpha)$ (note that for the said value of $l_0$, $W(R_{\mathrm{cusp}}) = 0$, where $R_{\mathrm{cusp}}$ is the location of the cusp, i.e.~the point in the equatorial plane where the critical equipotential surface crosses with itself). We will denote this particular value of $l_0$ as $l_{0, 1}$ and the corresponding value of $\Delta W_{\mathrm{c}}$ as $\Delta W_{\mathrm{max}}$. In this work we are going to use two values  of the parameter $l_0$ (namely, $l_{0, 2}$ and $l_{0, 3}$) for each set of parameters $(M, a, \alpha)$. These two values are chosen according to the following criteria: i) $l_{0,2}$ is such that $l_{0, 2} < l_{0, 1}$ and the corresponding value of the depth of the potential well is $\Delta W_{\mathrm{c}}  = \frac{1}{2} \Delta W_{\mathrm{max}}$; ii) $l_{0,3}$ is such that $l_{0, 3} > l_{0, 1}$ and the corresponding value of the depth of the potential well is also $\Delta W_{\mathrm{c}}  = \frac{1}{2} \Delta W_{\mathrm{max}}$. We chose these two criteria to be able to perform a comparison between models with different disk geometries, but eliminating the dependence that the thermodynamic quantities have on the potential well depth $\Delta W_{\mathrm{c}}$. In Fig.~\ref{angular_momentum_models} we show the morphology of the potential distribution at the equatorial plane for these two $l_0$ prescriptions (fixing the spin of the black hole and the exponent of the angular momentum distribution to $a = 0.9$ and $\alpha = 0.75$).

Finally, we discuss the third group of model parameters, those corresponding to properties of the black hole and of the disk. These parameters are fixed in the following way: we select three different values for the spin parameter of the black hole, namely a highly spinning counter-rotating Kerr black hole ($a = -0.9$), a Schwarzschild black hole ($a = 0$) and a highly spinning co-rotating Kerr black hole ($a = 0.9$). The mass parameter of the black hole, $M$, is fixed at 1. Furthermore, in most of this work, we will focus on a particular value of the magnetization parameter at the center of the disk, namely $\beta_{\mathrm{m,c}}=10^{-2}$, which corresponds to a strongly magnetized disk\footnote{This degree of magnetization may be too high for it to be realistic, but as we will see later, the most relevant conclusions of this work are unaffected by the particular value of $\beta_{\mathrm{m, c}}$.}. Lastly, we fix the polytropic exponents $\Gamma = \eta = 4/3$ and the rest-mass density at the center of the disk to $\rho_{\mathrm{c}} = 1$ (which fixes the polytropic constant $K$).

%=============================================
\section{Results}
\label{results}
%=============================================

Taking into account our 6-dimensional parameter space $(a, \alpha, l_0, \chi_0, \chi_1, \beta_{\mathrm{m,c}})$, we build in this work a total of 108 different accretion disk models (36 for each value of the black hole spin parameter considered). The most relevant physical information for each model is presented in Tables~\ref{results__09}, \ref{results_0} and \ref{results_09} for spins $a= -0.9$, $a=0$ and $a=0.9$, respectively.
Looking at these tables we can observe, in the first place, some generic trends that also appear when considering non-polarized magnetized accretion disks, namely: i) The value of the constant part of the angular momentum ansatz $l_0$ is smaller when the exponent is $\alpha=0.75$ (in contrast to the case $\alpha = 0$). This happens for both of the ways we use to prescribe the constant part of the specific angular momentum law (i.e.~$l_{0, 2}$ and $l_{0, 3}$) and it is irrespective of the value of the spin parameter of the black hole. ii) The value of the depth of the potential well $\Delta W_{\mathrm{c}}$ is always greater for the models with $\alpha = 0$ (while keeping the other parameters constant) and its value increases with the black hole spin parameter $a$. This is in agreement with the fact that the quantity $\Delta W_{\mathrm{c}}$ achieves its maximum value (for the Kerr spacetime) when $a \rightarrow 1$, $l = l_{\mathrm{mb}}$ and $W_{\mathrm{in}} = 0$, that value being $\Delta W_{\mathrm{c}} = \frac{1}{2} \log 3$ \citep{Abramowicz:1978}. iii) disks with non-zero values of the exponent $\alpha$ are, in general, slimmer but more radially extended in the sense that their characteristic radii $R_{\mathrm{in}}$ and $R_{\mathrm{c}}$ are greater than their corresponding values for the constant angular momentum models.

Moving on to the quantities affected by the magnetic susceptibility of the disk $\chi_{\mathrm{m}}$, we observe that for the models with a constant value of $\chi_{\mathrm{m}}$, the magnetization parameter at the inner edge of the disk $\beta_{\mathrm{m, in}}$ (which also corresponds with its maximum value) is, in general, smaller for the models with $l_0 = l_{0, 3}$ when compared to the corresponding values for $l_0 = l_{0, 2}$ (the only exception being model C1 for $a = -0.9$ and $\alpha = 0.75$). It is also relevant to mention that this difference between the two values is significantly smaller for the models with $\alpha = 0.75$ and is globally greater for greater values of the spin parameter $a$.
A similar trend is observed for the non-constant susceptibility models, with the particularity that, for models M1 and M3 (the models which are in the blue-shaded region of the left panel of Fig.~\ref{existence}) the value of the magnetization function at the inner edge of the disk $\beta_{\mathrm{in}}$ does not correspond to the maximum of $\beta_{\mathrm{m}}$ and it is achieved for a radial coordinate larger than $R_{\mathrm{in}}$. It can also be seen (especially for the case $a = 0.9$) that the models with $l_0 = l_{0,2}$ increase the magnetization at their inner region for increasing $\alpha$, when it is the opposite for the models with $l_0 = l_{0,3}$.

If we now turn our attention to the behaviour of the maximum of the rest-mass density $\rho_{\mathrm{max}}$, we can see that it is, in general, greater for models with $l_{0, 3}$ than for models with $l_{0, 2}$ and it is also greater for models with $\alpha = 0.75$ when compared with the same models with constant specific angular momentum. Moreover, $\rho_{\mathrm{max}}$ has a positive correlation with the difference $(R_{\mathrm{max}} - R_{\mathrm{c}})/R_{\mathrm{c}}$ for varying magnetic susceptibility  $(\chi_0, \chi_1)$ (and keeping constant the other parameters) and for varying values of the angular momentum parameters $(l_0, \alpha)$ (also keeping constant the other parameters).
Finally, looking at the locations of the maximum of the rest-mass density, $R_{\mathrm{max}}$, and of the magnetic pressure, $R_{\mathrm{m, max}}$, we can observe that the disks become less radially extended for increasing value of $\chi_{\mathrm{m}}$. This happens for both constant and non-constant $\chi_{\mathrm{m}}$ distributions because at $R_{\mathrm{c}}$, $\chi_{\mathrm{m}}$ decreases from model M4 to model M1 (see right panel of Fig.~\ref{existence}).

In Figures~\ref{density_2D_l02} and~\ref{density_2D_l03} we show the 2-dimensional distributions of the rest-mass density $\rho$ of the disk (top half of each panel) and magnetization parameter $\beta_{\mathrm{m}}$ (in logarithmic scale; bottom half) for a constant value of the exponent of the angular momentum ansatz $\alpha = 0.75$ and a black hole spin parameter of $a= 0.9$. The morphology for different values of $\alpha$ and $a$ is qualitatively the same. The color code is normalized to the maximum of each plot for the rest-mass density. Correspondingly, for the magnetization parameter, the deep green color corresponds to the greatest value of $\beta_{\mathrm{m}}$ of each row, the white color corresponds to $\beta_{\mathrm{m}} = \beta_{\mathrm{m,c}} $ and the deep purple color corresponds to the minimum value of $\beta_{\mathrm{m}}$ achieved in each row. 
These figures show that the high-density region of the disks moves towards the inner edge of the disks when $\chi_{\mathrm{m}}$ increases. This happens for all the magnetic susceptibility and angular momentum models we have considered. Moreover, looking at the distribution of $\beta_{\mathrm{m}}$, we can notice that the isocontours are almost vertical (as the vertical structure of the $\beta_{\mathrm{m}}(R,\theta)$ function is dominated by the function $\mathcal{L}$) and we observe that the models with greater $\chi_{\mathrm{m}}$ are less magnetized at the inner edge of the disk (higher values of $\beta_{\mathrm{m}}$). It can also be observed a non-monotonic behaviour for models M1 and M3 for both values of the angular momentum of the disk (the color of the inner region of the disk is whiter).

Figure~\ref{beta_r_fig} depicts the quotient $\beta_{\mathrm{m}}/\beta_{\mathrm{m,c}}$ versus the radial coordinate for our 54 models with $\alpha = 0.75$ (the results for the constant angular momentum solutions are qualitatively very similar).  In the left part of the figure we show the constant $\chi_{\mathrm{m}}$ models (C1 to C4) and in the right part, the non-constant $\chi_{\mathrm{m}}$ models (M1 to M4). It is apparent that, as we noted before, models with higher values of the susceptibility are less magnetized in the inner regions of the disk. In addition,  we can also observe in this figure that they are more magnetized in the outer regions of the disk (which also could be observed in Figs.~\ref{density_2D_l02} and~\ref{density_2D_l03} to a lesser extent). Both for the constant and non-constant susceptibility cases, models with $l_0 = l_{0, 2}$ are less magnetized than models with $l_0 = l_{0,3}$ in the inner region of the disk. It also can be seen that increasing the spin parameter of the black hole yields potentially higher values of the magnetization at the inner edge of the disk.
Focusing on the models with non-constant $\chi_{\mathrm{m}}$, we see that models M1 and M3 show a local maximum of the magnetization parameter at the inner part of the disk. We also observe that the slope of the $\beta_{\mathrm{m}}$ function is steeper than in the constant $\chi_{\mathrm{m}}$ cases. In the outer regions of the disk, as expected, models M1 and M3 depart more from the $\chi_{\mathrm{m}} = 0$ case than models M2 and M4 (which almost overlap with the $\chi_{\mathrm{m}} = 0$ curve) because for these two models, $\chi_{\mathrm{m}} \rightarrow 0$ when $R \rightarrow \infty$. It is very relevant to remark that the qualitative behavior of the magnetization function $\beta_{\mathrm{m}}$ for the different magnetic susceptibility $\chi_{\mathrm{m}}$ models is independent of the value of the magnetization at the center $\beta_{\mathrm{m,c}}$ (as it can be seen in the radial profiles for $\beta_{\mathrm{m,c}} = 1$ and $\beta_{\mathrm{m,c}} = 100$ shown in Figs.~\ref{beta_r_fig_1} and \ref{beta_r_fig_100}).

In Figure~\ref{logrho_r_fig} we plot the logarithm of the rest-mass density versus the logarithm of the normalized radial coordinate $R/R_{\mathrm{c}}$. The models selected are the same as in Fig.~\ref{beta_r_fig} and the profiles are shown at the equatorial plane. Focusing on the constant magnetic susceptibility models (leftmost two columns), we see that, as previously noted, the disks with $l_0 = l_{0, 3}$ are more radially extended than their $l_{0, 2}$ counterparts. Moreover, the diamagnetic models (C1 and C2) are less dense in the inner region of the disk, but are denser in the outer regions. The exact opposite happens for the paramagnetic models (C3 and C4), where matter is more concentrated in the inner regions of the disk. It is also apparent that the dependence of the rest-mass distributions on $\chi_{\mathrm{m}}$ is qualitatively the same for all values of $a$ and $\alpha$. Turning now to the non-constant susceptibility models (rightmost two columns), we see that models M1 and M4 behave in a very similar way compared with their constant $\chi_{\mathrm{m}}$ (diamagnetic and paramagnetic, respectively) counterparts. This happens because, for these two models, $\chi_{\mathrm{m}} \rightarrow \pm 0.4$ (i.e.~a non-zero value of $\chi_{\mathrm{m}}$) for increasing $R$, while models M2 and M3 (which have a significant value of $\chi_{\mathrm{m}}$ only in the regions close to the inner edge of the disk) have a radial rest-mass density distribution very close to the one of a $\chi_{\mathrm{m}} = 0$ model, even in the inner region where the magnetic susceptibility is most relevant for these two models. We also note that the presence of a relative maximum in the magnetization parameter $\beta_{\mathrm{m}}$ (cf.~Fig.~\ref{beta_r_fig}) does not seem to affect the radial distribution of rest-mass density in a significant way. It is also worth highlighting the similarity between the rest-mass density distributions presented here and the same kind of plots but for $\chi_{\mathrm{m}} = 0$ and different values of the magnetization parameter at the center of the disk $\beta_{\mathrm{m,c}}$ (see, for instance~\citet{Gimeno-Soler_etal_2019} and \citet{Gimeno-Soler_etal:2021}), with the paramagnetic models being similar to more strongly magnetized models, and the diamagnetic models being similar to more weakly magnetized models.

In Figure~\ref{h_r_chi_rmax} we show how variations of the magnetic susceptibility affect the value of the relative thickness of the disk $H/R$, which is defined as~\citep{shiokawa2011global}
\begin{equation}
\label{eq:HoverR}
\left( \frac{H}{R} \right) (r) := \frac{\int_{0}^{\pi}  \rho |\pi/2 - \theta| \ d\theta}{\int_{0}^{\pi} \ \rho \ d\theta} \,.
\end{equation}
Note that in this equation the radial coordinate is represented by $r$. The rows of Fig.~\ref{h_r_chi_rmax} correspond to the two different values for the exponent of the specific angular momentum ansatz (namely $0$ and $0.75$) while the columns indicate different values of the spin parameter $a$ ($-0.9$, $0$, and $0.9$, from left to right). In each plot, the dotted, dashed and solid (horizontal) grey lines join models with the same value of the magnetization parameter at the center of the disk $\beta_{\mathrm{m,c}}$ (respectively, $10^2$, $1$ and $10^{-2}$) and the solid and dashed (vertical) lines join models with the same kind of $\chi_{\mathrm{m}}$ distribution. Moreover, the data points located in the top part of each plot represent the models with $l_0 = l_{0,3}$ while those at the bottom part correspond to the models with $l_0 = l_{0,2}$. 
The relative thickness of the disk shown in Fig.~\ref{h_r_chi_rmax} is evaluated at the location of the maximum rest-mass density of each model. As expected, the relative thickness at such maximum density is almost constant for weakly and mildly magnetized disks i.e.~if the magnetic field is not strong enough, and the contribution of the magnetic polarization does not affect the disk. Dependence on the magnetic susceptibility can only be seen for strongly magnetized disks ($\beta_{\mathrm{m,c}} = 10^{-2}$). Moreover, as we already noticed in Figs.~\ref{density_2D_l02} and~\ref{density_2D_l03}, the models with $l_0 = l_{0,3}$ are consistently thicker than those with $l_0 = l_{0,2}$. It can also be seen that increasing the magnetization of the disk yields thinner disks. Finally, the relative thickness of the disk slightly decreases for increasing values of the spin parameter and, as expected, it depends strongly on the value of the exponent $\alpha$: models with a constant value of the specific angular momentum are significantly thicker than those with $\alpha = 0.75$.

In summary, in the case of constant angular momentum disks our models have values of the relative thickness $H/R$ in the range $(0.1, 0.22)$ while, in the case of non-constant angular momentum disks, the values are in the range $ (0.06, 0.15)$. We note that these upper values could be increased if we considered greater values of $l_0$. Conversely, the lower limit of the range could be further reduced considering values of the exponent $\alpha$ closer to $1$.

%=============================================
\section{Conclusions and outlook}
\label{conclusions}
%=============================================

In this work we have computed equilibrium solutions of magnetized, geometrically thick accretion disks with magnetic polarization endowed with a non-constant specific angular momentum distribution around Kerr black holes.
Our study is a generalization of the previous work on magnetized accretion disks with magnetic polarization carried out by~\citet{Pimentel:2018} in which only a constant angular momentum distribution (a ``Polish doughnut'') was inspected. The new disk solutions reported in this paper have been obtained by combining the two approaches reported by~\citet{Pimentel:2018} and~\citet{Gimeno-Soler_etal:2021}. On the one hand, we have followed the work of~\citet{Pimentel:2018} to consider a magnetically polarized fluid in the linear media approximation in which the magnetic susceptibility takes a specific functional form in order to fulfil the integrability conditions for Euler's equation. On the other hand, the specific angular momentum distribution of the fluid has been prescribed following the ansatz introduced by~\citet{Gimeno-Soler_etal:2021} in which the angular momentum at the equatorial plane is computed as a two-parameter function of the Keplerian angular momentum function which is, in turn, a function of the radial coordinate. From this solution the angular momentum distribution outside the equatorial plane is obtained by computing the so-called von Zeipel cylinders (following the approach employed in~\citet{Daigne:2004}).

We have studied the morphology and physical properties of a large number of equilibrium solutions which were obtained by selecting suitable parameter values within the substantial parameter space spanned by the models. Those include the spin of the black hole, the two parameters of the specific angular momentum distribution, the two parameters of the magnetic susceptibility of the fluid, and the magnetization parameter at the center of the disk. The latter was fixed to $\beta_{\mathrm{m,c }} = 10^{-2}$ for most of our models since the effects on the disk morphology induced by changes in the magnetic susceptibility of the disk are very small if the magnetization of the disk is weak. Our results show that the qualitative changes introduced in the morphology of the disks and in its physical quantities do not depend strongly on the angular momentum distribution. Despite our new set of models have an increased degree of realism as compared with those of~\cite{Pimentel:2018} the differences found are not large. We have also observed that the morphological changes do not seem to depend much on the black hole spin. 

Focusing on the effects of the magnetic susceptibility, we have seen that for constant $\chi_{\mathrm{m}}$, disks tend to be thicker and more radially extended for the case of diamagnetic models (i.e.~$\chi_{\mathrm{m}}<0$). 
Moreover, the magnetization in such models is stronger in the inner part of the disk ($R < R_{\mathrm{c}}$) and weaker in the outer region ($R > R_{\mathrm{c}}$). The exact opposite happens when the disk is paramagnetic (i.e.~$\chi_{\mathrm{m}}>0$). On the other hand, models with a non-constant distribution of magnetic susceptibility attain a qualitatively similar morphological appearance. However, for these models the distribution of the magnetization behaves in a different way: those with a positive value of the slope (and independently on the sign) of the $\chi_{\mathrm{m}}(R)$ distribution are less magnetized in the inner region and more magnetized in the outer part. The opposite happens for models with a negative slope of $\chi_{\mathrm{m}}(R)$. In addition, for such models the magnetization function exhibits a local maximum, which  cannot happen in standard Komissarov's disks~\citep{Komissarov:2006,Pimentel:2018}. The conditions for a maximum of $\beta_{\mathrm{m,c}}$ to appear are derived in the appendix~\ref{partial_beta_0} and only depend on the particular values of the susceptibility parameters $\chi_0$ and $\chi^*_1$. In particular, it must be highlighted that the qualitative behavior of the $\beta_{\mathrm{m}}(R, \theta)$ function does not depend on the particular value of $\beta_{\mathrm{m,c}}$, which means that the effects due to the magnetic susceptibility in the magnetization distribution of the disk are a feature present for all values of the magnetization, even in the cases when this does not impact the disk morphology. In summary, we have shown that the effects due to the magnetic polarization of the disk observed in~\citep{Pimentel:2018} are a robust feature of equilibrium configurations of magnetized thick accretion disks around Kerr BHs regardless of the spin parameter of the BH and the specific angular momentum distribution.

We close this paper by pointing out two immediate applications we plan to carry out as a result of the results reported here. On the one hand, the new behaviour of the magnetization function in disks with $d\chi_m/dR < 0$ is consistent with the stationary state of accretion found in numerical simulations~\citep{shiokawa2011global} (i.e. the radial profile of $\beta_{\mathrm{m}}$ at the equatorial plane) and could be potentially relevant in the context of jet generation. The main reason for this is because magnetic susceptibility enhances the development of the MRI, resulting in an accretion state with higher vertical stresses and a value of the alpha viscosity parameter close to the observed values~\citep{2007MNRAS.376.1740K}. Observations suggest that $\alpha_{\rm visc}$ should have a value around $0.1$ \citep{2007MNRAS.376.1740K} but MHD disk simulations yield typical values close to $0.02$ \citep{2011ApJ...738...84H}. Recently, a first effort to understand this problem has shown that paramagnetic disks reach values of this parameter close to $0.12$ \cite{Oscar2020}. Therefore, it is interesting to perform numerical simulations of the initial data reported here (considering weaker magnetic fields) for models with $d\chi_m/dR < 0$ to explore the possibility of obtaining jets from accretion disks with non-constant magnetic susceptibility. 

On the other hand, analytical solutions of a torus in hydrostatic equilibrium are commonly used to produce jets by introducing an ad-hoc, weak poloidal magnetic field~\citep{Porth2019,Cruz-Osorio2022,Fromm2021b}. One of the issues with this approach is that the initial state is not self-consistent (see discussion in~\cite{Cruz-Osorio2020}). A possible solution to this question is to start from magnetized disks with toroidal fields (i.e.~the Komissarov solution). There are already some indications of the appearance of outflows in such conditions but those are challenging to obtain due to numerical resolution limitations and computational requirements~\citep{liska2020large}. It is worth investigating whether non-constant magnetic susceptibility disks as the ones reported in this work, characterized by a more magnetized inner edge, could facilitate jet launching and reduce the computational requirements to produce it. Our findings on those two topics of research will be reported elsewhere.%Viscosity can be described by means of the parameter $\alpha_{\rm visc}$ motivated by the kinetic theory of turbulence \citep{1973A&A....24..337S}. 
%Observations suggest that $\alpha_{\rm visc}$ should have a value around $0.1$ \citep{2007MNRAS.376.1740K} but MHD disk simulations yield typical values close to $0.02$ \citep{2011ApJ...738...84H}. However, the inclusion of magnetic susceptibility in the disk model might have an impact on these values, as the recent simulations of paramagnetic disks reported by  suggest. Those yield values as high as $\alpha_{\rm visc} = 0.12$, closer to the values estimated from observations.

%=============================================

\begin{acknowledgements}
S.G-S is supported by The Center for Research and Development in Mathematics and Applications (CIDMA) through the Portuguese Foundation for Science and Technology (FCT - Fundação para a Ciência e a Tecnologia), references UIDB/04106/2020 (https://doi.org/10.54499/UIDB/04106/2020) and UIDP/04106/2020  
 (https://doi.org/10.54499/UIDP/04106/2020) and by the projects CERN/FIS-PAR/0027/2019 , https://doi.org/10.54499/PTDC/FIS-AST/3041/2020, https://doi.org/10.54499/CERN/FIS-PAR/0024/2021 and
https://doi.org/10.54499/2022.04560.PTDC. F.D.L-C is supported by the Vicerrectoría de Investigación y Extensión - Universidad Industrial de Santander, under Grant No. 3703. ACO gratefully acknowledges ``Ciencia Básica y de Frontera 2023-2024" program of the ``Consejo Nacional de Humanidades, Ciencias y Tecnología" (CONAHCYT, México), projects CBF2023-2024-1102 and 257435. JAF is supported by the Spanish Agencia Estatal de Investigaci\'on  (grant PID2021-125485NB-C21 funded by MCIN/AEI/10.13039/501100011033 and ERDF A way of making Europe). Further support is provided by the EU's Horizon 2020 research and innovation (RISE) programme H2020-MSCA-RISE-2017 (FunFiCO-777740) and  by  the  European Horizon  Europe  staff  exchange  (SE)  programme HORIZON-MSCA-2021-SE-01 (NewFunFiCO-10108625). 
\end{acknowledgements} 

%\nocite{*}

\begin{appendix}
\section{Conditions for the appearance of a local maximum of $\beta_{\mathrm{m}}$}\label{partial_beta_0}
As we observed, some of our models possess a maximum on the magnetization parameter function at the equatorial plane, $\beta_{\mathrm{m}}(R, \pi/2)$. In this appendix, we analyse the conditions required for that maximum to appear. For this derivation, we use the equation of state for the fluid $p = K w^{\Gamma}$ instead of $p = K \rho^{\Gamma}$ used in the main body of this paper. This choice greatly simplifies the proof and it is supported by the fact that it is a good approximation to $p = K \rho^{\Gamma}$ for magnetized disks in the Kerr spacetime (see~\citet{Gimeno-Soler_etal_2019}).
We start from
\begin{equation}
\frac{\partial \beta_{\mathrm{m}}(R)}{\partial R} = \frac{\partial}{\partial R}\left(\frac{p}{p_{\mathrm{m}}}\right)=\partial_R \left(\frac{K w^{\Gamma}}{K_{\mathrm{m}}\mathcal{L}^{\tilde{\lambda}}w^{\eta}\tilde{f}}\right) = 0\,.
\end{equation}
Taking into account that we consider $\Gamma = \eta$ and that $w \neq 0$ in the disk, we can simplify the previous expression to
\begin{equation}
\partial_R \left(\frac{1}{\mathcal{L}^{\tilde{\lambda}}\tilde{f}}\right) = 0\,.
\end{equation}
If we expand this expression, use Eq.~\eqref{eq:chi} and we take into account that, inside the disk, $\mathcal{L} \neq 0$, $\partial_R \mathcal{L} \neq 0$ 
and $\tilde{f} \neq 0$, we arrive at
\begin{equation}
(-1 + \eta) (1-\chi_0) + \mathcal{L}^{\sigma} \chi_1 (1 +  2\sigma - \eta) = 0.
\end{equation}
Then, if we consider that $\sigma = -1$ and solve for $\mathcal{L}$ we obtain
\begin{equation}
    \mathcal{L} = \frac{-(1+\eta)\chi_1}{(-1+\eta)(-1+\chi_0)}\,,
\end{equation}
and inserting $\mathcal{L}_{\mathrm{in}}$ in both sides and using the definition of $\chi_{1}^*$, Eq.~\eqref{eq:chi1star_def}, yields
\begin{equation}
    \frac{\mathcal{L}}{\mathcal{L}_{\mathrm{in}}} = \frac{-(1+\eta)\chi_1^*}{(-1+\eta)(-1+\chi_0)}.
\end{equation}
Since we are looking for extrema of the functions inside the disk (i.e.~$R>R_{\mathrm{in}}$, $\mathcal{L}/\mathcal{L}_{\mathrm{in}} \geq 1$), we choose $\Gamma = \eta = 4/3$ to arrive at the following inequality
\begin{equation}\label{eq:blue_line}
    \chi_1^* \geq \frac{1-\chi_0}{7}\,.
\end{equation}
This equation defines the region of the parameter space that allows for the existence of a local extremum of the magnetization function $\beta_{\mathrm{m}}$ within the disk.
To show that the previous expression yields indeed a maximum of the function, we can write the radial derivative of $\beta_{\mathrm{m}}$ as 
\begin{equation}\label{eq:max_or_min}
    \partial_R \beta_{\mathrm{m}} \propto (-1 + \chi_0) + 7\chi_1^* \frac{\mathcal{L}_{\mathrm{in}}}{\mathcal{L}}\,,
\end{equation}
where we have substituted the values of $\eta$ and $\sigma$ and we are omitting all positive multiplicative factors. It is relevant to mention that we are using here the fact that $\chi > 1/2$ and $\chi_0 > 1/2$ (see discussion in the main text). Considering these restrictions on the value of $\chi$, that $\mathcal{L}_{\mathrm{in}}/\mathcal{L}(R) \leq 1$ (and strictly decreasing), and that the first term of Eq.~\eqref{eq:max_or_min} is always negative, it is apparent that, if $\chi_1^* < (1-\chi_0)/7$, then $\partial_R \beta_{\mathrm{m}} < 0$ for $R\geq R_{\mathrm{in}}$, as in the standard Komissarov solution~\citep{Komissarov:2006}. In contrast, if $\chi_1^* > (1-\chi_0)/7$, then the derivative will be positive until the radial coordinate reaches a value such that $\mathcal{L}_{\mathrm{in}}/\mathcal{L} = (-1+\chi_0)/(7\chi_1^*)$ and then it will become negative again. Therefore, the extremum we found is a maximum.

\section{Radial profiles of the magnetization function for $\beta_{\mathrm{m,c}} = 100$ and $\beta_{\mathrm{m,c}} = 1$}\label{extra_plots}
We include here the radial profiles of the magnetization function $\beta_{\mathrm{m}}(R)$ at the equatorial plane to complement the results shown in Fig.~\ref{beta_r_fig}. It can be seen here that, apart from some slight changes in the values of $\beta_{\mathrm{m}}/\beta_{\mathrm{m,c}}$ for the different values of $\beta_{\mathrm{m,c}}$, the qualitative behavior of $\beta_{\mathrm{m}}(R)$ is the same. 
\begin{figure*}
\includegraphics[scale=0.08]{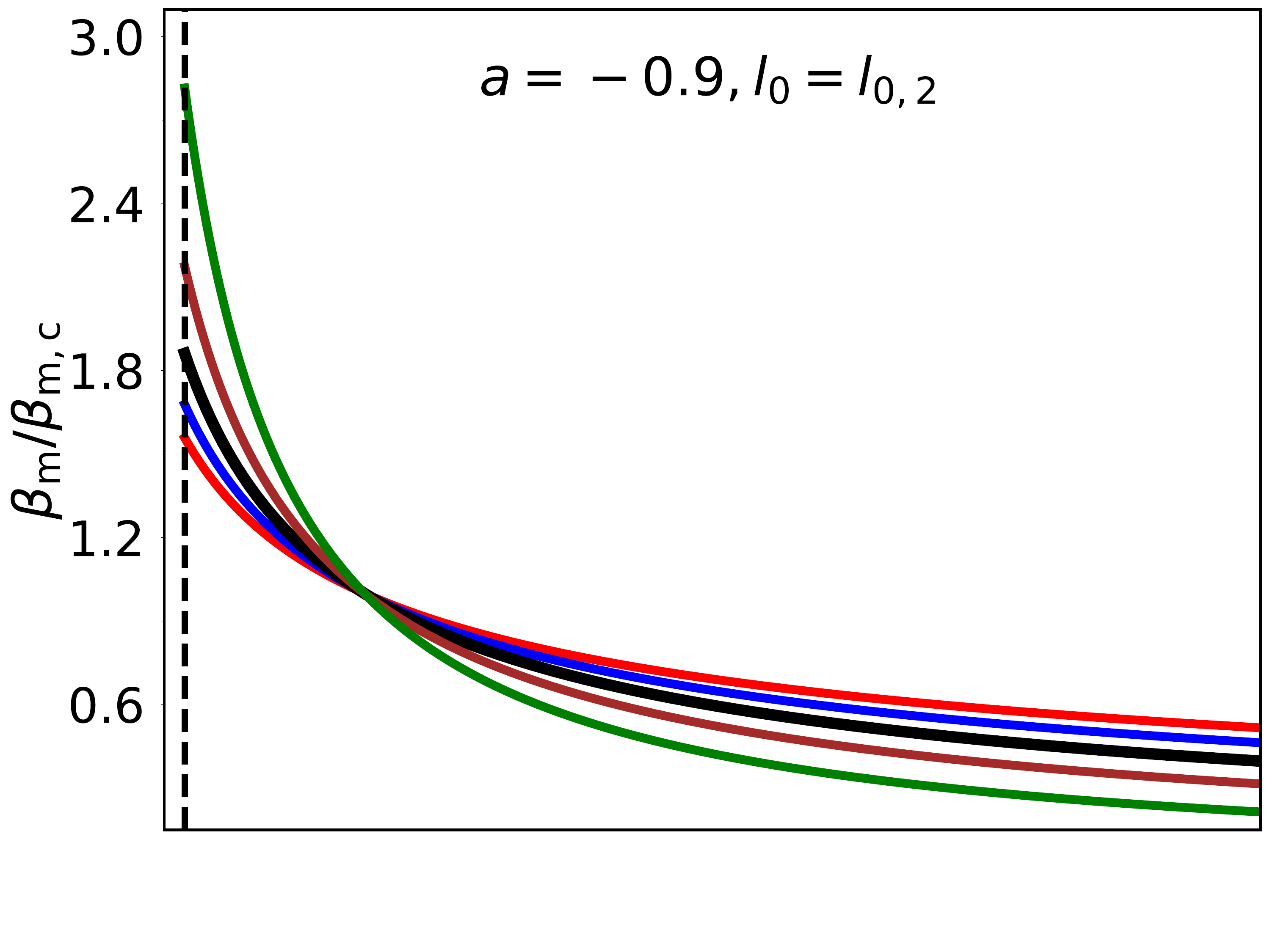}
\hspace{-0.76cm}
\includegraphics[scale=0.08]{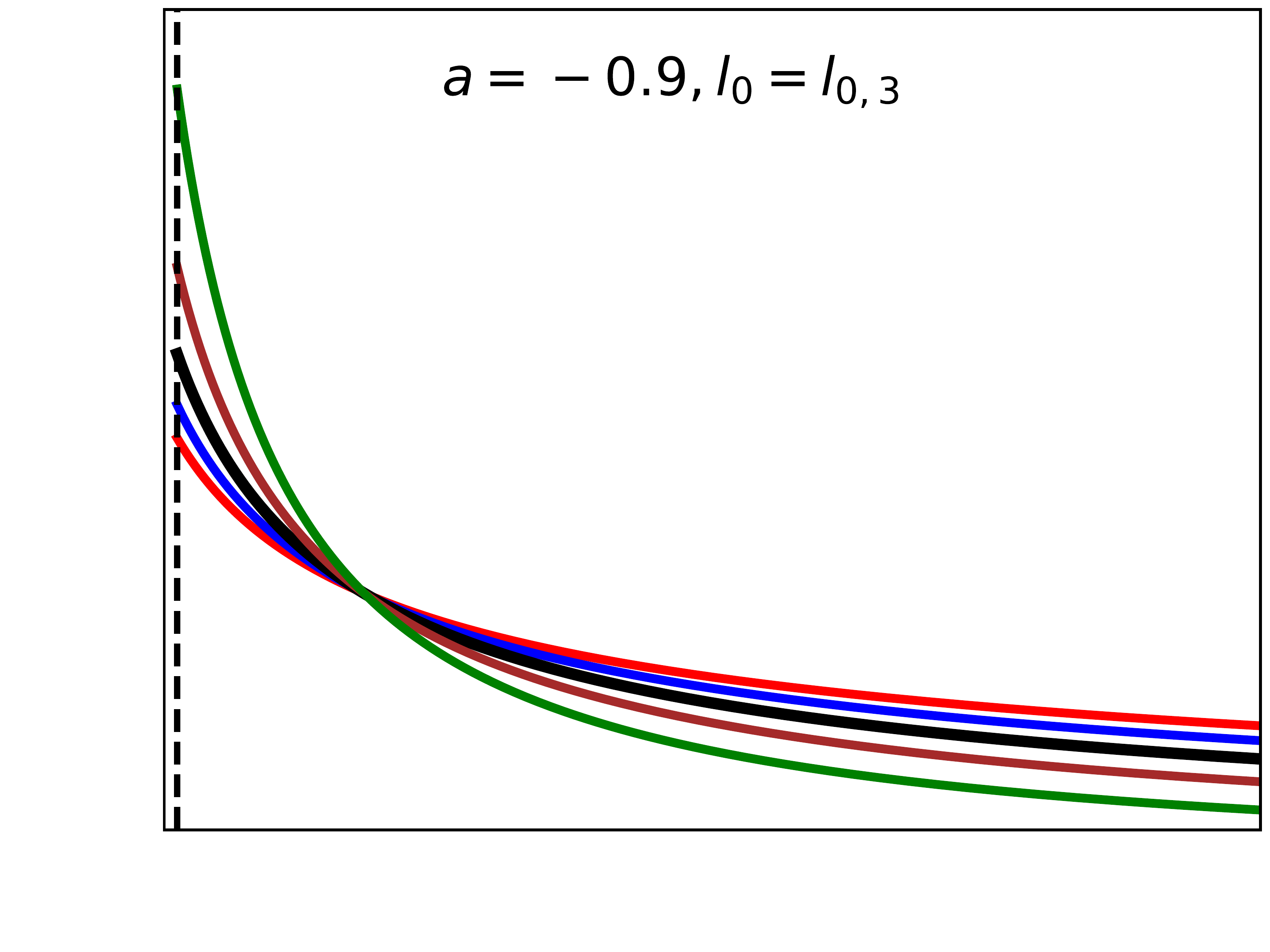}
\hspace{-0.2cm}
\includegraphics[scale=0.08]{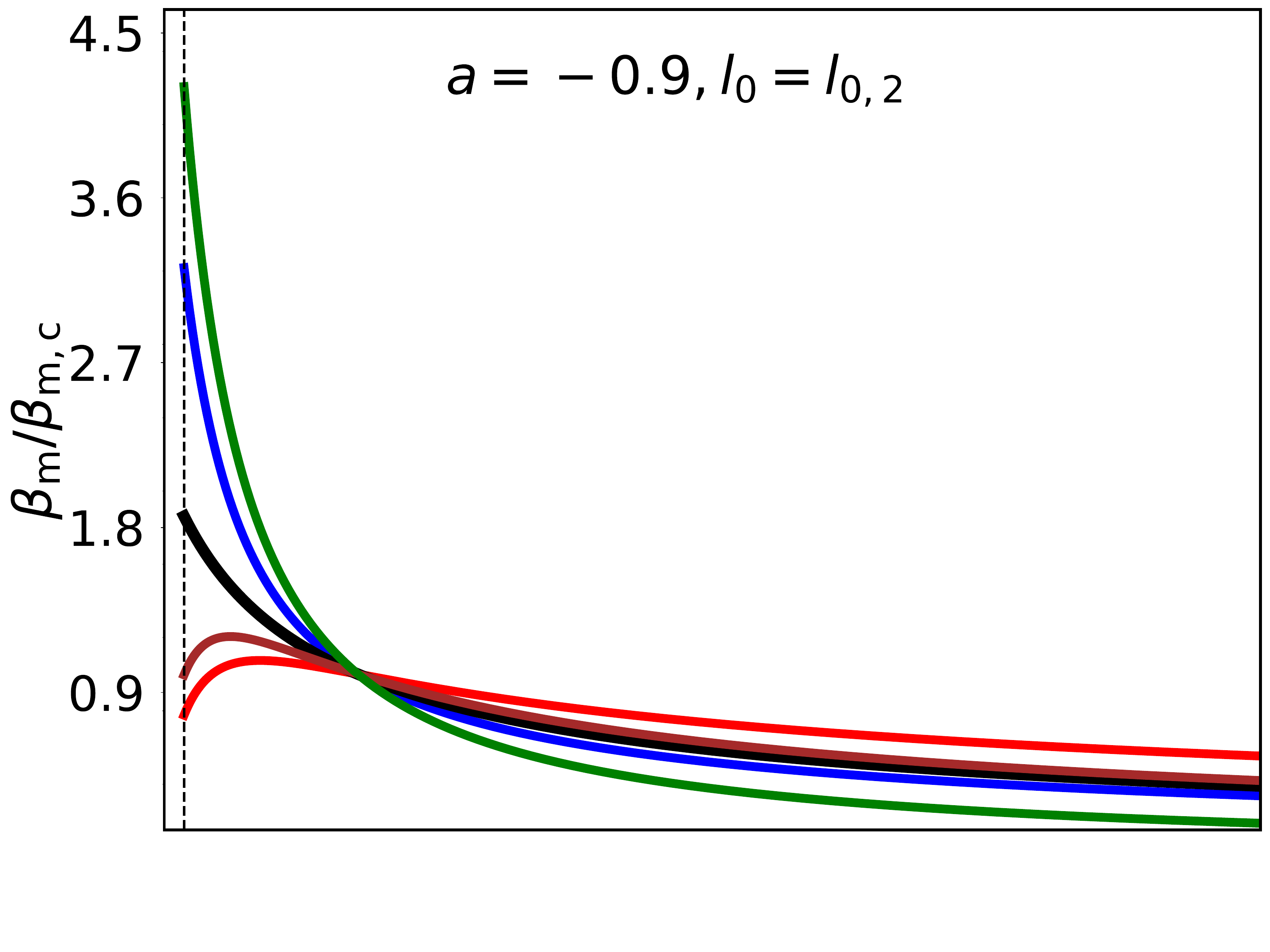}
\hspace{-0.76cm}
\includegraphics[scale=0.08]{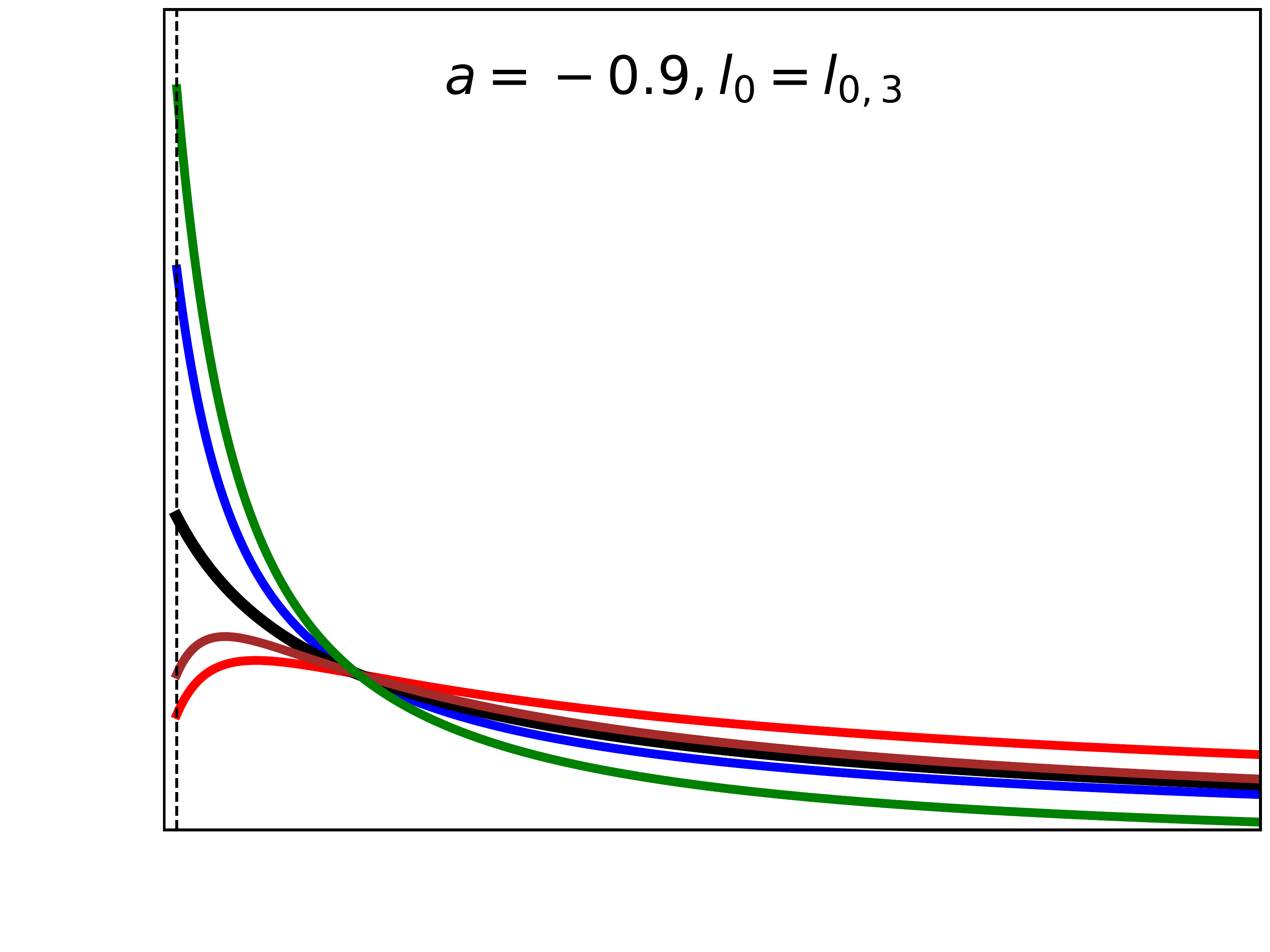}
\vspace{-0.3cm}
\\
\includegraphics[scale=0.08]{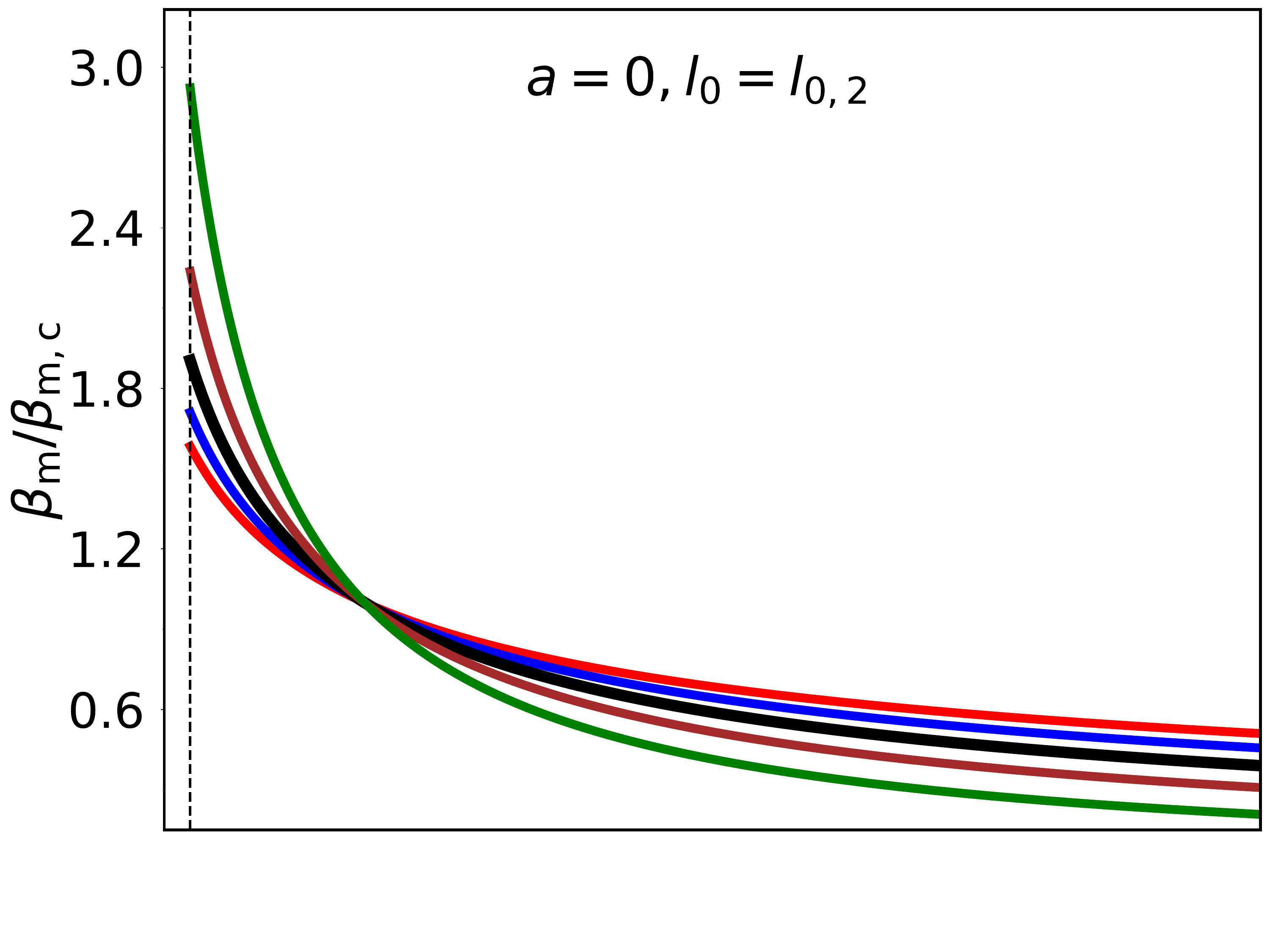}
\hspace{-0.76cm}
\includegraphics[scale=0.08]{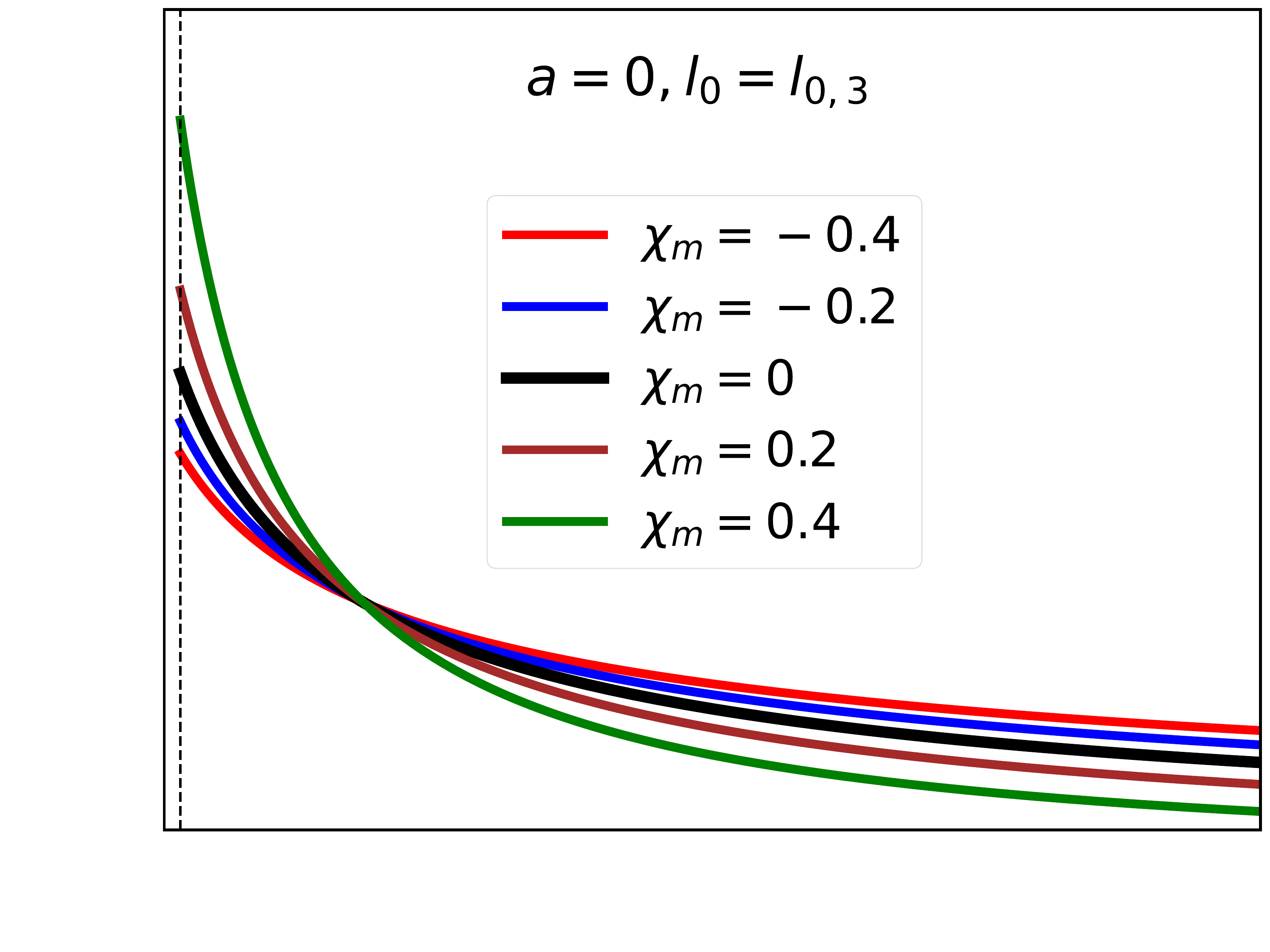}
\hspace{-0.2cm}
\includegraphics[scale=0.08]{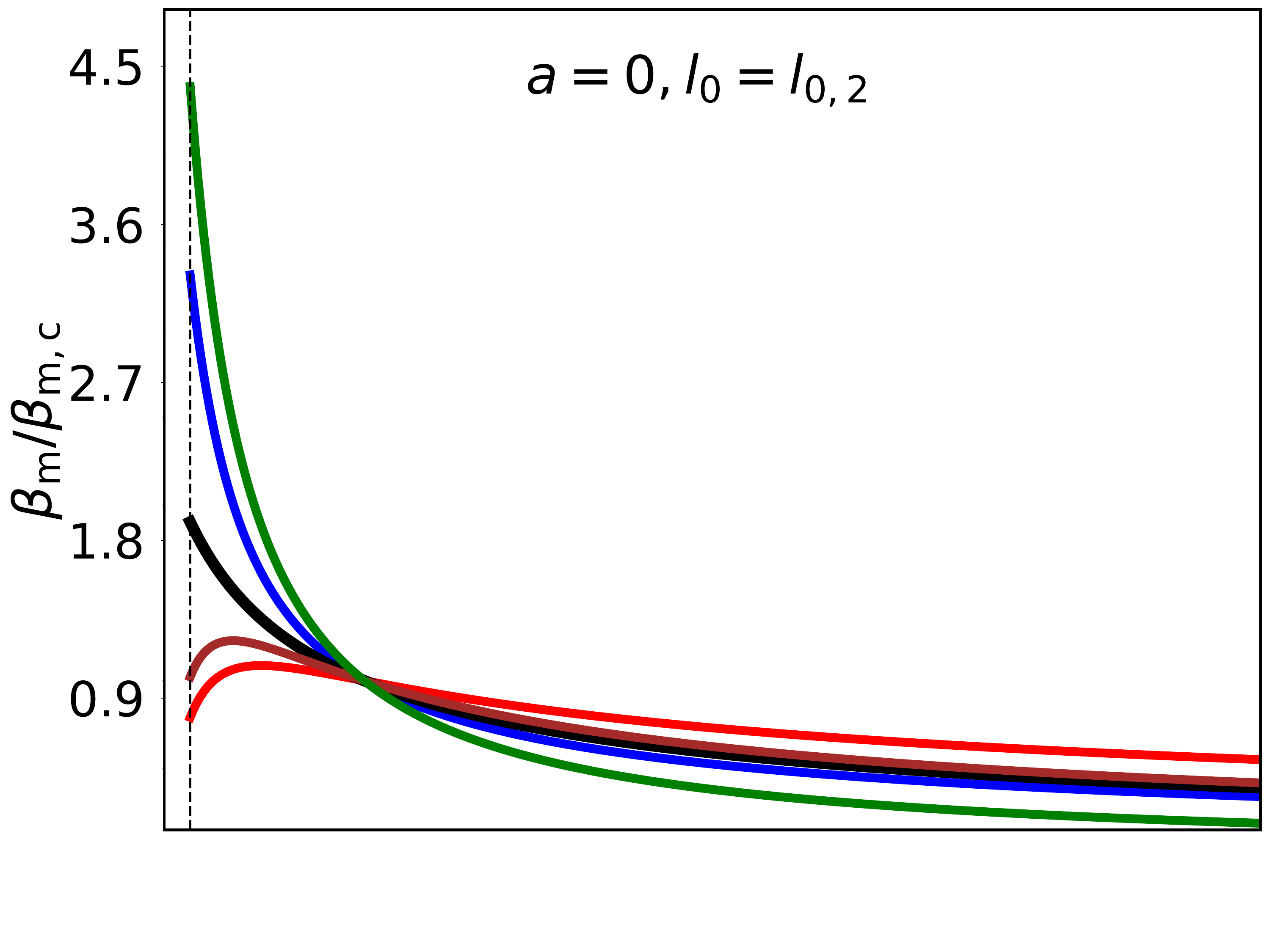}
\hspace{-0.76cm}
\includegraphics[scale=0.08]{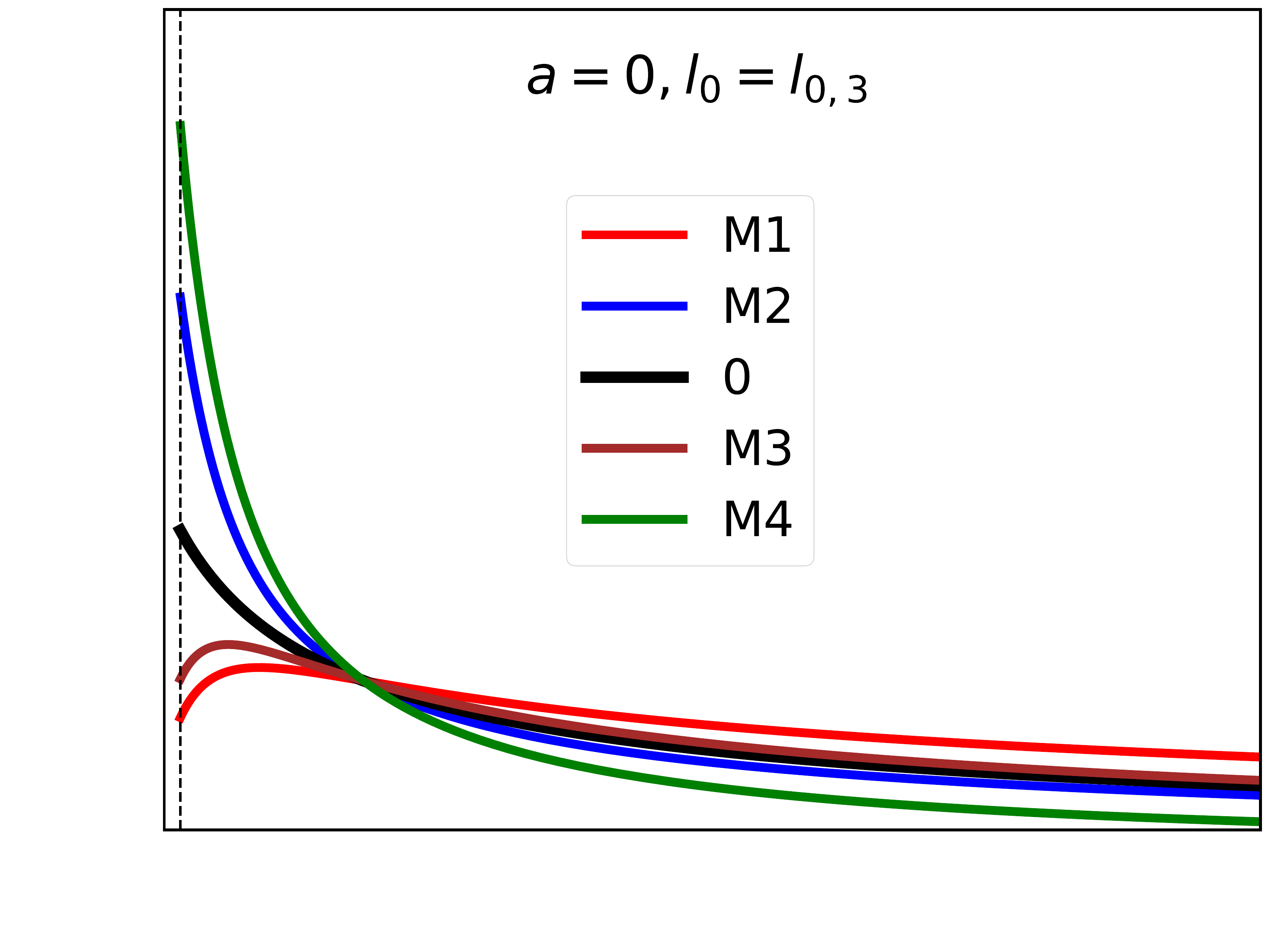}
\vspace{-0.3cm}
\\
\hspace{-0.2cm}
\includegraphics[scale=0.08]{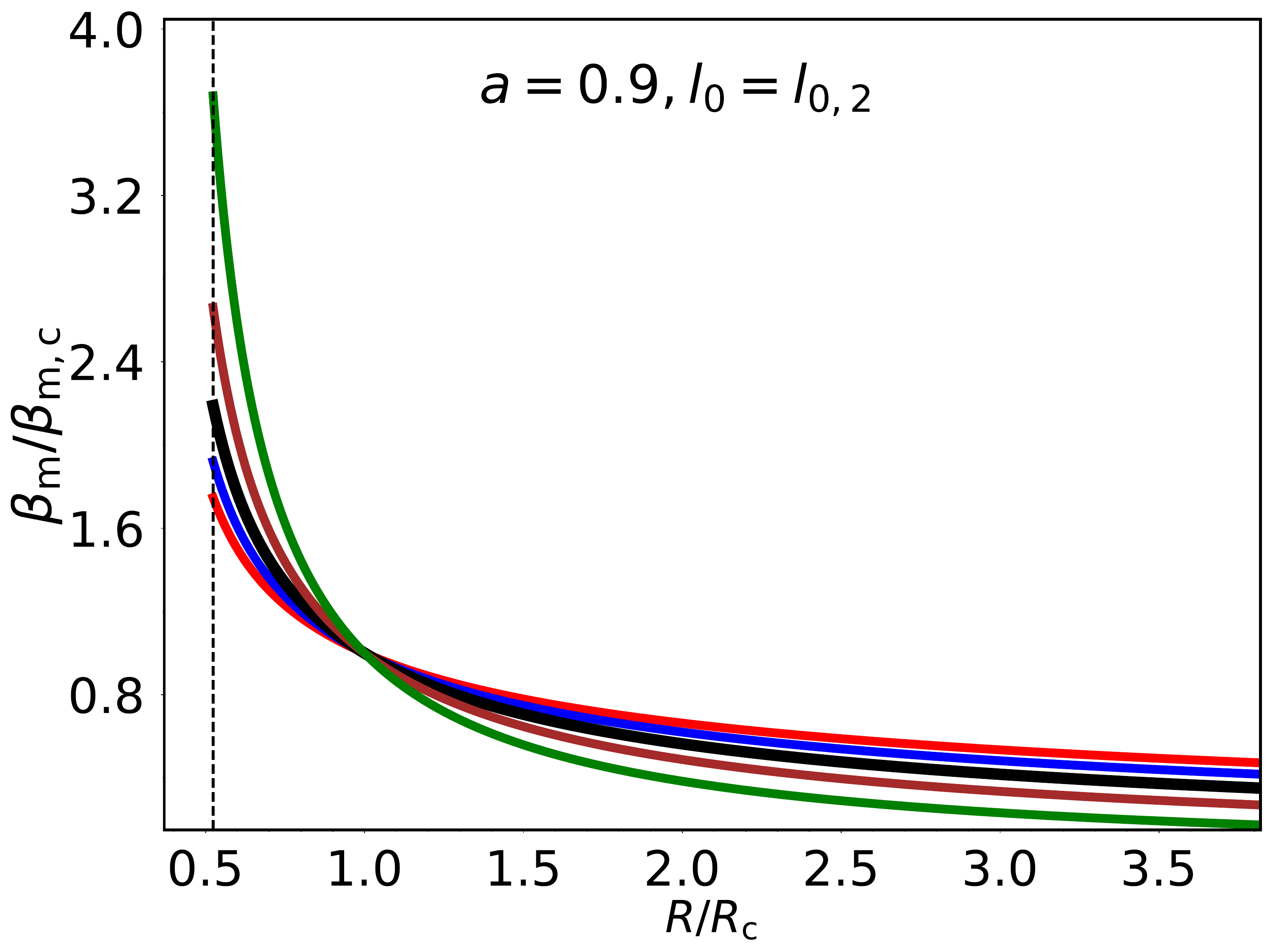}
\hspace{-0.76cm}
\includegraphics[scale=0.08]{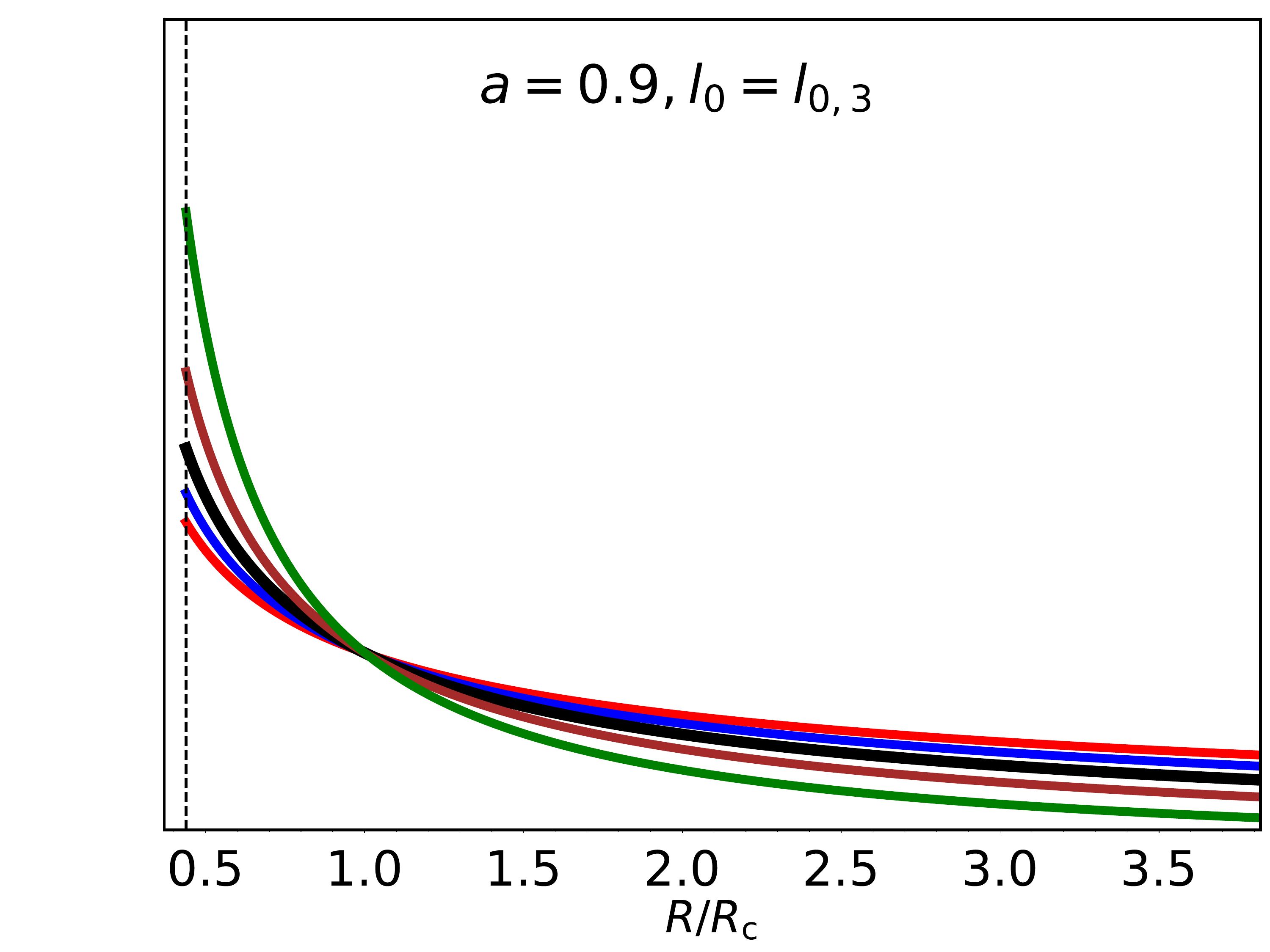}
\hspace{-0.2cm}
\includegraphics[scale=0.08]{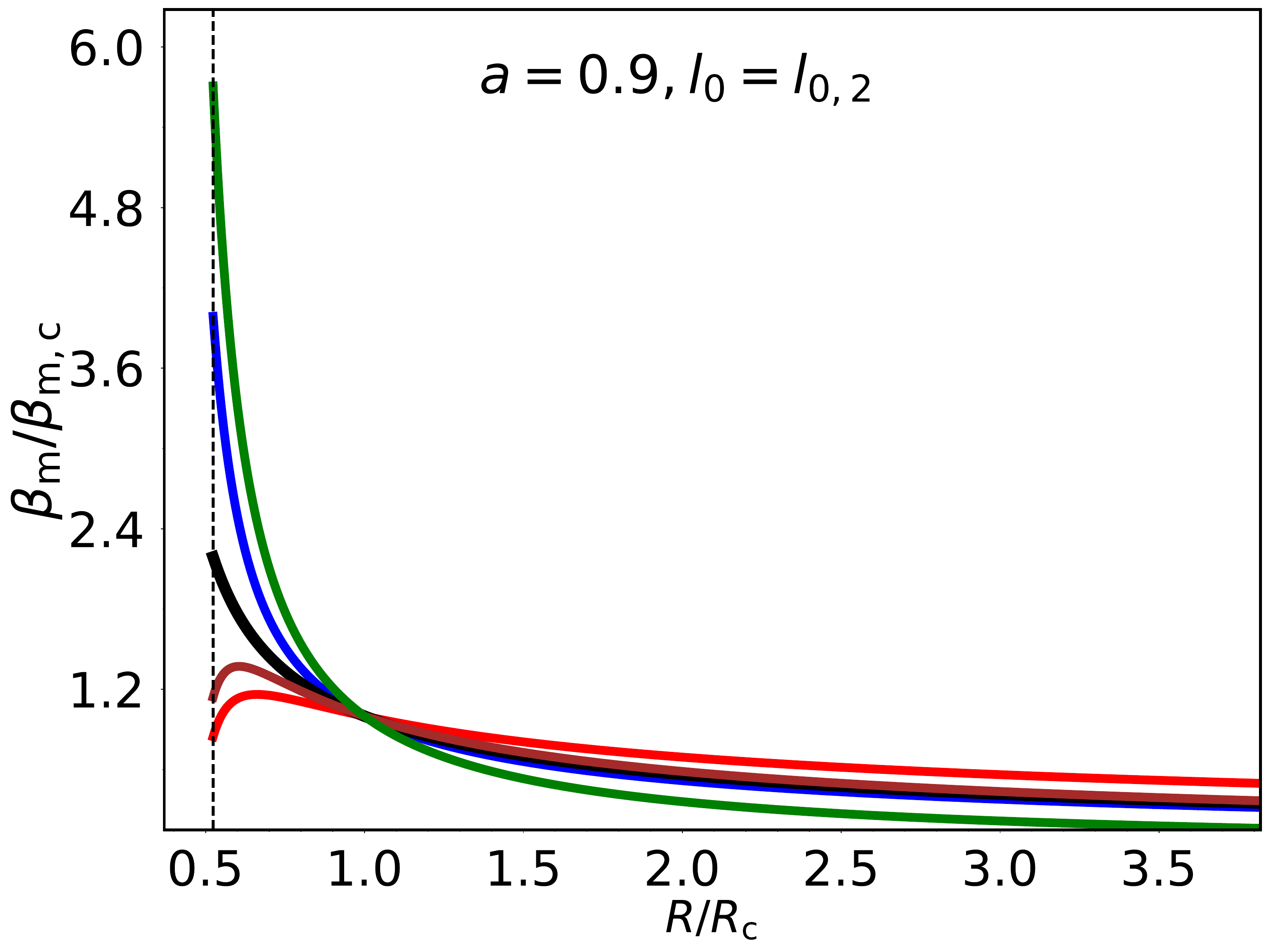}
\hspace{-0.76cm}
\includegraphics[scale=0.08]{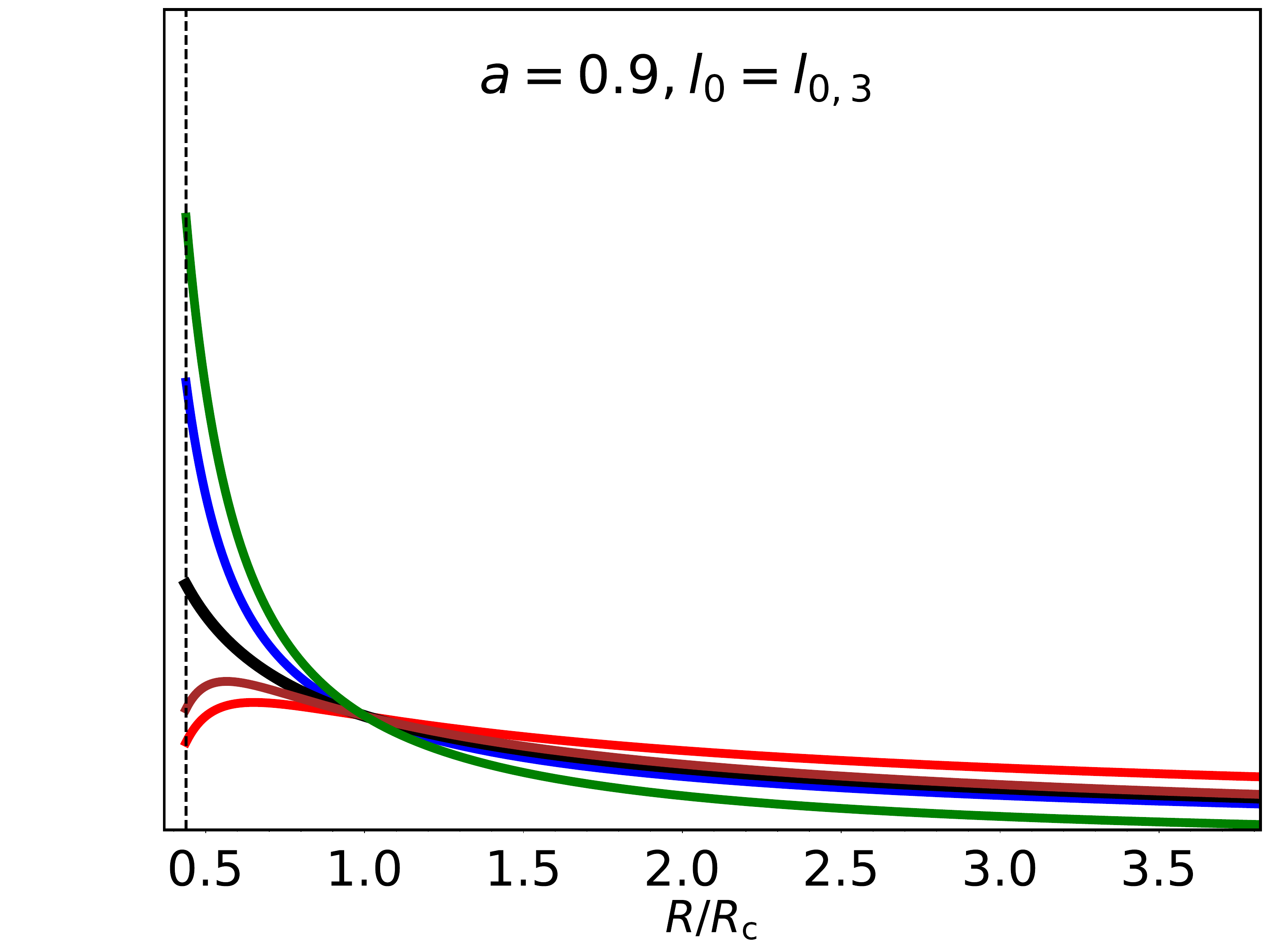}
\caption{Normalized radial profiles of the magnetization function at the equatorial plane, $\beta_{\mathrm{m}}(R)$, for magnetization parameter at the center $\beta_{\mathrm{m, c}} = 1$. The layout of this figure is the same as Fig.~\ref{beta_r_fig}.}
\label{beta_r_fig_1}
\end{figure*}

\begin{figure*}
\includegraphics[scale=0.08]{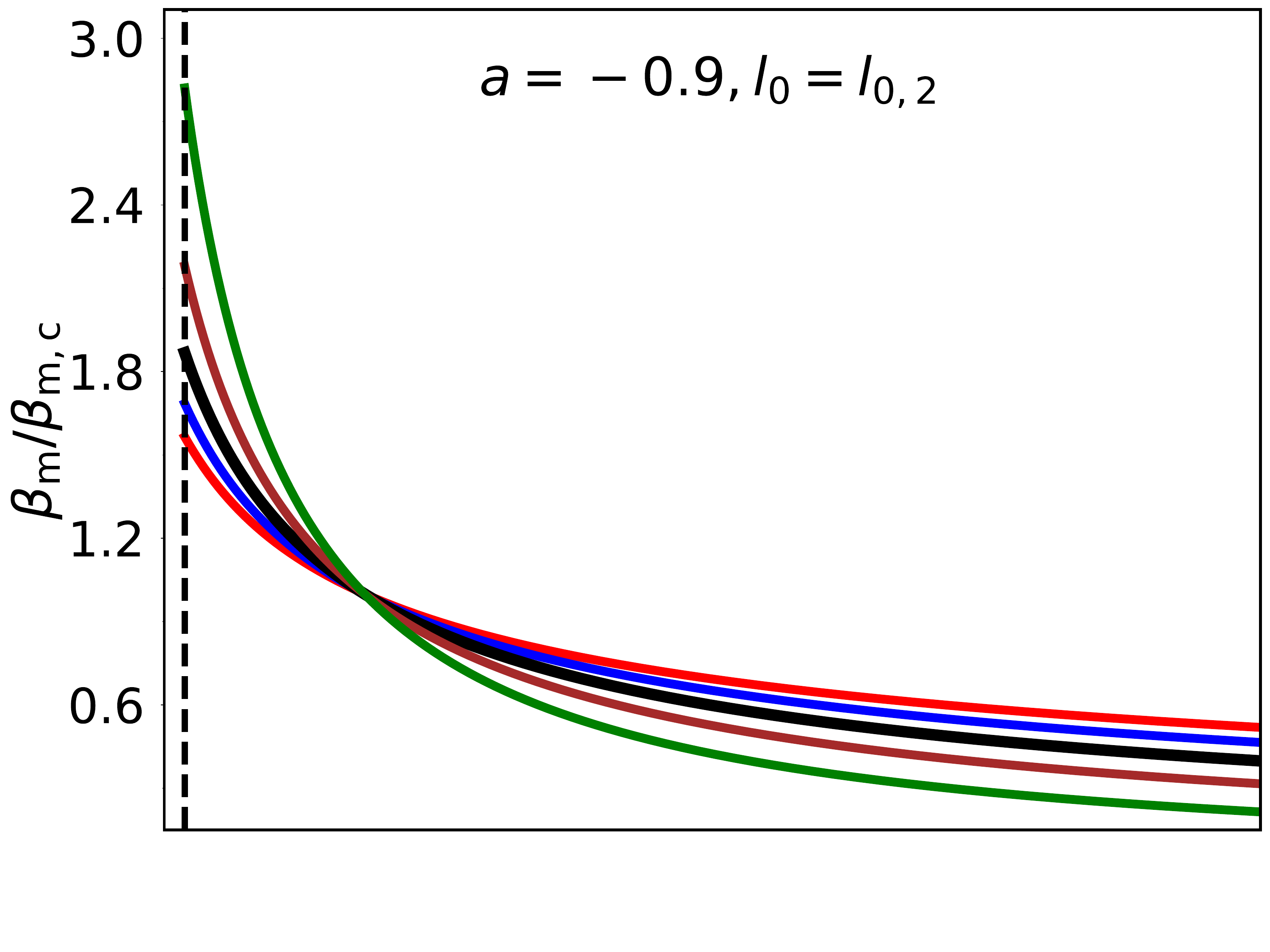}
\hspace{-0.76cm}
\includegraphics[scale=0.08]{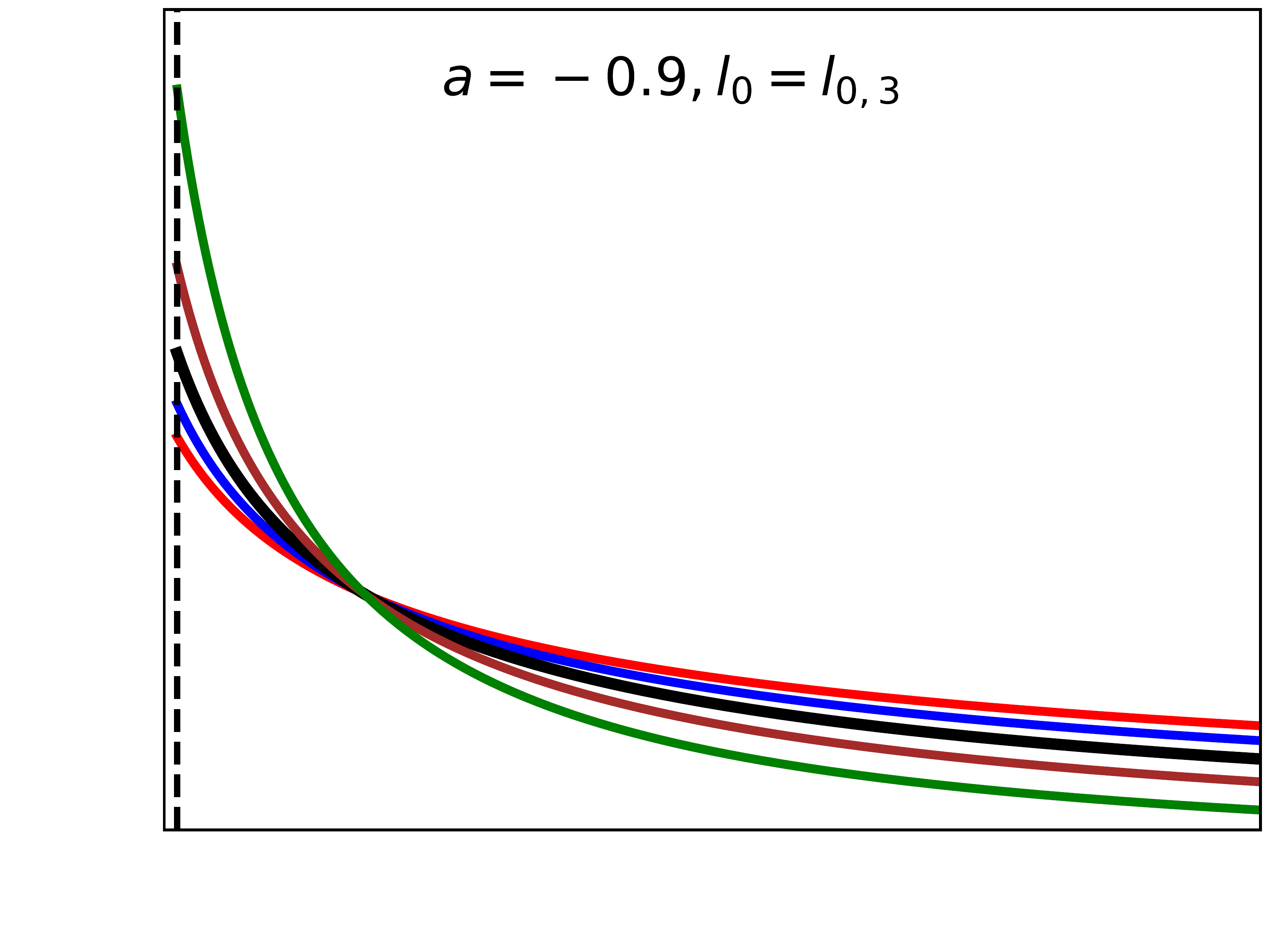}
\hspace{-0.2cm}
\includegraphics[scale=0.08]{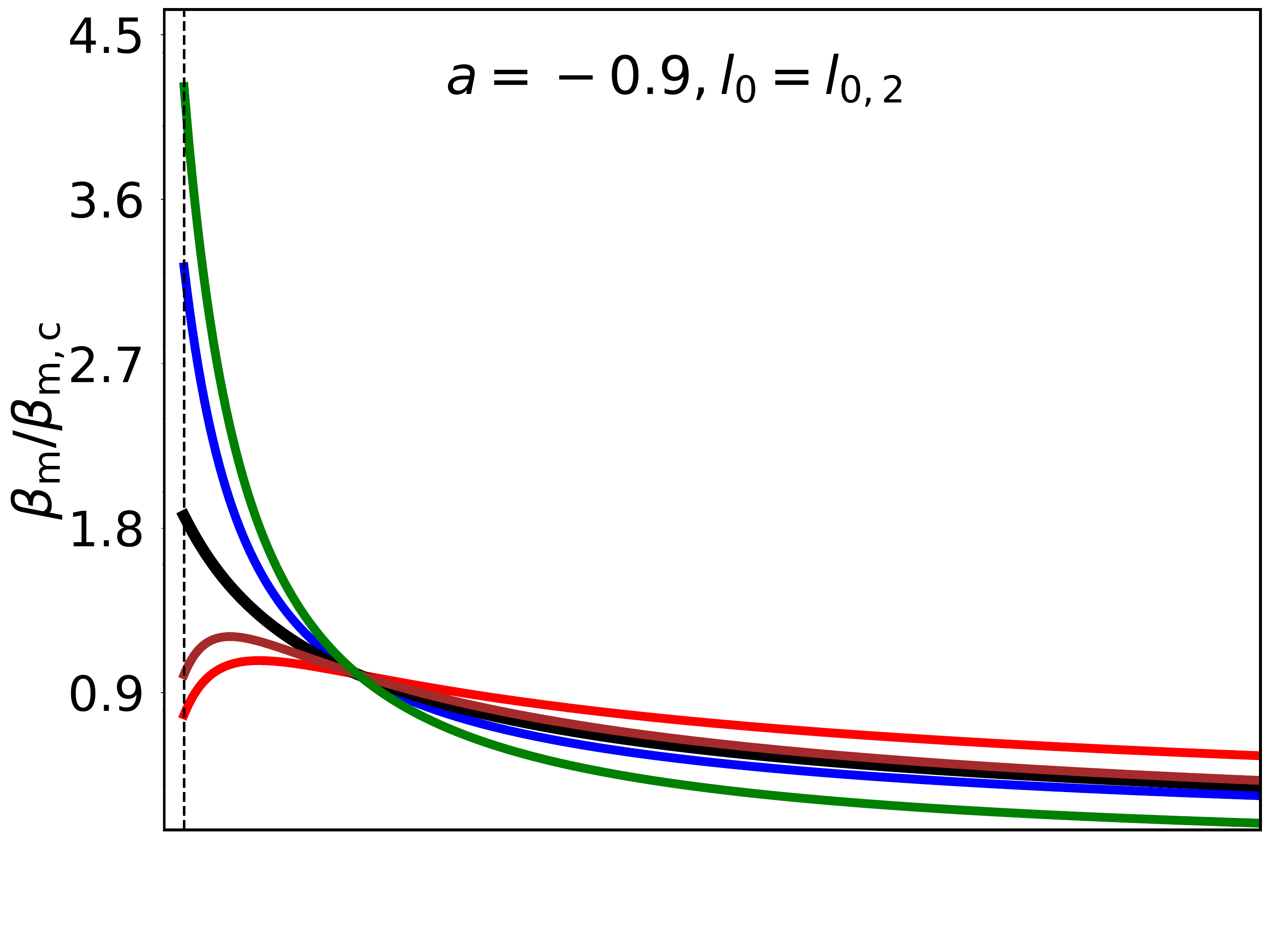}
\hspace{-0.76cm}
\includegraphics[scale=0.08]{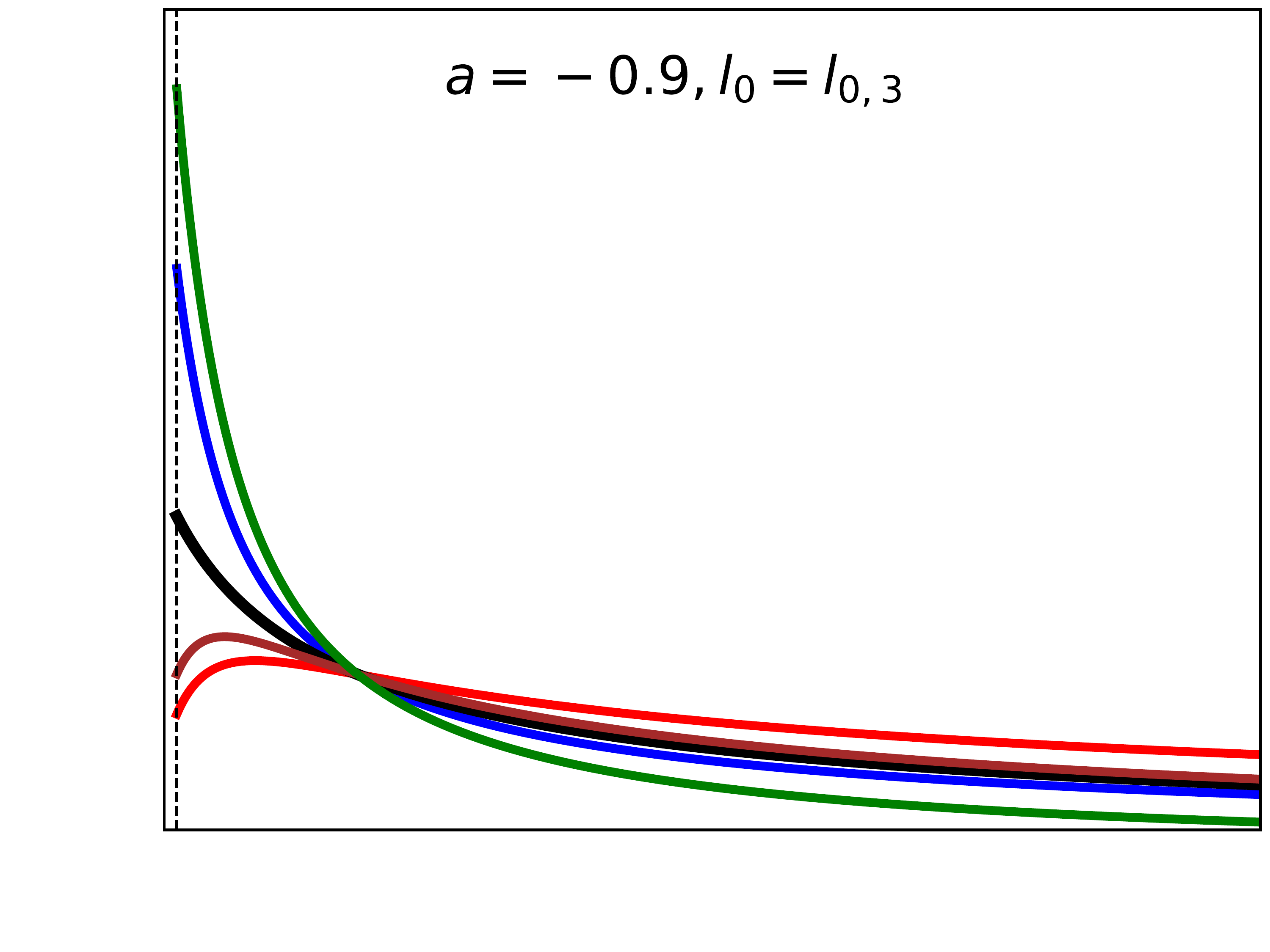}
\vspace{-0.3cm}
\\
\includegraphics[scale=0.08]{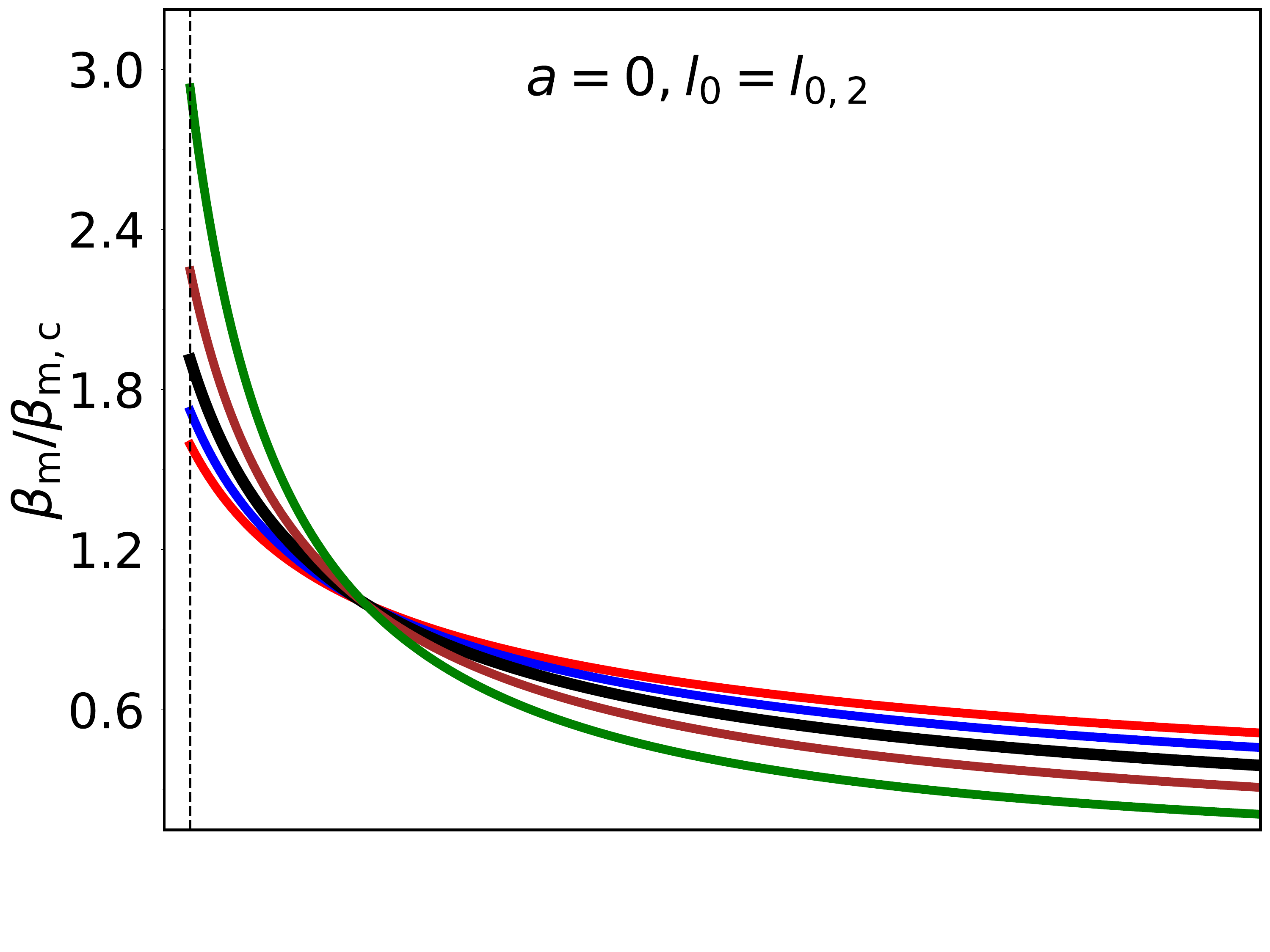}
\hspace{-0.76cm}
\includegraphics[scale=0.08]{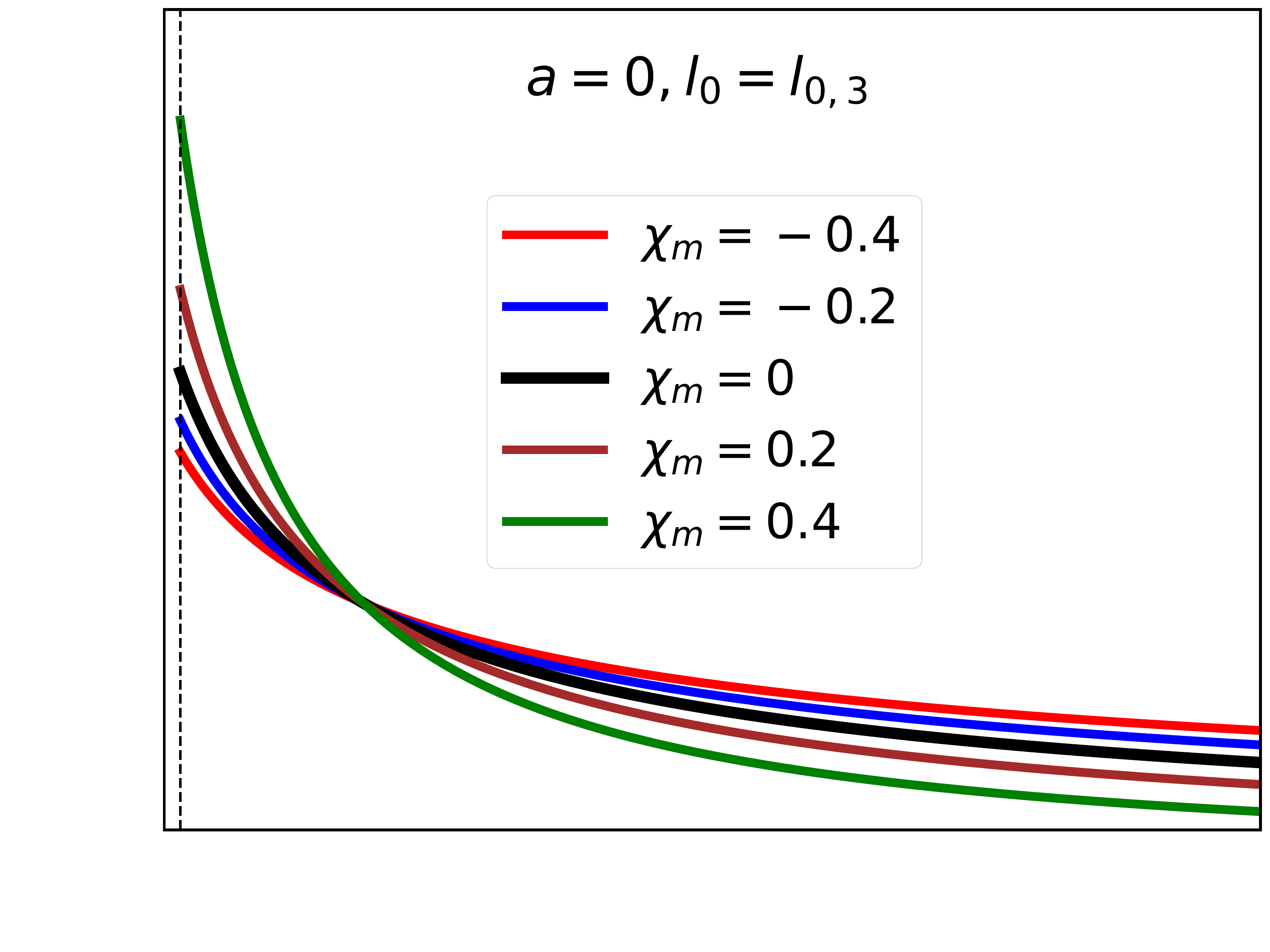}
\hspace{-0.2cm}
\includegraphics[scale=0.08]{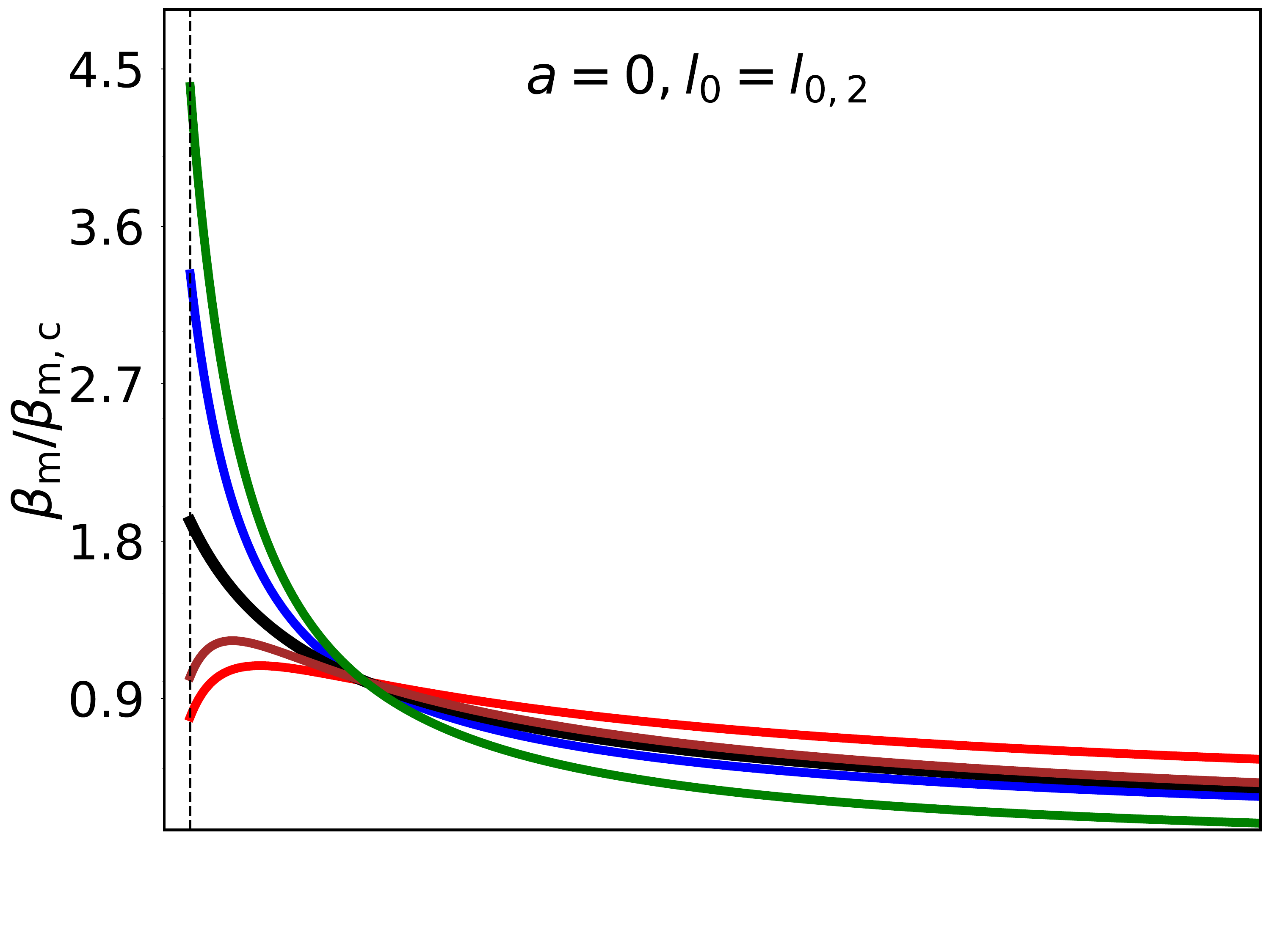}
\hspace{-0.76cm}
\includegraphics[scale=0.08]{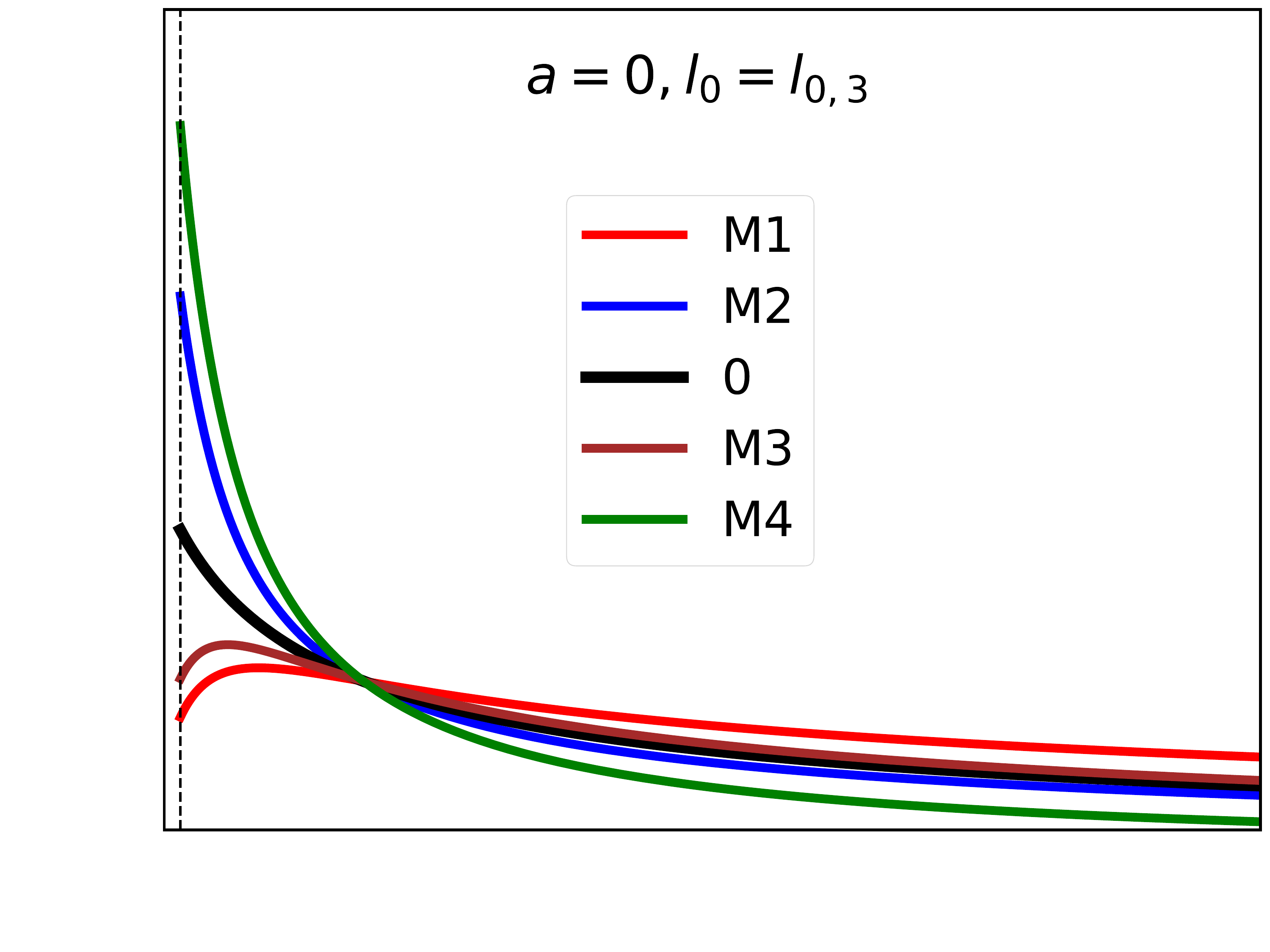}
\vspace{-0.3cm}
\\
\hspace{-0.2cm}
\includegraphics[scale=0.08]{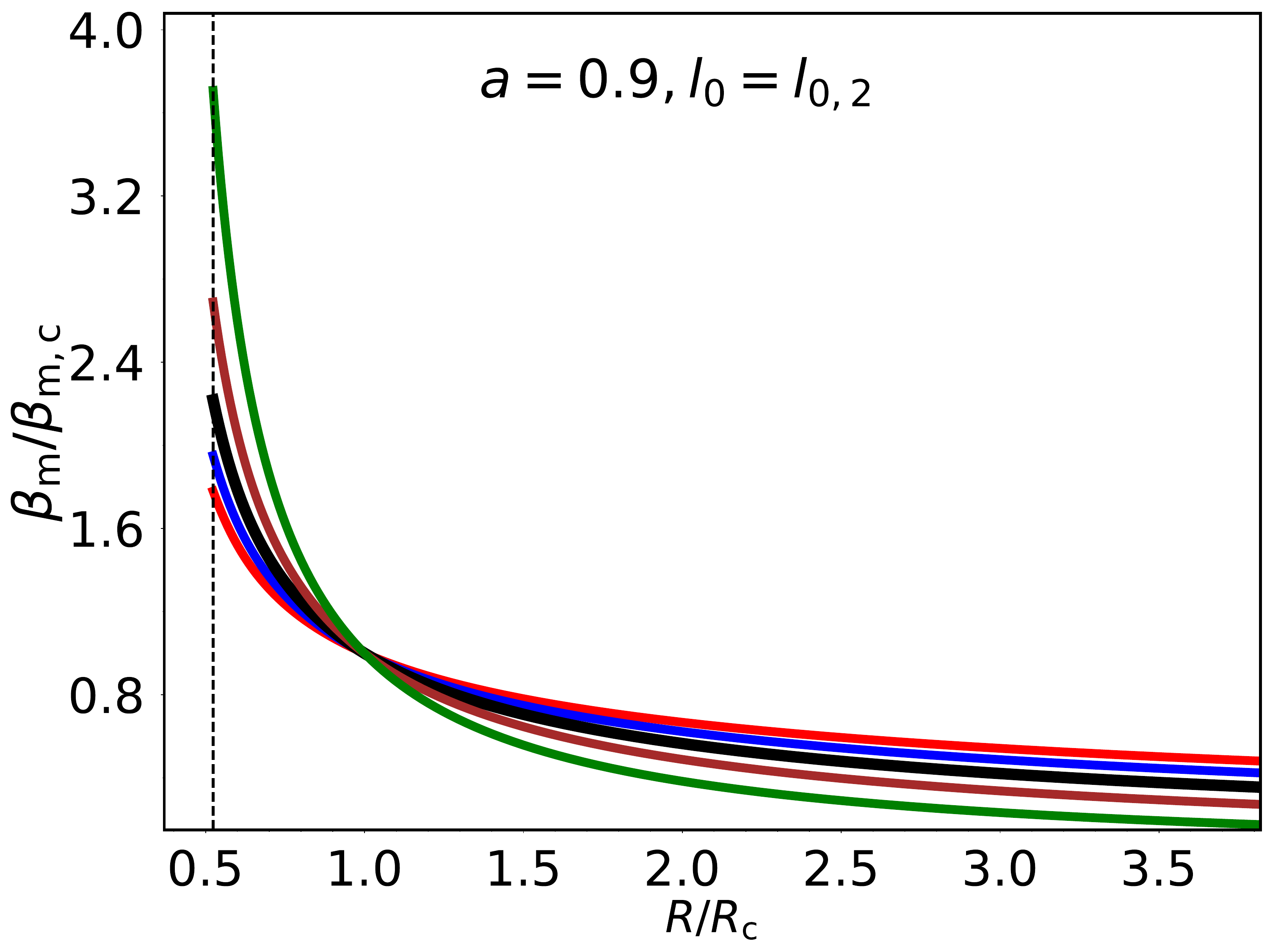}
\hspace{-0.76cm}
\includegraphics[scale=0.08]{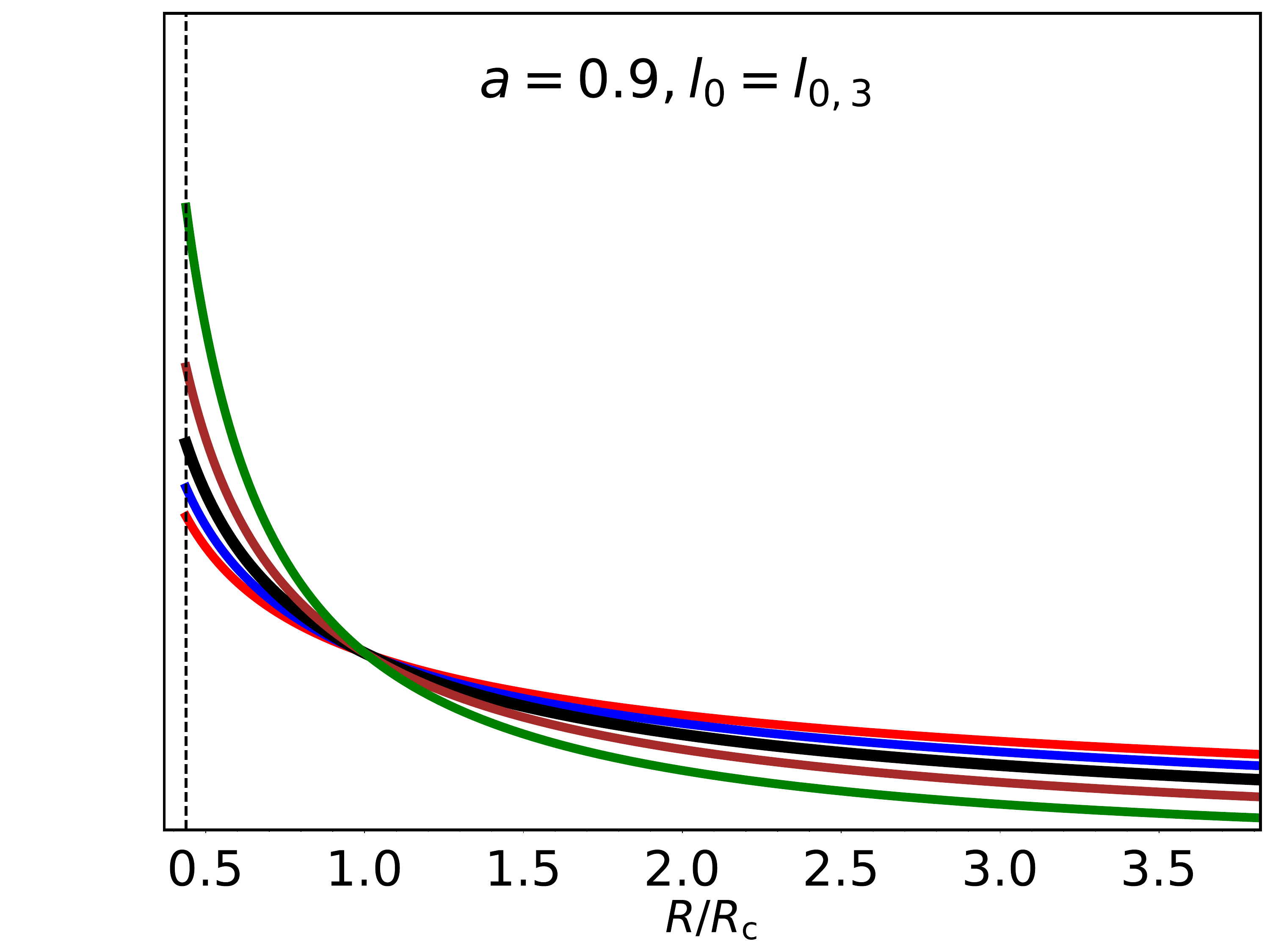}
\hspace{-0.2cm}
\includegraphics[scale=0.08]{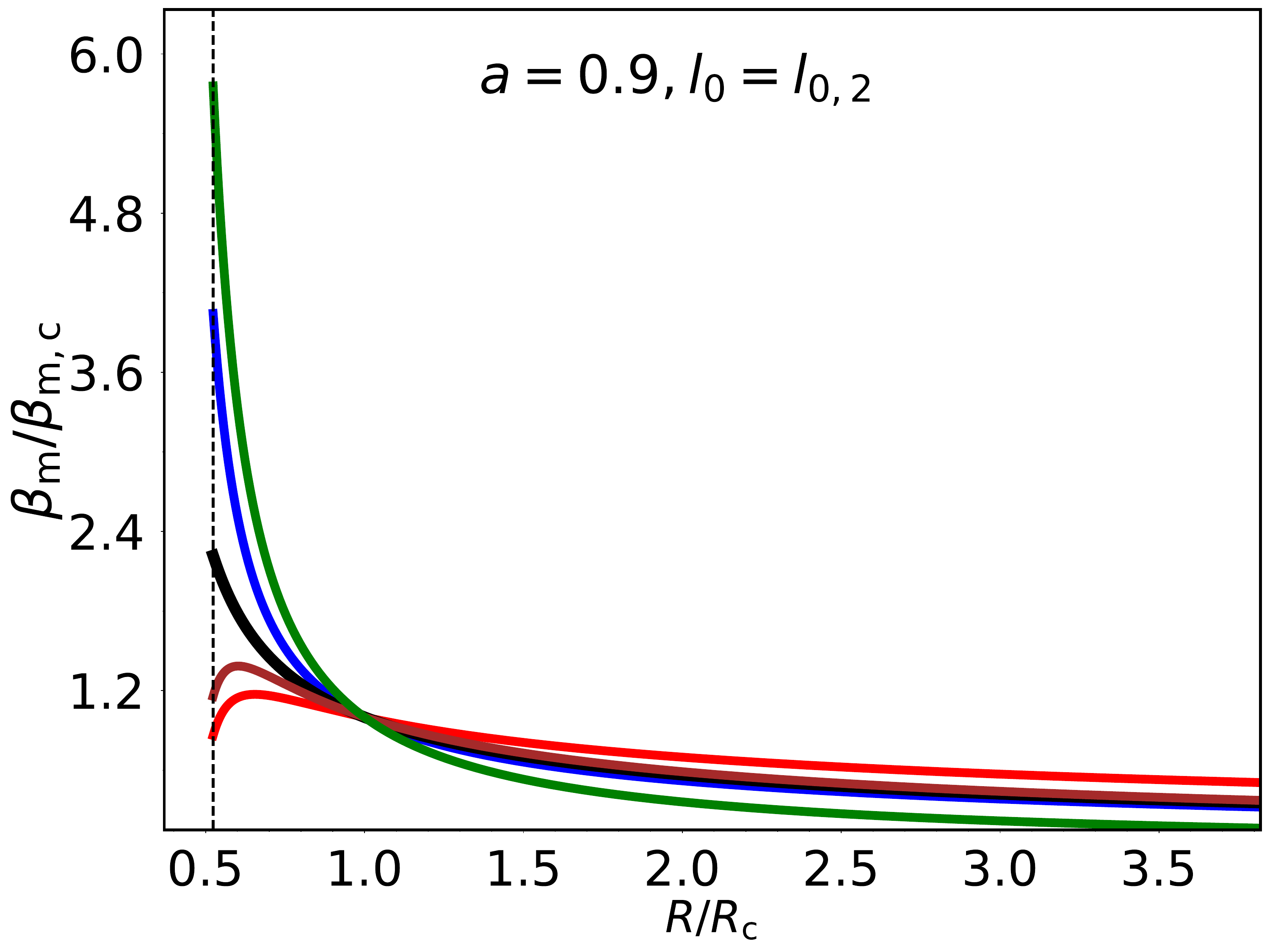}
\hspace{-0.76cm}
\includegraphics[scale=0.08]{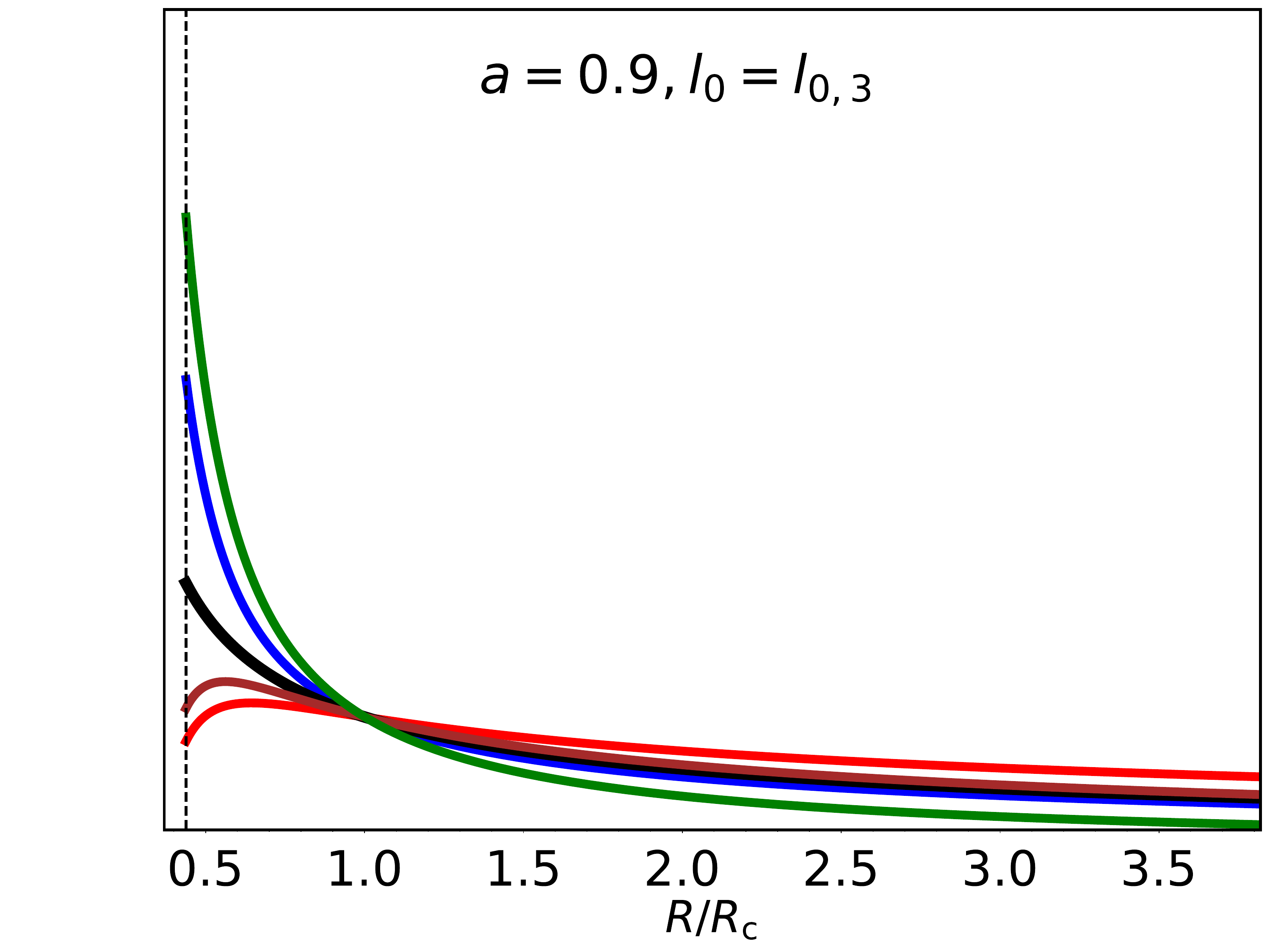}
\caption{Same as in Fig.~\ref{beta_r_fig_1} but for $\beta_{\mathrm{m, c}} = 100$.}
\label{beta_r_fig_100}
\end{figure*}

\end{appendix}

\bibliography{apssamp}% Produces the bibliography via BibTeX.

\end{document}